\documentclass[twocolumn, tighten, times]{aastex62}

\usepackage{amsmath,amssymb}
\usepackage{rotating}
\usepackage{enumitem}
\usepackage{graphicx}
\usepackage{CJKutf8}
\usepackage{float}

\newcommand{\sevenrm}{\rm\scriptsize}

\begin{document}

\title{\Large \bf
The Dusty Heart of NGC 4151 Revealed by \boldmath$\lambda\sim$1--40~$\mu m$ Reverberation
Mapping and Variability: A Challenge to Current Clumpy Torus Models 
}

\author[0000-0002-6221-1829]{Jianwei Lyu (\begin{CJK}{UTF8}{gbsn}吕建伟\end{CJK})}
\affiliation{ Steward Observatory, University of Arizona,
933 North Cherry Avenue, Tucson, AZ 85721, USA}
\author[0000-0003-2303-6519]{George H. Rieke}
\affiliation{ Steward Observatory, University of Arizona,
933 North Cherry Avenue, Tucson, AZ 85721, USA}

\begin{abstract}
	We probe the dusty environment of the archetypical type-1 AGN in NGC
	4151 with comprehensive IR reverberation mapping over several decades,
	in J ($\sim1.22~\mu$m), H ($\sim1.63~\mu$m), K ($\sim2.19~\mu$m), L
	($\sim3.45~\mu$m) and N-band ($\sim10.6~\mu m$), plus multiple
	measurements at 20--40~$\mu$m. At 1--4~$\mu$m, the hot dust
	reverberation signals come from two distinct dust populations at
	separate radii ($\sim$0.033 pc and $\sim$0.076 pc), with
	temperatures of $\sim$1500--2500 K and $\sim$900--1000 K, consistent
	with the expected properties of sublimating graphite and silicate dust
	grains. { The domination of the torus infrared output by carbon 
	and silicate grains near their sublimation temperatures and radii may account for the 
	general similarity of AGN near-IR spectral energy distributions.}
	The torus inner edge defined by the hottest dust remains at
	roughly the same radius independent of the AGN optical luminosity over
	$\sim$ 25 years.  The emission by hot dust warmed directly by the
	optical/UV AGN output has increased gradually by $\sim$ 4\%/year,
	indicating a possibly growing torus. A third dust component at $\sim$
	700 K does not seem to participate directly in the IR reverberation
	behavior and its emission may originate deep in the circumnuclear
	torus.  We find a reverberation signal at $\sim10~\mu m$ with an
	inferred radius for the warm dust of $\sim$2.2--3.1 pc.  The lack of
	variability at 20--40~$\mu$m indicates the far-IR emission comes from
	even more extended regions.  The torus properties revealed by dust
	reverberation analysis are inconsistent with predictions from pure
	clumpy torus models. Instead, the longer wavelength emission possibly 
	originates in a flared torus { or the polar wind}. 
\end{abstract}


\section{Introduction}

Circumnuclear dusty structures and their accompanying obscuration are essential
characteristics of the active galactic nucleus (AGN) phenomenon
\citep[e.g.,][]{Antonucci1993, Urry1995, Netzer2015} and are possibly tightly
linked to the growth and evolution of the accreting supermassive black holes
(SMBH) \citep[e.g.,][]{Hopkins2012, Wada2012,Wang2017}. Current models for AGN
obscuration are centered on an optically-thick torus-like circumnuclear
structure that absorbs optical/UV light from the central engine and reradiates
it at infrared (IR) wavelengths \citep[e.g.,][]{Fritz2006,
nen08a,nen08b,Stalevski2016} and have achieved some success in matching the
infrared observations \citep[e.g.,][]{nen08b, Ramos2009,Ramos2011, ah2011}. 
{ Dusty winds launched from this torus may also play a major role \citep{Honig2013}.}
However, due to the lack of spatially-resolved direct observations, the
inferred properties of the torus, such as its geometric structure and temporal
evolution, are highly ambiguous  \citep[e.g.,][]{Gonzalez-Martin2019}.

{Luckily, this compact dust structure can be probed by time-domain
observations. Depending on the AGN luminosity and observed wavelength,} the
circumnuclear torus  ranges from light-weeks to many light-years in size. {
When the energy output of the nucleus varies, the
signals we receive encode the torus structure as a result of the light travel
times from the central engine to the reradiating zone and then to us.} Thus,
cross-correlation analysis of the AGN optical and IR light curves, i.e. dust
reverberation analysis,  provides a powerful tool to pin down the torus
structures \citep[e.g.,][]{Clavel1989, Barvainis1992, Suganuma2006, kosh14,
Lyu2019}. Most such studies are focused on the K-band (at 2.2 $\mu$m), which
can be easily accessed with ground-based observations \citep[e.g.,][]{kosh14,
Minezaki2019}. Multi-band reverberation behavior can reveal additional aspects
of the circumnuclear torus structure (e.g., smooth, clumpy, some combination).
However, there are relatively few such explorations and the available studies show a
range of results \citep[e.g.,][]{glass04, okny14, Lyu2019}. It is also
hypothesized that the circumnuclear torus should change structure in reaction
to changing nuclear luminosity resulting in changes in the reverberation timing, but again there is a range of results
\citep[e.g.,][]{ okny14, Schnulle2015}. 

As one of the brightest and nearest type-1 AGN, NGC 4151 has been
extensively observed at multiple wavelengths, offering the opportunity to
explore these issues with  optical and IR data collected over 40 years. It has
been, by far, the most extensively studied AGN by means of IR  reverberation
mapping \citep[e.g.,][]{okny93, okny14, okny19,kosh14, Schnulle2015}.
These studies have generally used a subset of the available data and have often reached contradictory results \citep[e.g.,][]{Koshida2009,honig2011}. 
To clarify the inner torus properties in NGC 4151, we have conducted a
comprehensive dust reverberation mapping analysis over 20--30
years, combining all of the relevant data in the literature and archives.  Our more
comprehensive approach lets us reconcile some of the discrepant conclusions
reached previously.  More importantly, we will introduce and demonstrate new analysis approaches  to probe the circumnuclear
torus properties in greater depth than has been possible previously.

In addition, all the dust reverberation mapping studies published so far only
probe the very inner part of the torus at $\lambda\sim$1--4~$\mu$m. We will report the first IR reverberation analysis of the dusty component of an
AGN at $\sim10~\mu$m,  providing
 constraints on the warm dust component that dominates the IR energy
output. In fact, NGC 4151 is likely the only object that allows this
kind of study thanks to its significant UV/optical light variations and the
$\sim$30-year coverage of repeated  $\sim$10~$\mu$m observations. We will also
report the lack of long-term variability of this AGN at 20--24~$\mu$m and
34--37~$\mu$m, providing insights into the outer regions of the infrared source 
when combined with the 10 $\mu$m dust reverberation signals.

With a comprehensive analysis of the dust reverberation signals in the J, H, K,
L and N bands and constraints on IR variability at 20--40~$\mu$m over 20-30
years, this paper will: (1) give insights to the torus radial
and vertical structures as a function of wavelength; (2) explore possible
temporal evolution of the torus structures; and (3) provide tests of the
observed dust reverberation signals against the predictions of current torus
models. Integrating these results with observational constraints from other
wavelengths/methods and physical insights obtained from previous theoretical
studies, we can establish a complete picture of the dusty heart of NGC 4151 at
diverse physical scales.

This paper is organized as follows. In Section~\ref{sec:data}, we describe the
data collection and how the light curves have been constructed.
Section~\ref{sec:analysis} explores the physical origins of various features in
the AGN IR light curves and retrieves the dust reverberation signals as a
function of wavelength. We also explore the
possible time-evolution of the AGN torus properties. In Section~\ref{sec:discussion},  { by
integrating our reverberation results with other observations, we provide
constraints on the torus dust temperatures, vertical structures, and radial
sizes as a function of observed wavelength and the possible dust grain
properties. We also compare these results and popular torus models, specifically 
showing that the 10 $\mu$m reverberation behavior is not as predicted by current clumpy torus ones.}
A final summary is provided in Section~\ref{sec:insight}.

We adopt a distance of {15.8 Mpc} to NGC 4151 \citep{Yuan2020}.

\section{Light Curve Construction}\label{sec:data}

\subsection{Sources of Time-series Data}

Optical photometry of NGC 4151 was taken from the literature and as available in Vizier, i.e.  \citet{lyuty99, dor01, lyuty05,
ser05, Shapovalova2010, Roberts2012, okny13}\footnote{In some references, the author name ``Oknyansky'' 
has been transliterated as ``Oknyanskij''. We have adopted ``Oknyansky'' throughout the main 
text of the paper but used ``Oknyanskij" in the reference list when necessary so 
the citation can be readily found in the Astrophysics Data System (ADS).
} ; \citet{kosh14}\footnote{We transformed the
V photometry in \citet{kosh14} to B,  based on colors from the other
programs.}; \citet{okny18, shom19}\footnote{The errors on the nuclear B-band flux have been estimated by detrending (linearly) the dependence of B$-$V vs. 0.5 $\times$ (B+V), evaluating the scatter after detrending, and then correcting for the galaxy flux within the measurement aperture from \citet{doroshenko1998}. These steps are important because (1) the true errors in measuring the galaxy are likely to be significantly larger than the nominal photometric internal errors, particularly in the face of atmospheric seeing and its variations from night to night \citep[e.g.,][]{peterson1995}; and (2) although large measurement apertures significantly reduce the seeing issues, they also dilute the nuclear flux with that from the galaxy and the resulting net error in the nuclear flux needs to reflect this effect. The resulting errors (corrected to an aperture of 6$''$) range from $\sim$ 5\% for the SAI photometry when the nucleus is bright to $\sim$ 8\% for the photometry of \citet{shom19}. Given the amplitude of the variations in nuclear B-band output, these errors are not a significant issue in our analysis.}.

NGC 4151 was extensively monitored in the J, H, K, and L bands 
at the {Crimean Obervatory of the Sternberg
Astronomical Institute (SAI)} between 1994/05 and 2018/08; the measurements are available at 
\url{http://www.sai.msu.ru/basa/inf.html}
and the observations are described in
\citet{lyuty98, shen11, tar13, okny18}. NGC 4151 was monitored in the V- and
K-bands between 2001/03
and 2007/07 in the
Multicolor Active Galactic Nuclei Monitoring (MAGNUM) project  \citep[see][and references therein]{kosh14}. We collected
these data from \citet{kosh14}.  In addition, \citet{Schnulle2015} monitored the near-IR emission
of NGC 4151 from 2010 January to June and from 2012 February to 2014 June. We
collected this J, H, K band photometry from their paper. We also collected K-band photometry of NGC 4151 
from 1975 to 1980  \citep{allen76, stein76, kemp77,  odell78},
\citet{Lebofsky1980} and \citet{Cutri1981}.

To explore the variability behavior of NGC 4151 at $\sim10~\mu m$, we
collected observations from \cite{Rieke1981}, \cite{Ward1987}, \cite{soi03},
\cite{Radomski2003}, \cite{Gorjian2004}, \cite{weed05} and \cite{Asmus2014} and
the ISOPHOT-S spectrum \citep{Sturm1999}, the {\it Spitzer}/IRS spectrum, {\it
AKARI}/IRC S9W photometry and {\it WISE} W3 photometry, and converted them into
the same system, as detailed in Section~\ref{sec:nband-data}. These data cover
a length of 36 years from 1975 through 2010.

Besides the optical and IR data, NGC 4151 has also been intensively monitored
in the hard X-ray (15--50 keV) by the {\it Swift}/Burst Alert Telescope (BAT).
We collected its light curve from the {\it Swift}/BAT Hard X-ray Transient
Monitor that covers the time period between 2005/02 and 2020/07
\citep{Krimm2013}. We {combined all the measurements} and smoothed the
X-ray light curve with a 50-day window to improve the ratio of signal to noise.
In addition, the flux from NGC 4151 has also been tracked by the Monitor of
All-sky X-ray Image (MAXI) at 2--20 keV with the Gas Slit Camera (GSC) since
2009 August. With the MAXI on-demand process tool provided by the
team\footnote{\url{http://maxi.riken.jp/mxondem/index.html}}, we extracted the
2.0--6.0 and 6.0--20.0 keV light curves with a 30-day binned window to increase
the signal to noise ratio.  Finally, we have used the Band
A, 1.3--3 keV, data from the Rossi X-ray Timing Explorer (RXTE) All-Sky Monitor
(ASM) \citep{Levine1996} to track the low-energy X-ray output.
We co-added these data over $\sim$ 3 month time intervals to obtain sufficient
signal to noise.

\subsection{Subtraction of Host-galaxy Flux}\label{sec:host_sub}

Although it does not enter into our analysis of the variations, an accurate
determination of the galaxy contribution is necessary to determine if there is
a significant underlying non-variable excess flux above the stellar output.  We
will therefore subtract the galaxy contribution to put all the light curve 
data on the same scale.

We used B-band as a measure of the nuclear optical/UV luminosity, because it is relatively high contrast relative to the galaxy and has
been measured frequently and accurately.  We took
the galaxy component in this band from \citet{lyuty99} and subtracted it from all of the
{SAI} B-band photometry. Since we found from the K-band measurements that
determining galaxy components directly from limited sets of measurements is
challenging (as discussed below), we used a different approach for the rest of
the measurements. We adjusted all the  other photometry to match the {SAI}
measurements in regions of overlap by converting the data to a consistent flux
scale and then subtracting a constant term from the fluxes (without changing
their normalization).

\begin{figure}
\center
\includegraphics[width=1.0\hsize]{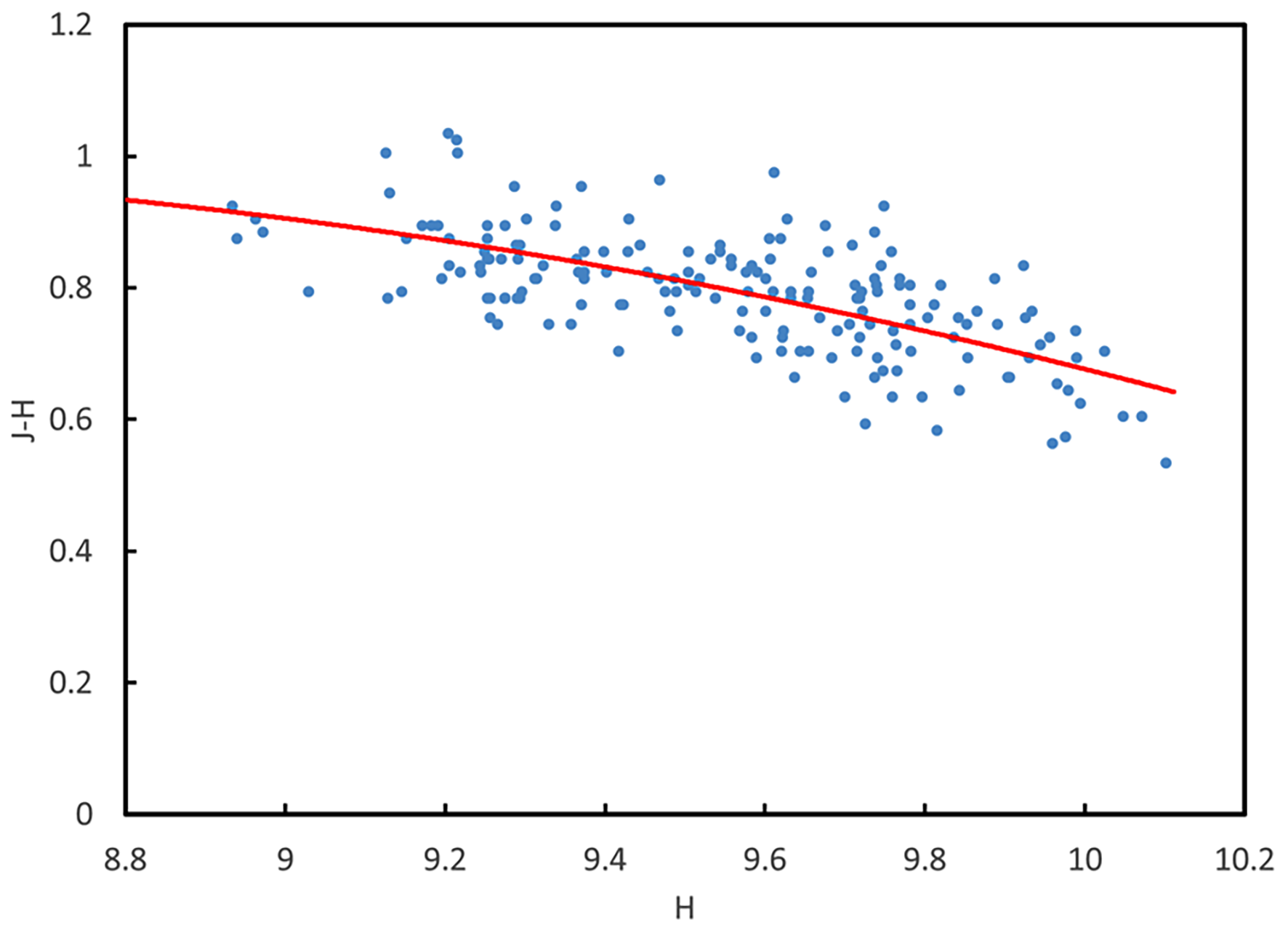}
	\caption{$J-H$ vs $H$ for the {SAI} photometry of NGC 4151. The blue dots are
	the measured values, and the red curve is the expected relation
	assuming that the variable component has a constant $J-H$ of 1.12,
	i.e., a ratio of frequency-unit flux densities of 1.75,  and the galaxy
	J and H magnitudes are 10.83 and 10.15, respectively,  corresponding to
	74.5 $\pm$ 5 and 89 $\pm$ 6 mJy, through a $12''$ aperture. }
\label{fig:jhhcolor}
\end{figure}

Estimates of the host galaxy K-band brightness in the literature differ
substantially. We have therefore used the J-H vs. H color-magnitude diagram to
obtain an independent estimate, with details provided in
Appendix~\ref{app:host}.  The behavior of NGC 4151 in J$-$H vs H is illustrated
in Figure~\ref{fig:jhhcolor}.  The blue dots are the {SAI} photometry
transformed to the 2MASS system (after rejecting four outliers). The line is
the predicted behavior assuming a flux density at H of 89 mJy for the host
galaxy and a ratio of flux densities for the excess variable component at H to
J of 1.75.  These constraints on both the galaxy flux density and the color of
the excess are reasonably tight; they translate at K$_S$ to $73 \pm 5$ mJy in
$12''$ and $59 \pm 4$ mJy in $8\farcs3$ apertures, respectively. There are
intervals where the flux at J above the galaxy contribution is only 4 mJy, or just 5 $\pm$ 6
\%; i.e., close to zero within the
errors. Since the J-H color at minimum light matches expectations for a normal
galaxy, the nuclear emission in the H band is also negligible at these times.
However, H-K and K-L were much redder than standard galaxy colors even at
minimum light; we used typical galaxy colors (see Appendix~\ref{app:host})
normalized to H to project the flux from the galaxy to K and L, and then
subtracted the galaxy component from all the {SAI} measurements.  

All the other infrared photometry sequences had significant time overlap with
the {SAI} one. As at B-band, we converted magnitudes to fluxes and then
subtracted a constant flux from all the measurements in a given photometry set
(but with no adjustment to the flux scale), which we adjusted to force
agreement for the times of overlap. For the \citet{kosh14} measurements,
we needed to increase the
subtracted value from their 44 $\pm$ 4 to 66 mJy, consistent within the mutual
errors with our value derived above.

We used the resulting generally smoothly varying light curve at K to search for
spurious outliers. We found that the photometry from the {SAI} group seems
unreliable on nights when results are reported for only a subset of the four
bands; for example, many of these measurements are outliers
from the K-band light curve, but also frequently report
 strange variations, e.g., by 0.3 magnitudes in one band
while an adjacent band does not change significantly. We therefore retained only those data that had
consistent behavior among the bands and with adjacent nights. We then combined
these infrared measurements with those from \citet{kosh14} and
\citet{Schnulle2015}.  The resulting very well sampled K-band
light curve plus the accompanying less-well-sampled light curves in J, H, and L
are the foundation for our analysis. 

Besides these data, we also converted the K-band measurements of NGC 4151 in
\citet{allen76}, \citet{stein76}, \citet{kemp77}, \citet{odell78},
\citet{Lebofsky1980} and \citet{Cutri1981} into a consistent flux scale and
built a single-band IR light curve covering $\sim1975$--1980.  This dataset
provides a comparison to the 10 $\mu$m measurements in 1975 and 1976.

\begin{figure*}
\center
\includegraphics[width=1.0\hsize]{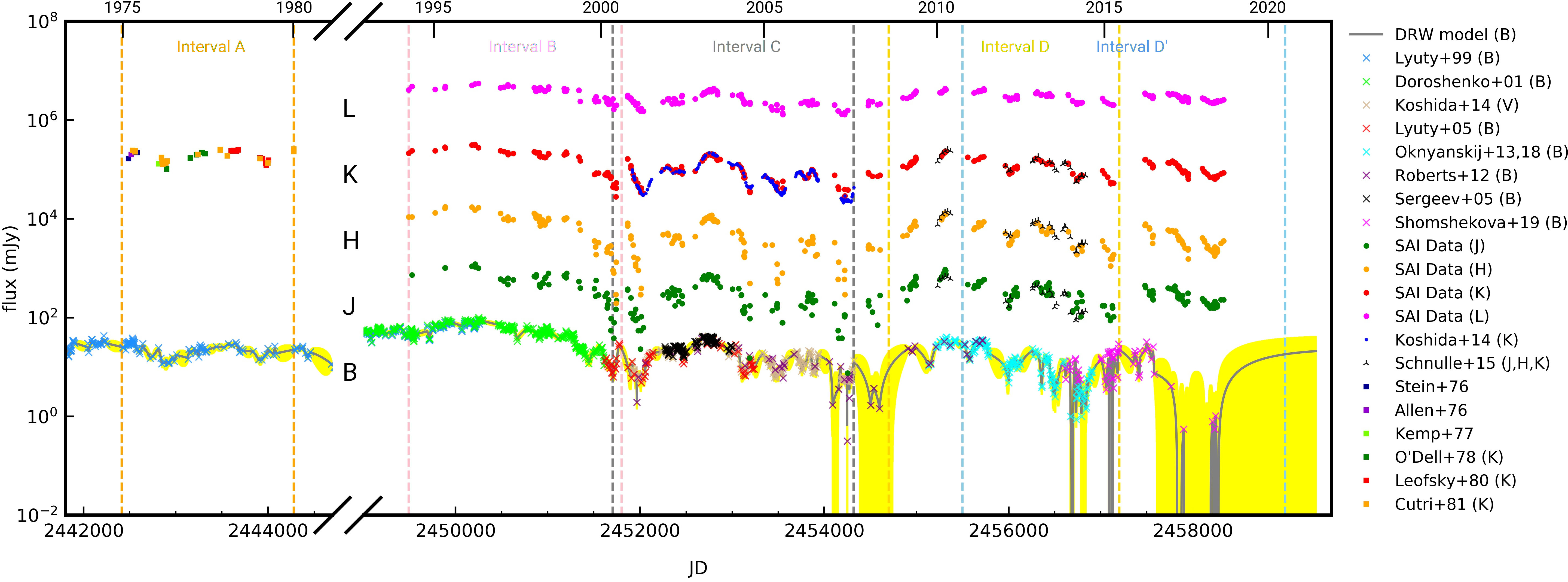}
\caption{ The B, J, H, K, L light curves of NGC 4151. The sources for and adjustments to individual light curves are discussed in the text. We also show a damped-random-walk (DRW) model (JAVELIN;
	\citealt{Zu2013}) to interpolate this light curve (the line).   
	We have
	arbitrarily scaled J, H, K, L light curves by factors of 10, 100,
	1000, 10000 to separate them for clarity. The time periods of Interval
	A, B, C, D, D' (see Section~\ref{sec:jhkl-review}) are indicated with
	vertical lines in different colors.
	}
\label{fig:lc_full}
\end{figure*}

Figure~\ref{fig:lc_full} shows the full B, J, H, K, L galaxy-subtracted light curves, which will be analyzed in
Section~\ref{sec:analysis}. { These data are provided in the online
journal and a small portion is given in Table~\ref{tab:lc_data} to show its basic format.}

\begin{deluxetable}{lccl}[hbt!]
\tabletypesize{\scriptsize}
\tablecaption{Galaxy-subtracted B, J, H, K, L light curves of NGC 4151}
\tablewidth{0pt}
\tablehead{
\colhead{JD}  &  \colhead{Band}  & \colhead{Flux (mJy)} & \colhead{Source}
}
\startdata
2439941.4 & B & 26.566 & Lyuty et al. (1999) \\
2439941.4 & B & 24.387 & Lyuty et al. (1999) \\
2439944.3 & B & 23.588 & Lyuty et al. (1999) \\
2439945.3 & B & 22.161 & Lyuty et al. (1999) \\
2439945.3 & B & 22.846 & Lyuty et al. (1999) \\
$\cdots$  &  $\cdots$  & $\cdots$ & $\cdots$
\enddata
\label{tab:lc_data}
\tablecomments{Table 1 is published in its entirety in the machine-readable format.
      A portion is shown here for guidance regarding its form and content.}
\end{deluxetable}

\subsection{Photometry at 10 $\mu$m}\label{sec:nband-data}

\begin{deluxetable}{llcl}[hbt!]
\tabletypesize{\scriptsize}
\tablecaption{NGC 4151 photometry transformed to 10.6 $\mu$m \label{Table 1}}
\tablewidth{0pt}
\tablehead{
\colhead{JD}  &  \colhead{flux(mJy)}  & \colhead{error(mJy)} & \colhead{Source}
}
\startdata
	2442417   & 2058$^1$   &  124 & \cite{Rieke1981} \\ 
	2442433  &  2083$^1$  &  87  &   \cite{Rieke1981}  \\
	2442468   & 2021$^1$  &  112 &  \cite{Rieke1981} \\ 
	2442494 & 2083$^1$  &  124 &   \cite{Rieke1981} \\
	2442518  &  1897$^1$  &  112  &  \cite{Rieke1981} \\
	2442466  &  2032 & 49 &   average of above five\\
	2442808   & 2046$^1$   &  87 & \cite{Rieke1981} \\ 
	2442846  &  2046$^1$  &  112  &   \cite{Rieke1981}  \\
	2442869  &  2108$^1$  &  87  &  \cite{Rieke1981} \\
	2442878   & 1922$^1$  &  62 &  \cite{Rieke1981} \\ 
	2442888 & 1972$^1$  &  112 &   \cite{Rieke1981} \\
	2442856 & 2002 &  38 &  average of above five \\
	2445116  &  1116$^1$  &  168  &   \cite{Ward1987}  \\
	2445118  &  1401$^1$  &  189  &  \cite{Ward1987} \\
	2445117 &  1259 &  90 & average of above two \\
	2445371   & 1724$^1$  &  149 &  \cite{Ward1987} \\ 
	2445498 & 1440$^2$  &  290 &  IRAS \\
	2445766  &  1275$^1$  &  109  &  \cite{Ward1987} \\
	2450211  &  1370  &  140  &   \cite{Sturm1999}  \\
  $\sim$ 2450401$^4$  &  1180$^3$  &  118  &  MSX \\
    2451628   & 2340$^5$  &    230  & \cite{Gorjian2004} \\
	2451680   & 2440$^6$  & 244 &  \cite{soi03} \\ 
	2452037   & 2324$^1$  & 232 &  \cite{Radomski2003} \\ 
	2453104   & 1900  & 133 &  \cite{weed05} \\
	$\sim$ 2454876$^4$   & 1820$^3$  & 130 &  {\it AKARI} \\ 
	2454990   & 1440$^7$  & 101 & \cite{Asmus2014} \\  
	2454990   & 1300$^7$  & 91 & \cite{Asmus2014} \\  
	2455285   & 1580$^7$  & 111 & \cite{Asmus2014} \\  
	2455285   & 1470$^7$  & 95 & \cite{Asmus2014} \\
	2455320   & 1353$^7$  & 95 & \cite{Asmus2014} \\   
        $\sim$ 2455321$^4$   & 1507$^3$  & 150 & {\it WISE} \\  
 \enddata
\tablenotetext{1}{Corrected via bandpass correction computed using Spitzer IRS spectrum \citep{weed05}}
\tablenotetext{2}{Average of values from \citet{sem87, san03}. Adjusted by 0.12 Jy for 
contribution of star formation in the host galaxy, based on fitting a template \citep{rie09} 
to the far infrared $\lambda \ge 100 \mu$m) measurements. Corrected to 10.6 $\mu$m using Spitzer/IRS 
spectrum \citep{weed05}}
\tablenotetext{3}{Corrected to 10.6 $\mu$m using Spitzer IRS spectrum \citep{weed05}}
\tablenotetext{4}{When measurements over an extended mission are combined, a typical date is shown as being approximate.}
\tablenotetext{5}{Reported value is corrected for flux outside central PSF by taking the average of the lower limit reported in \cite{Gorjian2004} and the correction derived by \citet{Radomski2003}.}
\tablenotetext{6}{Corrected for flux outside central PSF according to \citet{Radomski2003}. Corrected to 10.6 $\mu$m using Spitzer IRS spectrum \citep{weed05}}
\tablenotetext{7}{Rereduced to use aperture photometry of $\sim$ 4$''$ diameter. Corrected to 10.6 $\mu$m using Spitzer IRS spectrum \citep{weed05}}
\label{tab:nfluxes}
\end{deluxetable}

In the 36 years from 1975 through 2010, there are 28 measurements at $\sim$
10~$\mu$m useful for light curve tracking (excluding a few of low signal to noise, or with spectrometers with undetermined slit losses). Although these are unevenly spaced,
with gaps of 5--10 years, they are adequate to test for lags on decadal
timescales.

The different  spectral bands and calibration approaches must be understood to
construct an accurate light curve.  A simple demonstration of real variations
is, however, possible. The measurements of \citet{Ward1987} are with very
similar photometric systems as those of \citet{rie78, Rieke1981} and there are
enough sources in common with \citet{rie78} that the relative
calibration can be determined directly. The five galaxies other than NGC 4151
with measurements in common show excellent agreement in the reported flux
densities (average ratio of \citet{rie78} to \citet{Ward1987} values of $0.964
\pm 0.065$), while the earlier \citep{Rieke1981} measurements of NGC 4151 are
higher than the later \citep{Ward1987} ones by a factor of $1.363 \pm 0.042$.
This estimate includes non-statistical errors as already within the
\citet{Rieke1981} values, but at 7\% additional for the \citet{Ward1987} ones. The two N-band measurements early in the 21st century \citep{Gorjian2004, Radomski2003} are 1.58 $\pm$ 0.11 times higher than the ones reported in \citet{Ward1987}.

To generalize this result over the full baseline, we have assembled all the
relevant measurements and corrected them to flux densities at 10.6 $\mu$m, as
summarized in Table~\ref{Table 1}. The reconciliation of the measurements to a
common photometric basis is discussed in
Appendix~\ref{app:10um}.

\subsection{Accretion disk variability traced in the X-ray}\label{sec:X-ray}

\cite{Czerny2003} compared the optical and X-ray  light curves of NGC 4151 over several decades. They found 
an independent long-timescale (over $\sim$10 years) component that dominates
the optical variations, but that the optical variations also contain a
short-timescale component well-correlated with the X-ray ones over
5-1000 days.  \citet{Edelson2017} also confirmed the strong correlation between
the UV/optical and X-ray variability of NGC 4151, although they also found that the UV/optical band variation lags ~3-4 days behind the hard X-rays. Since the IR lag of NGC 4151
is less than 100 days, we have used the short-time X-ray variability to pin down the IR lag measurements when the optical data are
limited. This approach is successfully illustrated by \citealt{Noda2020}.
It is particularly useful for JD
2457000--2458500, when the optical data are too poor for our reverberation analysis.\footnote{Although supernova 2018aoq was discovered on 2018-04-01 (JD:2458208), 73\arcsec from the NGC 4151 nucleus (see \url{http://www.rochesterastronomy.org/sn2018/sn2018aoq.html}), 
we do not see its influences in
the X-ray light curves, possibly due to it being relatively faint compared to the AGN.}
To make a consistent comparison, we generate a synthetic B-band light curve by scaling and shifting the X-ray light curve and
adding a 4-day delay (as found by \citet{Edelson2017}).

\section{Dust Reverberation Analysis}\label{sec:analysis}

{ In Section~\ref{sec:jhkl-review}, we provide an initial evaluation of the optical and IR J, H, K, L
light curves and point out several issues that have been overlooked by the previous
dust reverberation analysis. Section~\ref{sec:retrieving} introduces our 
analysis methods and establishes that there are two main time lags that contribute to the AGN
reverberation signals at 1--4~$\mu$m. Section~\ref{sec:innertorus}
discusses the evolution of hot dust reverberation signals over 30-years.  
We compare these results with previous near-IR reverberation studies of NGC 4151
in Section~\ref{sec:comparison}. Finally, we report the detection of a IR time lag from the AGN warm
dust emission ($\sim$300K) in the N-band and the lack of IR variations of NGC~4151 
at wavelengths at 20--40~$\mu$m  in Section~\ref{sec:Nband}.}

\subsection{JHKL Reverberation Signals and Challenges for Simple Optical-IR Cross-correlation Analysis}\label{sec:jhkl-review}

Since our light curves come from various sources with different cadences, we
separate the time epochs into four intervals as shown in
Figure~\ref{fig:lc_full}, with Interval A from JD 2442414 through 2444281
(January, 1975 through February 1980), prior to our multi-band IR reverberation
analysis; Interval B from JD 2449490 through 2451800 (May, 1994 through
September 2000); Interval C from JD 2451704 through 2454319 (June, 2000 through
August, 2007); and Interval D from JD 2454700 through 2457200 (August, 2008
through June, 2015). In addition, we denote the time period covered by {\it
Swift} observations as Interval D', which spans JD from 2455500 through 2459000
(November, 2010 through May, 2020). 

{
\subsubsection{General variability behavior and time sampling}

Reverberation mapping has been applied extensively in the ultraviolet, optical,
and for emission lines. The requirements for success 
include a  measurement sequence that fully
samples the variations (e.g., Nyquist sampling) and to carry out this sequence
for at least three times the longest reverberation lag, with modest gaps in longer datasets not strongly affecting the results \citep[e.g.,][]{horne2004}.  Below we will find lags
of the infrared behind the blue signals of $\sim$ 35 and $\sim$ 90 days for NGC
4151. JHKL photometry is available from 1994 to 2018, but in general in
intervals with significant gaps that may in some cases be inconsistent with these goals. 

We initially focus on Interval C where we combine  campaigns by the
{SAI} and MAGNUM groups to improve the sampling at K.  During this period
there were also many campaigns at B-band, providing a very well sampled light curve.  We compare the B
and K-band light curves in Figure~\ref{fig01} during this interval, from JD
2451704 through 2454319  (June 8, 2000 through August 6, 2007), showing the very thorough data available in both bands. In favorable cases, the data extends
over 5--6 periods of the shorter ($\sim$ 35 day) lag. There is data for about six  times the longer ($\sim$ 90 day) lag with an average sampling of about four times Nyquist and just two gaps, each of roughly one period.

\begin{figure}
\center
\includegraphics[width=1.0\hsize]{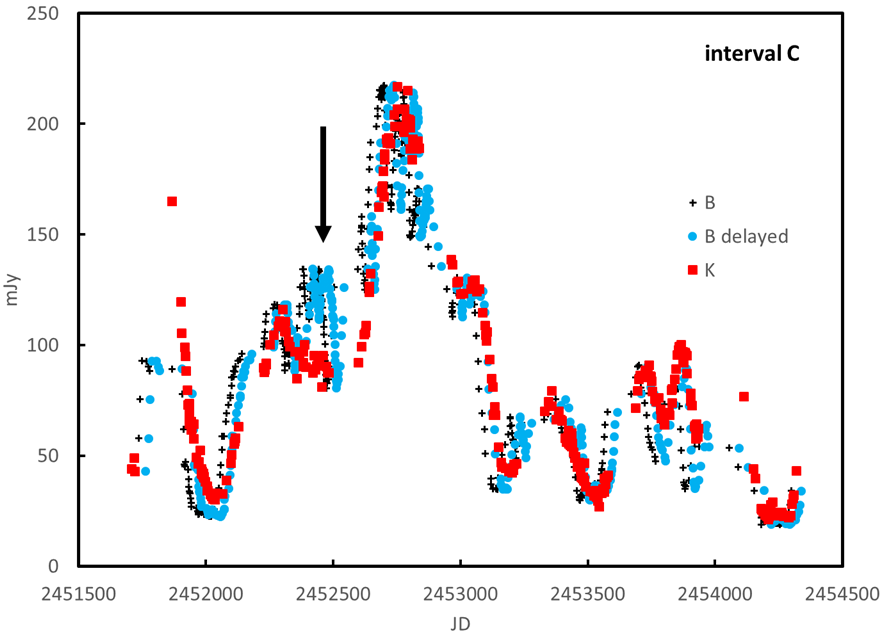}
\caption{Light curves in Interval C at B (blue circles) and K (red squares) (the former
	delayed by 35 days and smoothed by taking a seven-point running
	average). The undelayed B-band data is indicated by + symbols and
	clearly leads the K-band behavior. Overall, the two sets of data are in
	very good agreement suggesting a reverberation delay of order 35 days.
	However, there is a significant discrepancy in the range indicated by
	the arrow. }
\label{fig01}
\end{figure} 

The behavior during this period is largely consistent with the assumption that the K-band flux
results from a reverberation of the nuclear signal (characterized by B-band)
when it heats dust in the circumnuclear torus (but see next section). As a demonstration, in the figure we have smoothed the B-band data to reflect the light travel times for different zones of the torus and delayed it by 35 days (similar
to that found by \citet{okny18} of $37 \pm 3$ days and consistent with our
results reported below).  This general agreement is consistent with most other
previous work.

\begin{figure}
\center
\includegraphics[width=1.0\hsize]{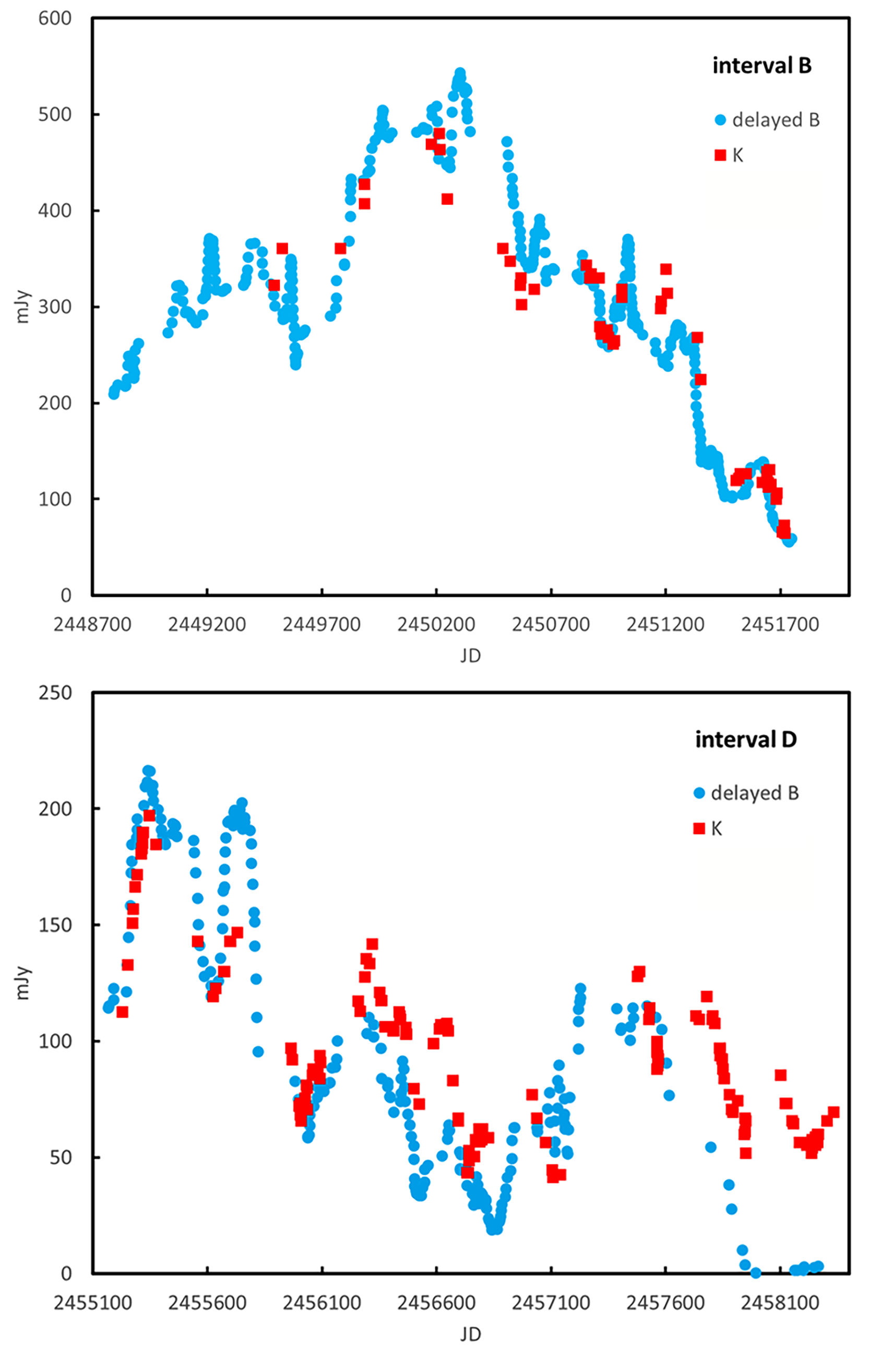}
\caption{Light curves for B-band and K-band in intervals B and D. To provide a
	rough match to the B-band curve, the K-band fluxes have been multiplied
	by 1.5 in interval B and by 0.81 in interval D. }
\label{fig05}
\end{figure}

In Figure~\ref{fig05}, we show the B-band and K-band light curves for Intervals
B and D. For Interval B, there are nominally sufficient K-band observations for
Nyquist sampling but their spacing results in the curve being significantly
undersampled. For Interval D, there is an on-off pattern of roughly 180 days and generally poorer sampling than in Interval C.

\subsubsection{ Uncorrelated Variations Between the Optical and IR }\label{sec:first-look}

Previous studies have assumed that the reverberations strongly
dominate the relationship between the optical and near infrared.  However, this is not always correct. As indicated with an arrow in Figure~\ref{fig01}, there was an outburst in the B-band that did not reach the torus, since the
IR emission seems to have little reaction. Figure~\ref{fig02} shows an expanded
view of this event. The B-band flare is verified by three independent
measurement series and the lack of a corresponding K-band feature  by two independent measurement series.

\begin{figure}
\center
\includegraphics[width=1.0\hsize]{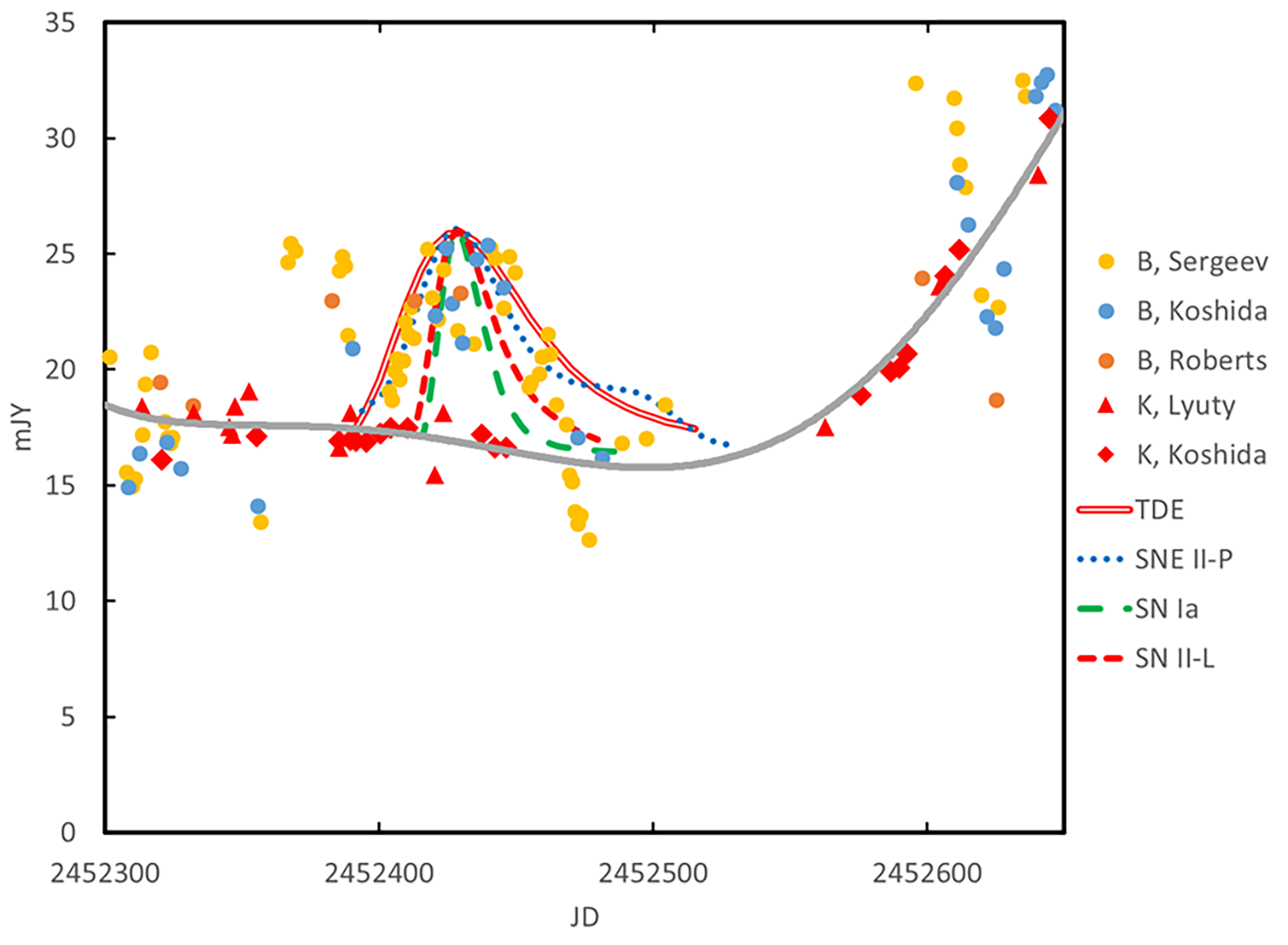}
\caption{Expanded view of the 35-day-delayed B and the K band light curves.
	The solid gray line is a polynomial fit to the K-band data. The peak in
	B-band near JD 2452430 is apparent in three independent sets of
	measurements, while the lack of this peak is shown by two independent
	sets of K-band ones. We show four possible light curves, for a SNE Ia,
	SNE II-L, SNE II-P \citep{Wheeler1990} or a tidal destruction event
	(TDE), although known TDEs are more luminous \citep{velzen2020}.}
\label{fig02}
\end{figure} 

Such behavior shows that, on top of the expected reverberation behavior, there
could be outbursts from the central source/accretion disk (e.g., jets) that are directed away from the torus  or other
events confused with central activity that do not
reach the circumnuclear torus  (e.g., nearby supernovae: {the B-band data are taken through an aperture of $\sim$ 2 kpc diameter at the galaxy}).   Figure~\ref{fig06} shows a
similar event in the IR light curve of another type-1 AGN, MCG 08-11-011 (the
data are from \citealt{kosh14}), indicating such behavior might be common.  The
various possibilities --  supernova, tidal disruption event and beamed jet
variability -- all can produce variable signals on a timescale of
$\sim$100--300 days.

\begin{figure}
\center
\includegraphics[width=1.0\hsize]{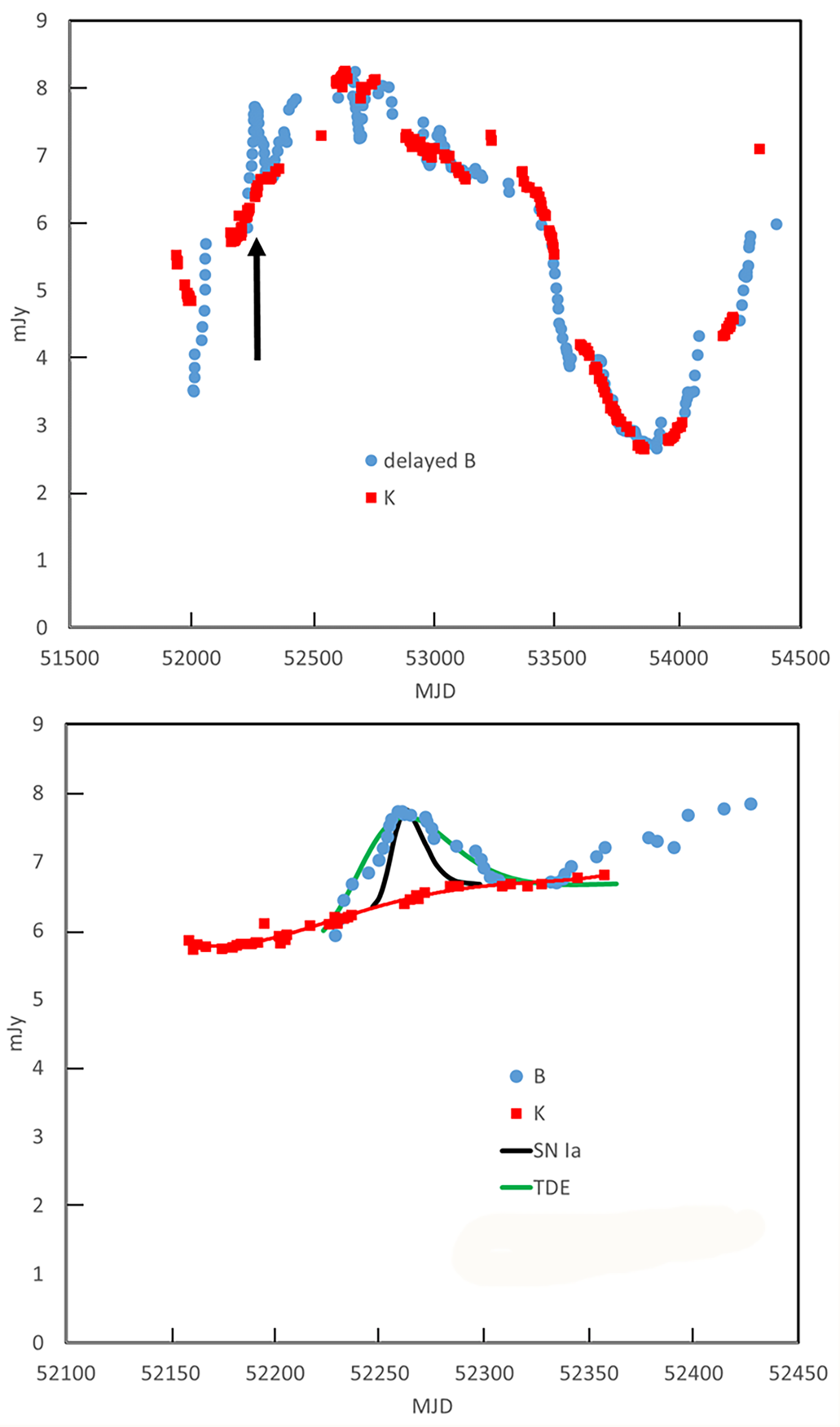}
\caption{A similar event to that in Figure~\ref{fig02}, but in comparing the
	V-band and K-band light curves of MCG 8-11-011, from \citet{kosh14}.
	Again, the behavior of the event is similar to that of a SNE II-P or a
	TDE. }
	\label{fig06}
\end{figure} 

For Interval D, there is a different behavior, namely a dramatic drop in the B-band curve beyond JD 2457900  that is not reflected at
K-band.  

That the B-band curve is influenced
by phenomena not reflected in the K-band adds uncertainty to conventional
reverberation analyses between the optical and IR.
To circumvent these issues, in our analysis described in Section~\ref{sec:retrieving} we will start with interval C both
because of its much better sampling in the infrared and because the B-band and
K-band seem to track each other much better than in interval D. To avoid completely any uncorrelated B-band variations, we will first
compare the J-, H-, and L-band curves with the K-band one. We then compare the
well-sampled K-band curve with the B-band one. In this latter analysis, we mask the IR-uncorrelated flare in 
B-band. This
approach gives us a robust lag of all three IR bands relative to each other and to
the B-band.  We will then carry out additional
analyses, e.g., for  interval D, to demonstrate that their behavior is
consistent within the significantly larger uncertainties. 
}

\subsubsection{Evidence for Multiple Dust Components}\label{sec:sed-fit}

\begin{figure}
\center
\includegraphics[width=1.0\hsize]{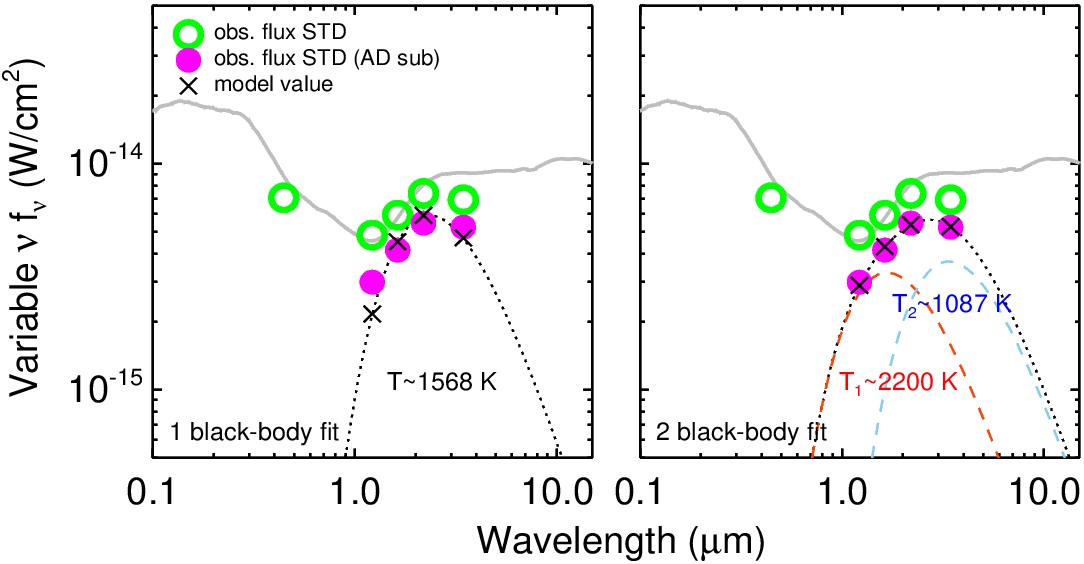}
\caption{Variability SED of NGC 4151. We show the standard deviations of the B, J, H, K, L light
	curves as green open circles. After subtracting the accretion disk
	variability contamination, the infrared dust SED (purple dots) is
	fitted with one single black body (left panel) and two black bodies
	(right panel). The model SEDs, shown as black crosses, only fit the observations for the two black body model. For
	comparison, we show the Elvis-like AGN normal template as grey lines.}
\label{fig:sed_decomp}
\end{figure}

Before progressing to reverberation modeling, we compute the standard deviations of the 
light curves as an unbiased measure of the wavelength dependence of
the IR variability. This
approach should give a good approximation
for the SED of the dust components most directly heated by the AGN. The results are presented in Figure~\ref{fig:sed_decomp}. They are reasonably matched by the Elvis normal AGN template, as is consistent with our SED decomposition of NGC 4151 based on high
spatial-resolution data in \cite{Lyu2018}.

After subtracting the accretion disk variability in the infrared bands
derived from B band with a $\Delta f_\nu\propto\nu$ dependence (see
Section~\ref{sec:method}), we fitted the near- to mid-IR dust-only variability
SED with black-body models. Such models should be applicable for wavelengths
short of $\sim$ 4 $\mu$m; the longer wavelength behavior is not constrained.
As demonstrated in Figure~\ref{fig:sed_decomp}, we needed two,
instead of one, black body components to match the observations:  one with a
dust temperature of $\sim$1600--2500~K\footnote{Due to the lack of
constraints at shorter wavelengths and the strong contamination from the
accretion disk variability, the temperature of this component cannot be
accurately determined. We adopt an upper value of 2200 K.} 
("Component A") and another with a dust temperature
of $\sim1000$K  ("Component B").

\begin{figure}
\center
\includegraphics[width=1.0\hsize]{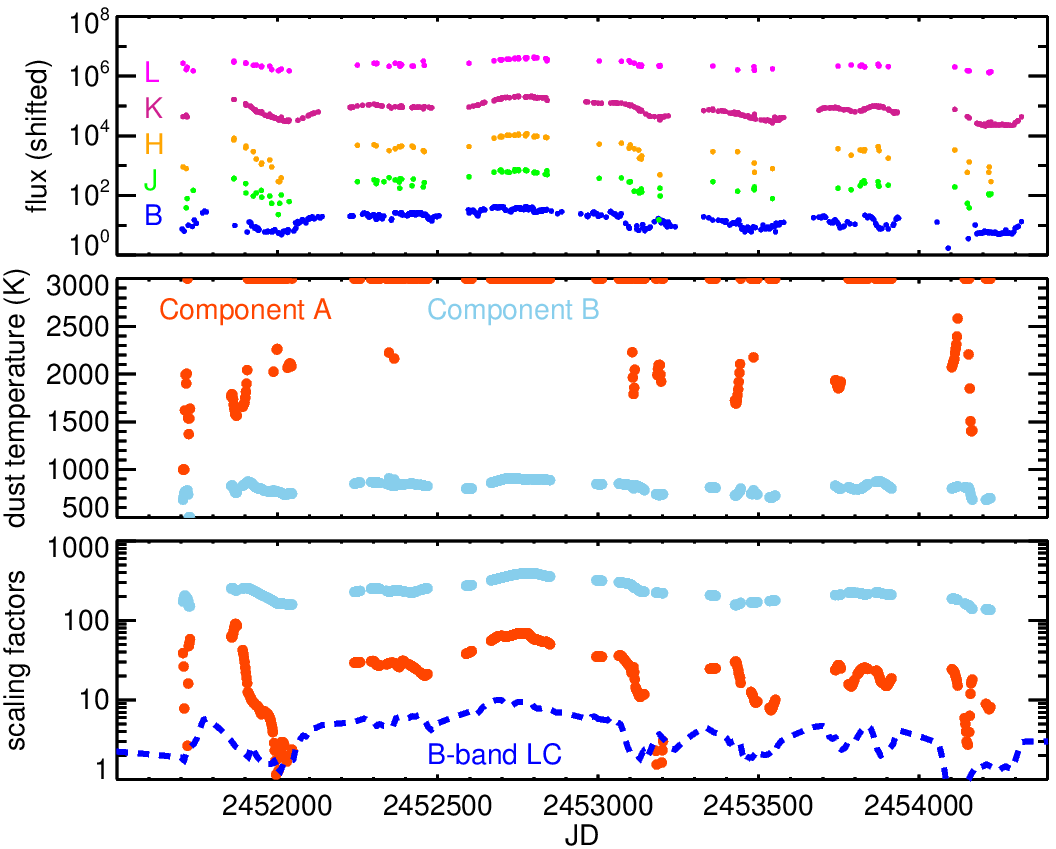}
\caption{NGC 4151 light curves and the time dependence of the double-black-body
	best-fit parameters in Interval C.}
\label{fig:sed_para_evo}
\end{figure}

We use a 
double black-body to investigate the time-dependent behavior of the SED as 
summarized in Figure~\ref{fig:sed_para_evo}. The dust temperatures of the two
 components are  reasonably stable\footnote{Small variations
similar to those found by \citet{Schnulle2015} are not ruled out by this simple
analysis.} while their emission strengths correlate with the B-band light curve. This confirms that two dust
components with different temperatures are likely to be responsible for the variable
 $\sim$1--4~$\mu m$ emission.

The presence of two dust components with different temperatures is not a surprise. 
Carbon and silicates are the dominant species of interstellar
grains, and presumably also for the grains in the AGN torus. The
sublimation temperatures, $T_\text{sub}$, of these grains are 
1500--1800~K for graphite and 800--1000 K for silicates; consequently only
graphite grains would survive at the innermost regions and silicate grains
would be  distributed to larger radii.  Assuming an optically-thin environment,  the sublimation radius, $R_\text{sub}$, for graphite grains is

\vspace{-4mm}

{\footnotesize
\begin{equation}\label{eqn:sub_c}
    \frac{R_{\rm sub, C}}{\rm pc} = 1.3 \left(\frac{L_{\rm UV}}{10^{46} {\rm erg~s}^{-1}} \right)^{0.5} \left( \frac{T_{\rm sub, C}}{1500 {\rm K}} \right)^{-2.8} \left( \frac{a_{\rm C}}{0.05~\mu m}\right)^{-0.5}~,
\end{equation}
\normalsize
}
\vspace{-1mm}
\noindent
based on the relation in \cite{Barvainis1987} 
with a $a^{-1/2}$ term to approximate the  dependence on the
grain size $a$ \citep{Kishimoto2007}. Here $L_{\rm UV}$ is the UV luminosity in the direction of the grains. Similarly for
silicate dust grains, adopting the absorption efficiency of astronomical
silicate \citep{Draine1984, Laor1993}, we have

{\footnotesize
\begin{equation}\label{eqn:sub_s}
    \frac{R_{\rm sub, S}}{\rm pc} = 2.7 \left(\frac{L_{\rm UV}}{10^{46} {\rm erg~s}^{-1}} \right)^{0.5} \left( \frac{T_{\rm sub, S}}{1000 {\rm K}} \right)^{-2.8} \left( \frac{a_{\rm S}}{0.05~\mu m}\right)^{-0.5}.
\end{equation}
\normalsize
}

Thus the two temperatures inferred from our variability analysis imply that the
inner structure of the dust torus is governed  by grain sublimation
\citep[e.g.,][]{Rieke1981, Barvainis1987} and thus might have two characteristic radii. We will test this
hypothesis by seeing if the reverberation signals show two distinct lags in the
IR light curves.

\subsection{ Near-IR Reverberation by Two Dust Components}\label{sec:retrieving}

\subsubsection{Subtraction of Accretion Disk Variability in the IR}\label{sec:ad-var}

{ To study the dust reverberation signals of NGC~4151, we need to remove the IR variations of the accretion disk itself from the IR light curve. 
The infrared continuum from the accretion disk is not well-determined.} Although
theoretically it might be expected to go as $\nu^{1/3}$, it has been difficult
to confirm such behavior in the optical \citep{Gaskell2008} and thus there is
no obvious reason to adopt it in the infrared.  One approach is to do profile
fitting to remove the contribution of the galaxy and deduce a nuclear spectrum
\citep[e.g.,][]{Garcia-Bernete2019}. However, we found in
Section~\ref{sec:host_sub} surprisingly large discrepancies in such analyses.
Therefore, we used the variability to constrain the nuclear spectrum, using the observations of \citet{Schnulle2015}.
We assumed that the $z$ band (0.9 $\mu$m) is dominated by the accretion disk
and the K band by the thermal reradiation and then adjusted the mixture of
signals from each to match the variations of the J and H bands. The preferred mixture is determined by ratioing to the true J and H band fluxes and minimizing the standard deviation of these ratios over all 29 epochs.  This
analysis indicates that the SED falls more rapidly than $\nu^{1/3}$, although
it does not have sufficient wavelength baseline to derive a full IR  spectrum
independently. 

We also used the wavelength dependence of the variability amplitude of
NGC 4151 from the $\sim$10~years of BVRI
photometric observations by \cite{Roberts2012}; these data cover a similar time period to Interval C, where our analysis is focused.
We smooth each light curve with a window of three data points and then take the
variation amplitude as the difference between the max and min
flux values.  We find $\Delta f_{\rm B}:\Delta f_{\rm V}:\Delta f_{\rm R}:\Delta
f_{\rm I} = 1.88: 1.66: 1.22: 1$, which can be described as $\Delta
f_\nu\propto\nu$.  Since there is no obvious physical reason for a
turn-over of the $\Delta f_\nu$ slope just entering the near  IR, we adopted
the same $\Delta f_\nu\propto\nu$ through the near-IR, i.e. 
\begin{equation}
	F(t)_\text{IR, dust} = F(t)_\text{IR} - F(t)_\text{OPT}\left(\frac{\nu_\text{IR}}{\nu_\text{OPT}}\right)^{1.0}
\end{equation}
\noindent
where $F(t)_\text{IR}$ is the observed light curve. Because the spectrum of the infrared excess rises steeply through 2 $\mu$m  (in
Section~\ref{sec:host_sub},  we showed the excess has an equivalent power law index
$< -2$.), the possible range in the nonthermal SED we subtract should have little
effect on our results. To  confirm this, we repeated our
analysis assuming a $\nu^{1/3}$ nonthermal spectrum, finding no significant
change. 

\subsubsection{Basic Dust Reverberation Fitting Method}\label{sec:method}

 Our lag analysis is a two-step progress, as introduced in
 \cite{Lyu2019}: we first interpolate the driving  light curve (usually in the optical) with a damped
random walk (DRW) model and smooth it when necessary, and then this light curve
model is shifted along the time axis and scaled along the amplitude axis to
match the driven light curve (usually in the IR). Besides quantifying the time lag between the IR and
optical variability, we can constrain their relative variation amplitudes.
For two lags, the reverberation signal is

\begin{equation} \label{eqn:drm-model-2lag}
\begin{split}
	F(t)_\text{IR, dust} = ~~& {\rm AMP}_1 \times <F(t-\Delta t_1)_\text{OPT}> +
	\\ &  {\rm AMP}_2 \times  <F(t-\Delta t_2)_\text{OPT}>  +\\ &
F_\text{const.} \\ 
\end{split} 
\end{equation} 
The first two terms on the right of the equation represent two reverberation signals; AMP$_1$, AMP$_2$ are the ratios between the optical
light curve and the corresponding IR reverberation flux  amplitude,
and $<F(t-\Delta t_1)_\text{OPT}>$, $<F(t-\Delta t_2)_\text{OPT}>$ denote the
smoothed optical light curve model with different time lags. The third term
$F_\text{const.}$ is the sum of all non-variable
components.  A single lag is described by setting AMP$_2$ = 0. 

{ There are a number of approaches to evaluating the lag between the light curves. 
In \cite{Lyu2019}, we used a high degree of smoothing of the driving light curve to retrieve the dust reverberation
signals. Despite sampling with WISE data generally well below the Nyquist rate, we obtained consistent time lag measurements compared with the classical cross-correlation analysis and with the expectation that the lags would scale with the square root of the AGN luminosity. In this approach, a $\chi^2$ minimization can be used (see Appendix C2 in \citealt{Lyu2019}).  However, we have much better sampling for the near-infrared bands (JHKL) in NGC 4151 and can go beyond the high degree of smoothing we used previously (and in this paper at N-band). We smooth only very modestly with the DRW technique to fill in gaps, leaving the potential lag times intact. To find the best-fit parameters and their uncertainties for these data, we use the Dynamic
Nested Sampling package, {\it Dynesty}, to estimate the  fitting parameters
and Bayesian posteriors \citep{dynesty}.} Compared with other methods such as
Markov chain Monte Carlo or the Levenberg-Marquardt technique, the fitting of
{\it Dynesty} does not require any initial guess of the parameter values and
can detect model degeneracies in the parameter space efficiently. {As the
maximum a posteriori (MAP) values of the parameters provide the maximum
conditional probability of matching the observed data given a model weighted by
a prior, we mainly quote such parameters  as the ``best fit'' to the data.
To describe the parameter uncertainties, the 1-D marginalized posteriors, i.e.,
the median parameter value with 2-$\sigma$ ``uncertainties'' (i.e., 2.5\%, 50\%
and 97.5\% quantiles), are also used when necessary.
{  To check the validity of these results independently, we have also used the interpolated 
cross correlation function (ICCF) method recommended by \citet{Peterson1998} and 
the improved ICCF (I$^2$CCF of Guo et al. (2021, in prep.) for the cross-correlation 
analysis between the IR light curves. }}

{ 
In Section~\ref{sec:IRcorr}, we first use a one-lag model
to determine the lags of the infrared J, H, and L bands relative to the K-band, confirming the existence of a secondary dust time
lag at the longer wavelengths (as previewed in Section~\ref{sec:first-look}). Next, in Section~\ref{sec:fullreverb} we apply a
two-lag analysis between the optical and IR light curves and put constraints
on the time lags, dust temperatures and variablity amplitudes of the 
IR reverberation signals from the two dust components. 
}

\subsubsection{Correlation Analysis between IR Bands}\label{sec:IRcorr}

Since the K-band light curve in Interval C has relatively high cadence, we use
it as the base curve to correlate against other IR light curves. To do so, we
first subtracted the accretion disk contamination in all the IR light curves,
following the description in Section~\ref{sec:ad-var}. We then interpolated the
K-band dust emission light curve with the DRM model to bridge gaps. We finally
carried out one-lag dust reverberation mapping against the J, H, and L bands with {\it Dynesty}.
As shown in Table 2 (further details in Appendix C), the J and H light curves are strongly
correlated with the K-band light curve with lags close to zero. These results
indicate that the  reverberation signals in these three bands are dominated
by the same dust component. This result is not surprising since all three bands
fall close to the expected SED for carbon dust at its sublimation temperature.

As the L-band observes a longer wavelength, its variability signal can be a
mixture of dust components with different temperatures (thus at different
radii), as indicated by the variability SED fitting in
Section~\ref{sec:sed-fit}.  Adopting a temperature of $\sim$ 1800 K for the hot
dust that is probed at J and H (consistent with the flux ratio derived in
Section~\ref{sec:host_sub}), its flux levels in $f_\nu$ in the K and L bands
are similar. To bring out any additional lag component, we subtract the K-band light curve from the L-band
one, so that the signal from any other dust component will be more prominent in the L-band residual signal. After this subtraction, the remaining L-band signal is half due to the persistent (non-varying)  component and half (peak-to-peak) due to the variable one. We correlate the K-band light curve with the L-band one after this
subtraction; the results are provided in Table~\ref{tab:cc-ir} { (with details in Appendix C)}. The K-subtracted L band has an indicated  delay of about 53 days
relative to K. As 
shown in the next section, this lag difference is also consistent with the two-lag 
dust reverberation analysis between the optical and
IR light curves.

The cadence in Interval B is inadequate to repeat this analysis. However,  we
have carried out a similar analysis for data in Interval D, where we improved
the K-band cadence by combining the {SAI} light curve and the data from
\citet{Schnulle2015}. The results can be seen in Appendix C. 
Similarly to Interval C, the J and H bands are strongly correlated with the
K-band with negligible lags ($\sim-$10 days for J, $\sim-$1.7 days for H). In
contrast, the L-band residual signal lags behind the K-band variability by
$\sim$34 days, consistent with the result for Interval C within the larger
uncertainties due to the lower cadence compared to Interval C. Interestingly,
\citet{Schnulle2015} fitted two black bodies to their z, J, H, K light curves
and derived dust lags at $\sim$29 and $\sim$67 days with a relative lag about
38 days in agreement with our result, but they rejected this fit.

\begin{deluxetable*}{@{\extracolsep{4pt}}ccccccc}
    \tabletypesize{\footnotesize}
    \tablewidth{1.0\hsize}
    \tablecolumns{7}
    \tablecaption{1-lag best-fit parameter values of the J, H, L against K band light curves\label{tab:cc-ir}
    }
    \tablehead{
	\colhead{Band} &
	\multicolumn{2}{c}{$\Delta t$} &
	\multicolumn{2}{c}{log(AMP)} &
        \multicolumn{2}{c}{$F_{\rm const}$} \\
	\cline{2-3}
	\cline{4-5}
	\cline{6-7}
	\colhead{} &
	\colhead{Median\tablenotemark{1}} &
	\colhead{MAP} &
	\colhead{Median} &
	\colhead{MAP} &
	\colhead{Median} &
	\colhead{MAP} 
}
\startdata
\multicolumn{7}{c}{Interval C: 2000/06--2007/05} \\
\hline
	J(K)  &  $-0.70^{+0.63}_{-1.01}$    & -0.55 & $0.25^{+0.01}_{-0.01}$ &  0.25   &  $-8.49^{+0.29}_{-0.30}$ & -8.45    \\
	H(K)  &  $2.44^{+0.14}_{-0.21}$     &  2.45  & $0.35^{+0.01}_{-0.01}$ &  0.36   &  $-10.33^{+0.44}_{-0.40}$  & -10.31   \\
	L(K)  &  $52.78^{+14.56}_{-26.14}$  &  53.55 & $0.50^{+0.09}_{-0.08}$  & 0.52    &   $97.96^{+8.97}_{-8.68}$ &  97.10   \\
	\hline
\multicolumn{7}{c}{Interval D: 2008/08--2016/01} \\
\hline
	J(K)  &  $-9.84^{+3.71}_{-4.76}$ & -9.38  &  $0.35^{+0.02}_{-0.02}$  &  0.35  & $-14.01^{+0.91}_{-0.86}$  & -13.96    \\
	H(K)  &  $-2.47^{+2.00}_{-3.49}$ & -1.67  &  $0.56^{+0.03}_{-0.03}$  &  0.56  & $-18.22^{+1.88}_{-1.82}$  & -18.24   \\
	L(K)  &  $31.94^{+75.97}_{-37.12}$ & 34.11 &  $0.35^{-0.14}_{-0.12}$  &  0.36   &  $132.34^{+15.01}_{-19.45}$ &  131.09  
\enddata
\tablenotetext{1}{The ``uncertainties'' of these median values are 2 $\sigma$ (i.e., 2.5\%, 50\% and 97.5\% quantiles).}
\end{deluxetable*}

\begin{figure*}[htp]
\center
\includegraphics[width=1.0\hsize]{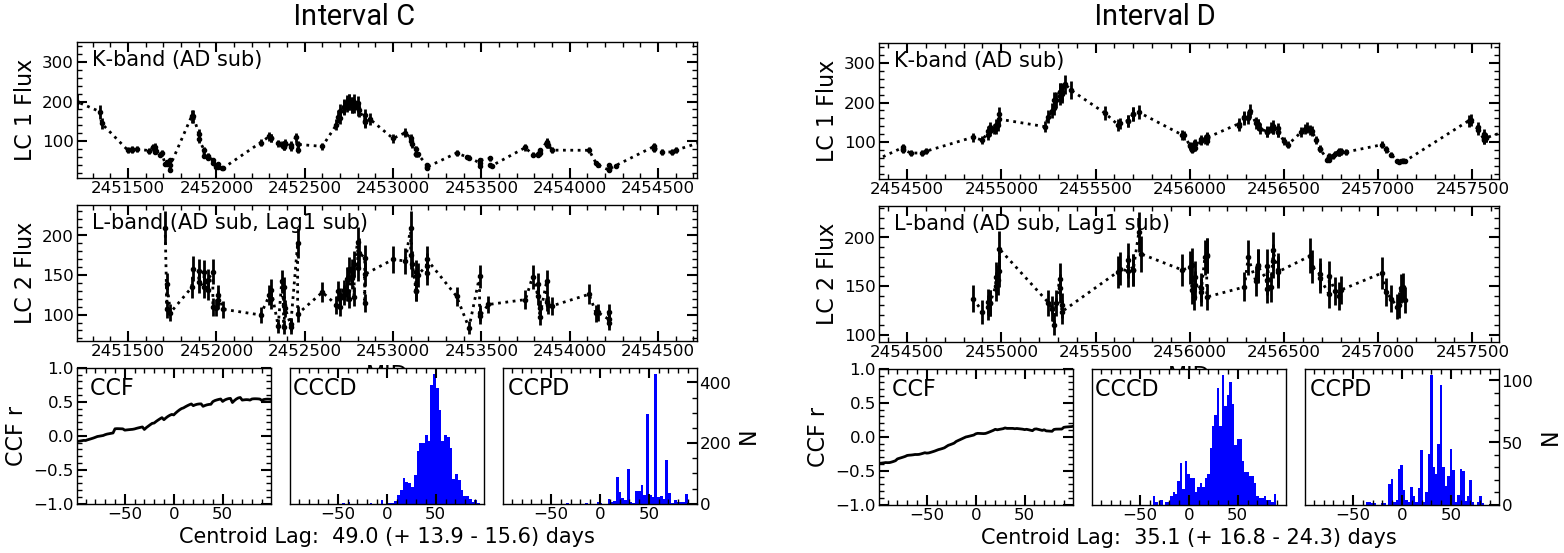}
\caption{ Interpolated cross-correlation analysis of the secondary lag in the L-band for Interval 
C (left) Interval D (right). The top and middle panels show
the K-band light curve and the K-subtracted L-band light curve after removing the accretion disk (AD)
variability, which have been used as input for the PyCCF code. The bottom three panels show 
the cross-correlation function (CCF), the cross-correlation centroid
distribution (CCCD) and the cross-correlation peak distribution (CCPD). The breadth of the CCF indicates that there are additional components besides the variable one; for example, half of the signal in Interval C comes from a persistent component discussed later in the paper. The quoted centroid lags are for the CCCD cases; see text for further details. The secondary peak at zero lag in Interval D indicates that our simple subtraction of the K-band light curve from the L-band one left a residual signal with similar behavior to that at K-band. }
\label{fig:ir_corr_check}
\end{figure*}

{ We applied two additional independent approaches to test the {\it Dynesty} result of a lag of $\sim$ 50 days of L relative to K.
\citet{Peterson1998} discuss issues with the standard cross correlation function (CCF) analysis both in accurate lag determination and in error estimation. We used their  recommended model-independent Monte Carlo methods to overcome these issues.} We 
carried out an  interpolated cross-correlation function (ICCF) analysis with the 
PyCCF code \citep{Peterson1998, Sun2018} on the K-band (with the accretion
disk variability removed) and L-band (with the accretion disk variability and
the first lag signals, i.e. K-band, removed) light curves. Our approach follows closely the recommendations of  \citet {Peterson1998}. Adopting the flux randomization 
technique, we made 5000 Monte Carlo simulations
to determine the cross-correlation centroid distribution (CCCD) and the
cross-correlation peak distribution (CCPD) as well as the corresponding uncertainties. The results are shown in Figure~\ref{fig:ir_corr_check}. The centroid and peak lags for L-band behind K-band are $48.97^{+13.90}_{-15.59}$ and $51.0^{+17.0}_{-20.97}$ days
for Interval C, respectively for CCCD and CCPD. They are  $35.14^{+16.82}_{-24.30}$ and $34.00^{+16.86}_{-26.63}$ days for Interval D, i.e., consistent within the errors (the quoted uncertainties indicate where 15.87\% of the trials fall above and below the indicated lags, respectively, i.e., $\sim$ 1 $\sigma$). Both results agree well with the {\it Dynesty} analysis.

{ For another test, 
we used an Improved ICCF (I$^2$CCF) tool (Guo et al. 2021, in prep) \footnote{\url{https://github.com/legolason/PyIICCF/}}.}
The basic methodology of this code has been introduced in 
\cite{Uttley2003, Arevalo2008, Chatterjee2008} and \cite{Max-Moerbeck2014}. Briefly speaking, $\sim$10$^3$ realizations 
of Monte Carlo simulations were made to produce optical and IR light curves with the DRW models. 
The simulated light curves  are segments randomly selected from the DRM model predictions after a 100-times longer time coverage with the same 
observational constraints. The same cadence and uncertainties as the observations
are then assigned to each simulated light curve  (i.e., we independently evolve the optical and IR light curves and check their 
behavior over a long time period with the same observational conditions.)
The CCFs of all the simulated light curves are then measured. The number of positive lags with 
the peak CCF value $r_{\rm max}$ higher than the observed $r_{\rm max}$  relative to the number of
total simulations yields a $p$-value that describes the probability that a correlation signal between the two light curves could occur by chance. 
For the L-band data in the Interval C, we
got a value of 59$^{+11}_{-19}$ days for the lag of L behind K. The $p$-value is 0.019, i.e., the null hypothesis that this lag detection (i.e., no lag) arises from noise can be rejected at the 98.1\% (=1-$p$) confidence level. This third method therefore is also in good agreement with the $\sim$ 50 day lag of L relative to K found previously.

In summary, we { confirm that there are two dominant lag times in the JHKL data, consistent with the separate dust components} as predicted in
Section~\ref{sec:sed-fit}.

\subsubsection{ Reverberation Analysis of the Optical and IR Light Curves}
\label{sec:fullreverb}

To obtain further constraints on the torus structure, we now apply a two-lag
dust reverberation model as described in Section~\ref{sec:method} to the 
optical and IR light curves of NGC 4151 together.
{ First, we fitted the B-band and {individual IR light curves} in pairs
in Interval C,
where the K-band light curve is best-sampled (details provided in Appendix C).}  Two components have been
robustly detected with IR-to-optical lag values of $\sim$40 days and $\sim$90
days. The lag difference is consistent with the analysis of the K-
and L-band light curves in the previous section. In addition, the relative
strength of the long-lag component to the short-lag component increases as a
function of wavelength, consistent with the SED fitting analysis in
Section~\ref{sec:sed-fit}.

Besides fitting each IR light curve individually, we also combined all the IR
light curves in Interval C and fitted them with an integrated model that
includes two variable black body spectra with separated time lags and that fits all four infrared bands simultaneously. For this model, the dust
emission at a particular IR wavelength  $F(t, \lambda)_{\rm IR, dust}$  is
related to the AGN optical light curve $F(t, \lambda)_{\rm OPT}$ by
\begin{equation}
\begin{aligned}
    F(t, \lambda)_{\rm IR, dust} = & F(t-\Delta t_1)_{\rm OPT}\times B(T_1, \lambda)\times {\rm AMP}_1  +\\
    & F(t-\Delta t_2)_{\rm OPT}\times B(T_2, \lambda)\times {\rm AMP}_2 + \\
    & F_{\rm const}(\lambda) ~~,
\end{aligned}
\end{equation}
where $B(T_1, \lambda)$, $B(T_2, \lambda)$ are the two black body spectra with
different dust temperatures $T_1$ and $T_2$; AMP$_1$, and AMP$_2$ are the
corresponding time- and wavelength-independent scaling factors and  $\Delta
t_1$, $\Delta t_2$ are the time lags. { Although the emission spectrum is likely to be more complex than just the two blackbodies included, the analysis in Section~\ref{sec:sed-fit} shows that such an approximation provides a good fit and probably reflects first-order reality.}

\begin{figure*}[htbp]
\center
\includegraphics[width=0.7\hsize]{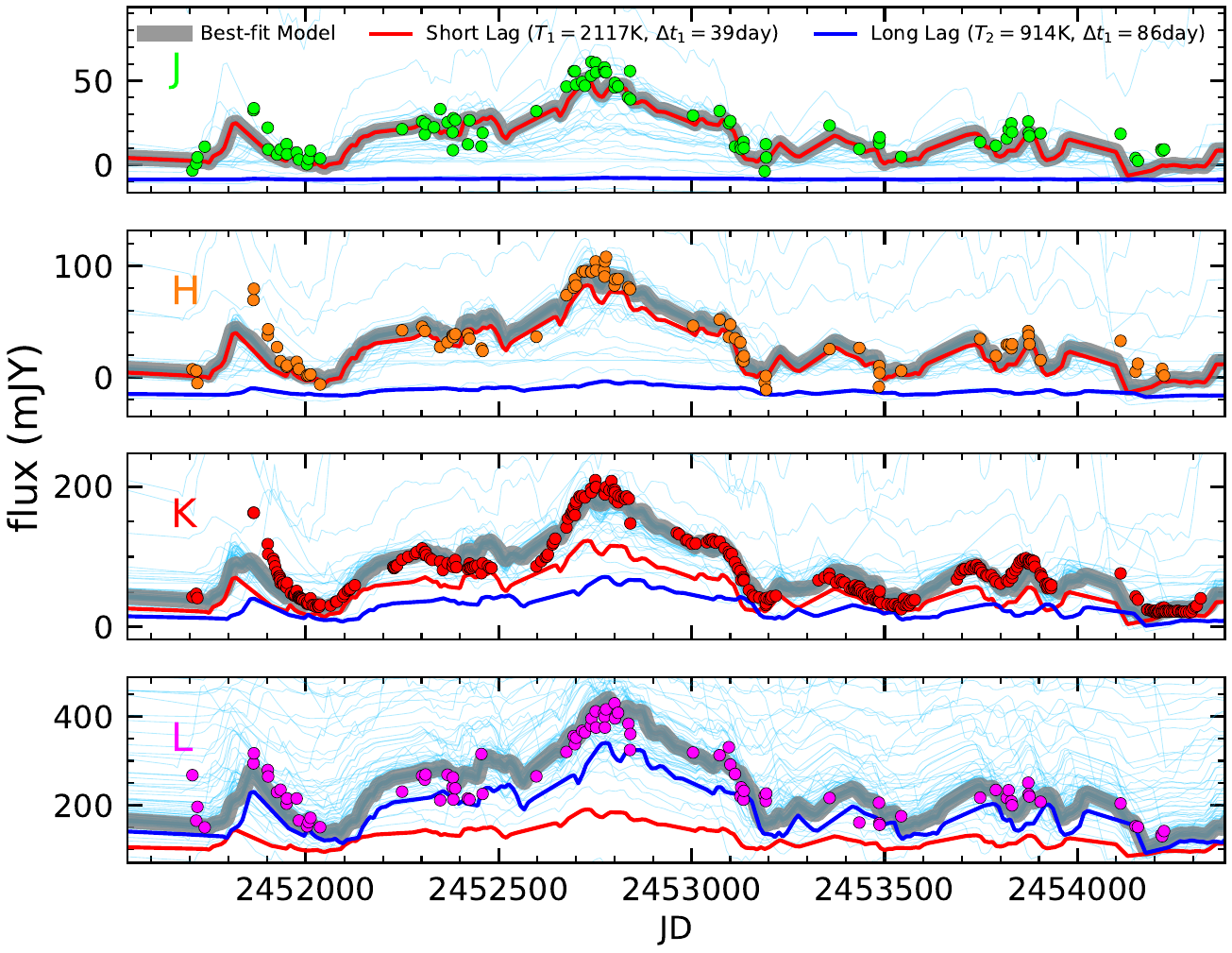}
\caption{Best-fit results of all the IR light curves in Interval C with an
	integrated two-black-body dust reverberation model (see text for the
	model details). The observed light curves are shown as dots with
	different colors. We plot the best-fit model as a thick grey line and
	the contribution from the two different dust components as red and blue
	lines. The relative contribution of the long lag (blue line) gradually
	increases towards the longer wavelength and finally dominates the
	L-band variability.}
\label{fig:newmodel-event-C}
\end{figure*}

\begin{figure*}
\includegraphics[width=1.0\hsize]{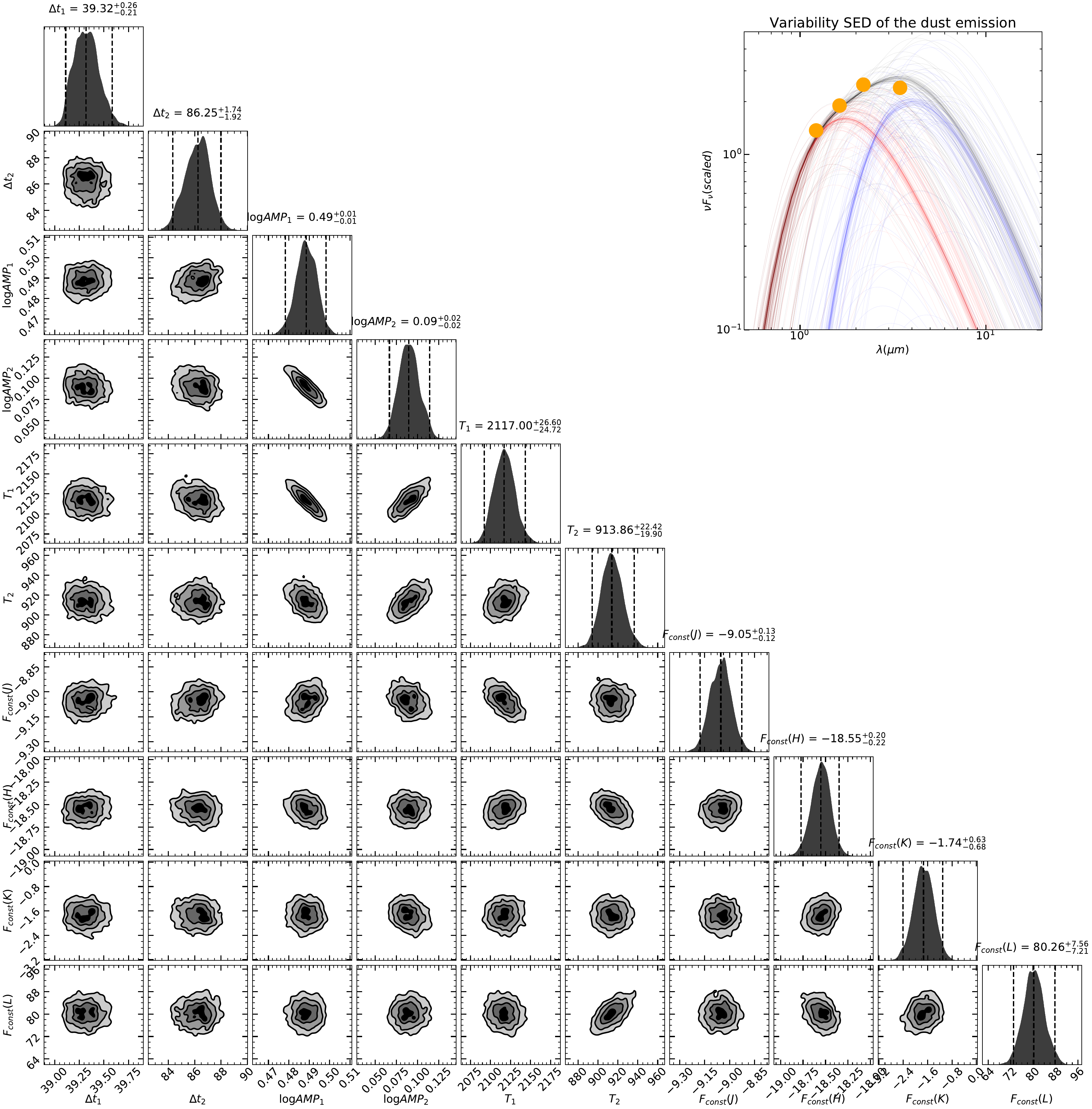}
\caption{Marginalized posterior probability distributions of the fitting
	parameters of the integrated two-black-body dust reverberation model
	for light curves in Interval C. On the top of each histogram,
	we denote the median value of the fitted parameters with 2-$\sigma$
	``uncertainties'' (i.e., 2.5\%, 50\% and 97.5\% quantiles) that cover
	95\% credible intervals. In the top-right corner, we present the
	SEDs of the two best-fit black bodies (red and blue lines) and their
	combination (grey lines). The standard deviations of J, H, K, L light
	curve flux are shown as orange dots and they generally match the model
	prediction despite the fact that the model does not try to fit this SED
	directly.}
\label{fig:newmodel-event-C-corner}
\end{figure*}

{ We used {\it Dynesty} to sample the model parameter space, deriving the best-fit model in Figure~\ref{fig:newmodel-event-C}, and we summarize the best-fit parameters of this
model in Table~\ref{tab:newmodel-c}\footnote{We arbitrarily increased the
fitting weights of the J, H, and K data at JD 2452700--2452900 by a factor of
five to mitigate the fitting bias of the AMP parameter due to the relatively
short duration and resulting modest sampling of the high-flux epochs. This
adjustment did not significantly change the best-fit values of other
parameters.}. Figure~\ref{fig:newmodel-event-C-corner} provides marginalized posterior probability distribution of
the fitting parameters, showing that the model is well-constrained,
The reported time lags are $\Delta t_1=39.3^{+0.3}_{-0.2}$~days 
and $\Delta t_2=86.3^{+1.7}_{-1.9}$~days, consistent with the cross-correlation
analysis and the individual IR light curve fitting. The best-fit temperatures
are $T_1=2177^{+27}_{-25}$~K and $T_2=914^{+22}_{-20}$~K, also in good agreement 
with the variability SED analysis in Section~\ref{sec:sed-fit}. }

\begin{deluxetable}{@{\extracolsep{4pt}}cccc}
    \tabletypesize{\footnotesize}
    \tablewidth{1.0\hsize}
    \tablecolumns{4}
    \tablecaption{Best-fit parameter values of the model fitting all light curves in Interval C \label{tab:newmodel-c}
    }
    \tablehead{
	\colhead{Parameter} &
	\colhead{Prior\tablenotemark{1}} &
	\colhead{Median\tablenotemark{2}} &
	\colhead{MAP} 
}
\startdata
	$\Delta t_1$/day        & $[-50, 150]$      &  $39.32^{+0.26}_{-0.21}$       &  39.37   \\
	$\Delta t_2$/day        & $[-50, 150]$      &  $86.25^{+1.74}_{-1.92}$       &  85.98   \\
	$\log(AMP_1)$       & $[-2, 2]$         &  $0.49^{+0.01}_{-0.01}$        &  0.49   \\
	$\log(AMP_2)$       & $[-2, 2]$         &  $0.09^{+0.02}_{-0.02}$        &  0.09   \\
	$T_1$/K               & $[1000, 2500]$    &  $2117.00^{+26.60}_{-24.72}$   &  2118.45 \\
	$T_2$/K               & $[500, 1500]$     &  $913.86^{+22.42}_{-19.90}$    &  914.43  \\
	$F_{\rm const}(J)$/mJy  & $[-20.1, 20.1]$   &  $-9.05^{+0.13}_{-0.12}$       &  $-$9.02  \\
	$F_{\rm const}(H)$/mJy  & $[-29.4, 29.4]$   &  $-18.55^{+0.20}_{-0.22}$      &  $-$10.54  \\
	$F_{\rm const}(K)$/mJy  & $[-24.9, 74.6]$   &  $-1.74^{+0.63}_{-0.68}$       &  $-$1.62 \\
	$F_{\rm const}(L)$/mJy  & $[-234.3, 702.9]$ &  $80.26^{+7.56}_{-7.21}$       &  80.51 
\enddata
\tablenotetext{1}{All these priors have been sampled linearly.}
\tablenotetext{2}{The ``uncertainties'' of these median values are 2 $\sigma$ (i.e., 2.5\%, 50\% and 97.5\% quantiles).}
\end{deluxetable}

Despite the poorer time sampling, we have applied the two-lag B-band and JHKL dust reverberation mapping model to the individual IR light curves in intervals A, B,  D and D' and summarize the
best-fit parameters in Table~\ref{tab:2lag-fit}. In all cases, the shorter lag
is similar to that found for interval C, ranging from 30--40 days. The second
lags for the intervals other than C have a wide scatter, presumably because of
the various shortcomings in the data such as poorer time sampling and features in B-band that are not reflected in the IR, such as those
discussed previously for Intervals C and D. We therefore do not include the
results in the table, since we believe that the infrared-to-infrared lags
reported in Table~\ref{tab:cc-ir} are much more reliable indications of the
behavior.

\begin{deluxetable}{cccccc}
    \tabletypesize{\footnotesize}
    \tablewidth{1.0\hsize}
    \tablecolumns{6}
    \tablecaption{MAP values of the best-fit parameters from the 2-lag model fitting of individual IR light curves in different epochs\label{tab:2lag-fit}
    }
    \tablehead{
	\colhead{Band} &
        \colhead{$\Delta t_1$ (day)} &
	\colhead{log(AMP$_1$)} &
        \colhead{$\Delta t_2$ (day)} &
	\colhead{log(AMP$_2$)} &
        \colhead{$F_{\rm const}$}
}
\startdata
\multicolumn{6}{c}{Interval A: 1976/03--1980/02\tablenotemark{1}} \\
	K  & 30--40  &  ..  &  .. & .. & ..\\
\multicolumn{6}{c}{Interval B: 1994/05--2000/09} \\
	H  &  32.41 &  -0.024 & ..  & .. &  -21.75 \\
	K  &  27.76 & 0.29    &  ..  & ..  &    5.89 \\
	L  &  19.52 & 0.49    & ..  & ..  &  146.95 \\
\multicolumn{6}{c}{Interval C: 2000/06--2007/05} \\
	J  &  39.14 & -0.30 & 93.78 & -0.10 & -8.67  \\
	H  &  41.97 & 0.16 & 89.45 & -0.69 & -9.42 \\
	K  &  36.43 & 0.43 & 92.28 &  0.29 &  3.43 \\
	L  &  31.17 & 0.48 & 92.84 &  0.64 & 107.15 \\
\multicolumn{6}{c}{Interval D: 2008/08--2016/01} \\
  H  & 14.14\tablenotemark{2} & 3.11e-2 &  .. & .. &  -4.00e-1 \\
  K  & 40.36 & 6.50e-1 & .. & .. &  2.80e+1 \\
  L  & 36.22 & 0.63 & .. & .. & 180.58 \\
	\multicolumn{6}{c}{Interval D': 2010/11--2018/08 (X-ray)\tablenotemark{3}} \\
  H  & 34.72 & 0.48 & .. &  .. &  -53.05 \\
  K  & 36.13 & 0.82 & .. &  .. &  -72.77 \\
  L  & 32.71 & 1.02 & .. &  .. &  56.79
\enddata
\tablenotetext{1}{As the Interval A data is composed by different datasets with very limited time sampling, the 2-lag 
model is poorly constrained. We only provide the possible range of the short time lags that dominate the K-band signal.}
\tablenotetext{2}{These small lags at the shortest fitted wavelength may indicate contamination 
by the direct central engine signal.}
\tablenotetext{3}{For Interval D', we have approximated the accretion disk B-band light curve by shifting and scaling the X-ray light curve to match the observed B-band light curve data and fitted the IR light curves with this synthetic B-band light curve.}
\label{tab:2lag}
\end{deluxetable}

{ The results reported in this section strongly 
validate that the AGN near-IR emission is
dominated by two dust components with different locations and 
temperatures corresponding to the sublimation of silicate and carbon grains. If the 
torus is perfectly face-on to the observer, the two time lags as best
measured in Interval C correspond to physical scales (radii) of about 0.033 pc and
0.076 pc.}

{ 
\subsubsection{Effect of sampling cadence on lag results}
\label{sec:cadence}

To understand the errors in IR reverberation mapping, we investigated the dependence of the derived lags on the level of
light curve sampling by fitting only the  {SAI} K-band photometry (omitting
that from \citet{kosh14}) for Interval C. The sampling in K-band is then for the best long sequences only roughly at the Nyquist rate for the expected $\sim$ 35 day lags, with some intervals falling below this rate (and annual gaps). Although the derived  temperatures for the two
dust components were similar to those using the full set of data, the lags were
30 and 48 days, compared with 41 and 90 days using all the data. This result
suggests that previous studies, which were generally based on relatively lower
sampling cadences, may have systematic errors as a result. Such errors may
account for some of the apparent changes in behavior between different time
intervals \citep[e.g.,][]{okny14}. The same concern applies to our analysis of
Intervals B and D. 

Where Nyquist sampling is not available, the simpler approach in \citet{Lyu2019} is preferable, namely smoothing and delaying the driving light curve to the expected behavior of the driven one, and comparing the results directly. This method strongly constrains the free parameters of the fits; e.g., it is unlikely to extract multiple lags correctly. However, the consistency of the lags derived in that work with the expected square root of luminosity scaling indicates its validity with the undersampled data available in that study. 
}

\subsection{Temporal Evolution of the Torus Inner Properties}\label{sec:innertorus}

Taking advantage of the $\sim$30-year time coverage of the IR and optical light
curves of NGC 4151, we now explore the long-term behavior of the infrared emission as an indication of possible evolution of the torus properties.
As pointed out in Section~\ref{sec:host_sub}, even at minimum light the $H-K$
and $K-L$ colors are significantly redder than expected for a normal galaxy.
This behavior reveals a persistent near infrared source component that does not
partake in the reverberation-mapped variability. 
Determining its nature is made more complex by the
long-term trends in the variable flux. 

\begin{figure}[htp]
\center
\includegraphics[width=1\hsize]{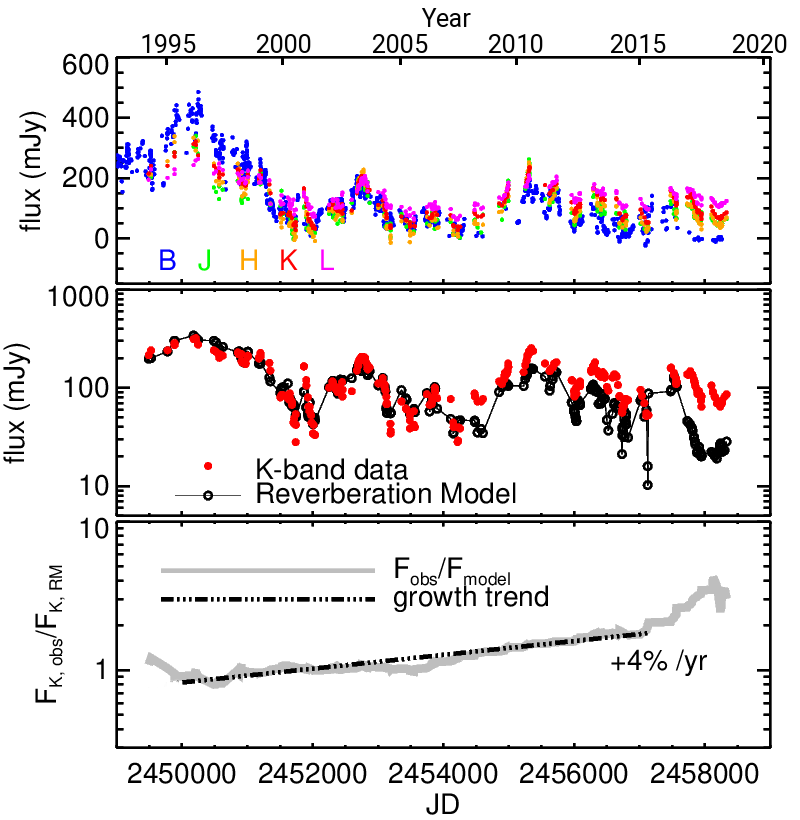}
\caption{Top Panel: Comparison of the B, J, H, K, L light curves of NGC 4151.
	We normalized these light curves over Interval C.  Middle Panel--the
	best-fit dust reverberation model (black open circles connected with
	black lines) to the observed K-band light curve over the whole observed
	interval (red filled dots).  Bottom Panel: the ratio of the observed
	K-band flux to the prediction from the best-fit dust reverberation
	model (grey line), which assumes a constant ratio of K-band to B-band
	fluxes. { After a  period (to the extreme left) where the ratio decreases (discussed in   \citet{okny99}), the ratio increases steadily.} We fit this trend with an exponential  (black
	dot-dash line) and find a growth by $\sim$ 4\% per year.}                   
\label{fig:ngc4151_k_growth}
\end{figure}

We discuss this latter behavior first. Although we have successfully matched the variable component of 
the IR light curves with the two-lag reverberation mapping model on a 6-8 year
scale, there is also a systematic trend over the full 25
year span: the variable component in B-band fades
substantially more than the variable component in K-band.  The top panel
of Figure~\ref{fig:ngc4151_k_growth} compares the optical B and IR (J, H, K,
L) light curves by normalizing to the average flux during Interval C. The
B-band flux is systematically higher than the scaled IR flux in the earlier
epochs but lower in the latter epochs.  To extract the systematic trend between
the optical and IR light curves, we fit a single lag model to the whole K-band
light curve and present the best-fit results in the middle panel of the same
figure. The residuals between the best-fit dust reverberation model and IR
light curves show a systematic trend as a over the whole
period, i.e., the model over-predicts the IR flux before Interval C
but under-predicts it after. { The bottom panel shows the ratio
of the K-band observed flux to the best-fit K-band reverberation model, 
showing the increasing trend with time. An exponential fit indicates 
that the ratio of variable K-band to B-band flux grows by $\sim$ 4\% per year between 1995 and 2015.}

{ 
Changes in the overall levels in the K-band relative to the 
optical been reported previously by \citet{okny99}. They find that 
the K-band was reduced relative to the optical in 1995--1998 compared with 1988--1994; 
the end of this trend can be seen to the extreme left in Figure~\ref{fig:ngc4151_k_growth}. 
They suggest that the change is evidence for enhanced grain sublimation during the peak 
in nuclear luminosity in $\sim$ 1995. Our finding of changes over an extended time not 
associated with extraordinary activity in the optical suggests that there must be 
some other cause.
}

\subsubsection{The Persistent Near Infrared Component}\label{sec:non-var-comp}

{ Having characterized the long-term trend in the variable emission component, 
we turn to the persistent one.} We have assumed two subcomponents: (1) a variable one that has
the same relative amplitude of variations in each band as the blue light; and
(2) one that is constant in each of H, K, and L bands. We have fitted this
simple model separately for each of Intervals B, C, and D, using the {SAI}
photometry (since our method of reconciling the other sets could damp out a
constant component). We required that the SED of the persistent source be
similar as derived individually for each of the intervals. 

The result of this modeling is that the persistent source fluxes at H, K and L
can be fitted by a blackbody of $\sim$ 700K. This modeling also confirms the
$\sim$ 4\% per year growth in the variable component { after 1995}.

\subsubsection{Long-term Growth of Torus Hot Dust Component}\label{sec:torus-growth}

The gradually increasing near-IR flux partaking in the dust reverberation
signals, as just derived, indicates either: (1) the B-band flux is not truly
representative of the heating of the variable infrared component; or (2) more
and more dust grains have been heated by the AGN. 

We first explore whether the trend  arises because the B-band has
faded significantly more than the shorter wavelength (hard UV, soft X-ray)
continuum that contributes to the dust heating. Interstellar dust has a broad
absorption peak between $\sim$ 1 nm and 700 nm \citep{corrales2016}. It is
believed that the heating of the circumnuclear torus is from a single source
component dominating the emission from the optical to the soft X-ray
\citep{schurch2004}.  The nuclear continuum over this range falls roughly as
$\nu^{-1}$, or $\nu f_{\nu} \sim$ constant \citep{alexander1999}. Assuming all
of the nuclear flux is absorbed over the relevant spectral range, the heating
of the circumnuclear torus is therefore roughly constant per logarithmic
frequency interval. 

\begin{figure}[htp]
\center
\includegraphics[width=1\hsize]{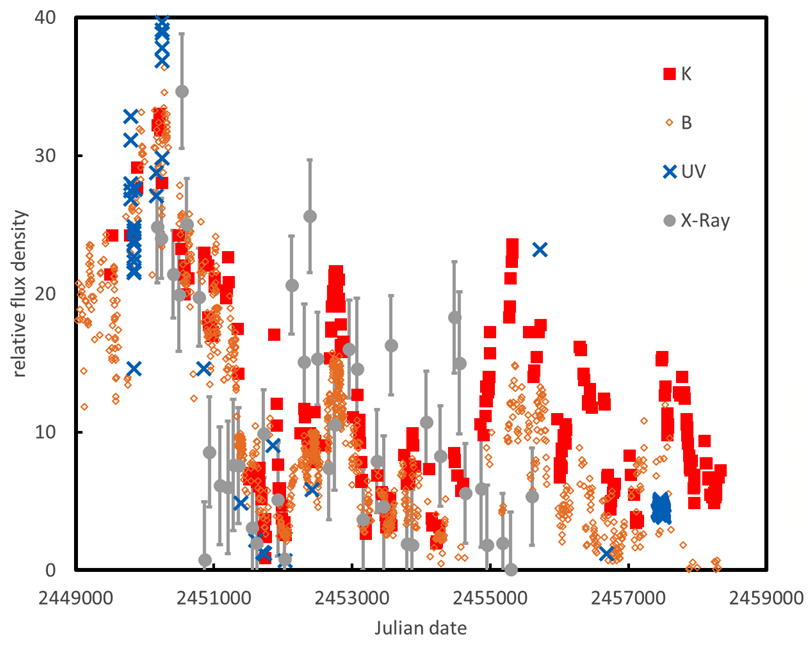}
\caption{Comparison of the long-term behavior of the variable output of NGC
	4151 at K, B, in the UV, and in the soft X-ray. The K-band and B-band
	data are as discussed previously; we have subtracted the persistent
	component from the K-band data and the galaxy from both. The U-band
	data are from \citet{couto2016} at 0.135 $\mu$m, except for the final
	complex of measurements that are from \citet{Edelson2017} at 0.193
	$\mu$m but are corrected to 0.135 $\mu$m assuming a $\nu^{-1}$ spectrum
	in frequency units. The X-ray measurements are from the Rossi XTE
	all-sky-monitor, band A, from 1.3--3 keV.}
\label{fig:fig7}
\end{figure}

Much of this frequency range is inaccessible due to interstellar absorption,
but we can probe both the UV and soft X-ray, as shown in Figure~\ref{fig:fig7}.
All the measurements are normalized to be of similar strength during the strong
peak in emission between JD 2449000 and 2451000. With a single exception (near
JD 245750, presumably an outburst), all the UV measurements appear to fade in
synchronism with the B-band ones. This result is demonstrated in greater detail
for NGC 5548 by \citet{Gaskell2008}, providing a useful analogy to NGC 4151.
The soft X-ray measurements to some extent have their own individual excursions
but still appear also to fade as in B-band. Although we cannot constrain the
behavior between 0.135 $\mu$m = 135 nm and 1.3 keV (1 nm), the available
evidence suggests that the B-band behavior is characteristic of that of the
heating source of the torus hot dust emission in general, since both the UV and
soft X-ray behave similarly to B-band.  

We therefore attribute the
increase in the variable K-band to B-band relative flux to growth in the amount of exposed dust associated with the 
circumnuclear torus. From the K-band emission, we estimated the hot dust mass
of about $7\times10^{-4}~M_\odot$. With a $\sim$ 4\%/yr flux growth rate, the
corresponding dust mass growth would be $\sim 2.7 \times 10^{-5}~M_\odot$/yr.
One possible source for these dust grains is the ultra-fast outflows in NGC
4151 (see references in \citealt{Mou2017, williamson2020}).  If there is torus dust lifted up
by such outflows or some dust forms in-situ \citep{Elvis2002}, such excess IR
emission can be expected.  In fact, the typical outflow rate in the NGC 4151
nucleus is calculated to be $3\times10^{-3}$-- $8\times10^{-2}~M_\odot/{\rm
yr}$ \citep{Mou2017}. Once the gas to dust mass ratio of these clouds is below
300-8000, the required dust mass growth can be provided.  The possibility of
changes in the inner structure of the circumnuclear torus on decadal timescales
due to winds and turbulence thus deserves further consideration.

Finally, the disappearance of the nucleus in the B-band at the end of the
sequence we have collected (JD 2457890 and beyond in Figure~\ref{fig05}) is
highly suggestive of the hypothesis that rapid changes in the appearance of an
AGN nucleus can occur due to a small BLR cloud intervening along our line of
sight \citep{wang2012, gaskell2018}. The near infrared flux does not seem to
react to this event. The situation can give interesting insights to the AGN
behavior and furthermore emphasizes the hazards in using B-band as a reference
for infrared reverberation mapping at some epochs. 

\subsubsection{Lack of Evidence for a Receding Torus}\label{sec:receding-torus}

Since the optical emission of NGC 4151 has varied by a factor of
seven between 1970 and 2018, the torus inner size is expected to vary by a
factor of 2.6 if the dust sublimation radius follows the expected relation,
$r_\text{sub}\propto\sqrt{L_\text{AGN}}$.  However, we find an inner lag of $\sim$30-40 days for all  four broad time intervals, corresponding to very limited variations.

\begin{figure*}[htp]
\center
\includegraphics[width=0.6\hsize]{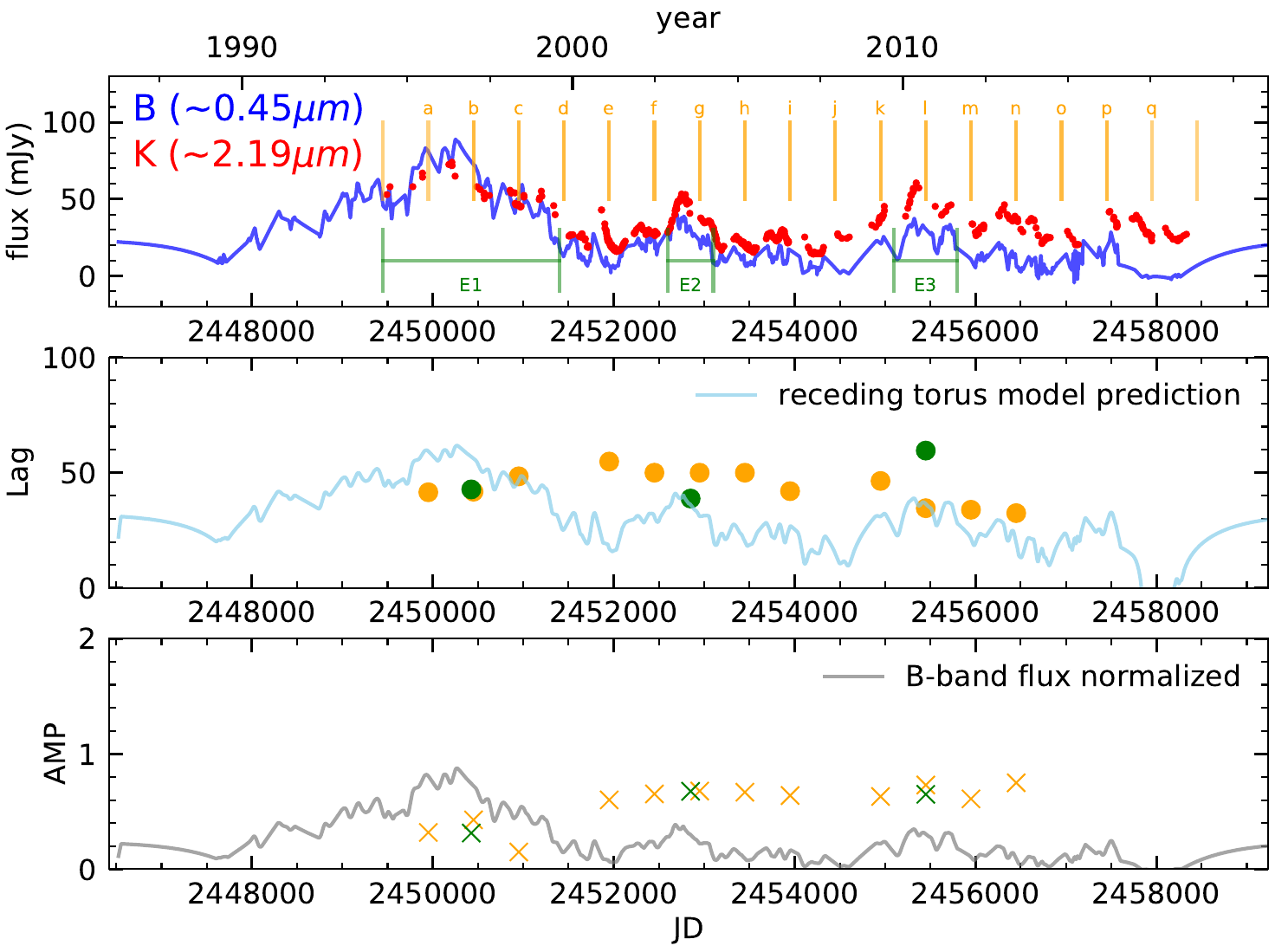}
\caption{Time evolution of the best-fit single-lag dust reverberation model of
	NGC 4151.  Top panel: the optical B-band (blue) and near-IR K-band
	(red) light curves and the different epochs in which the fitting is
	carried out. Middle panel: the evolution of the K-to-B band time lag
	for epoch a, b, ... q (orange dots) and E1, E2, E3 (green dots). The
	light blue is the expected trend if the dust lag is following a $\Delta
	t\propto \sqrt{L_{\rm AGN}}$ relation; we have normalized this relation
	assuming a K-band single lag of 37 days for the mean AGN B-band flux at
	Interval C.  Bottom panel: the evolution of the relative variation
	amplitude (AMP).  The grey line is a scaled version of the optical
	light curve.}
\label{fig:lag_evolution}
\end{figure*}

To further explore if the torus size  evolves with time, we divide the K-band light curve into 17 epochs with a time window of 1000
days and a step size of 500 days.  Due to the relatively low-cadence 
sampling in these epochs, we fit a single dust lag 
independently to the IR data within each epoch. The results are summarized in Figure~\ref{fig:lag_evolution} and
Table~\ref{tab:1lag-fit}\footnote{Since the K-band variability
includes two time lags, the lag from this single lag model
is systematically larger than the shorter lag (and smaller than the longer lag)
reported in Section~\ref{sec:fullreverb} and Table~\ref{tab:2lag}.}.  The best-fit K-band time lag has a
range of 30-60 days and does not follow the evolution of the
optical light curve, irrespective of the smoothing window size we use
(1/4$\times$, 1/2$\times$ or 1$\times$lag). These results indicate very weak,
if any, evidence for a receding torus, contrary to previous conclusions in
\citet{Koshida2009} and \citet{Kishimoto2013} but consistent with the results
reported in \citet{Schnulle2015}.

\begin{deluxetable*}{ccccccccc}[hbt!]
\tabletypesize{\scriptsize}
\tablecaption{Best-fit Parameters of Single-lag Model Fitting at K band \label{tab:1lag-fit}}
\tablewidth{0pt}
\tablehead{
 \colhead{epoch ID}  & \colhead{JD range}  &  \colhead{JD (median)} &  \multicolumn{2}{c}{1/2 lag smooth} & \multicolumn{2}{c}{1/4 lag smooth} & \multicolumn{2}{c}{1 lag smooth} \\
               &                     &                         & \colhead{Lag} &  \colhead{log(AMP)} &  \colhead{Lag}  & \colhead{log(AMP)} & \colhead{Lag} & \colhead{log(AMP)} 
}
\startdata
a &  2449450--2450450 &  2449950 &    41.52  &   0.32  & 43.35  &  0.32  &   40.92  &  0.339     \\ 
b &  2449950--2450950 &  2450450 &    41.77  &   0.43  & 43.93  &  0.43  &   42.66  &  0.434     \\ 
c &  2450450--2451450 &  2450950 &    48.42  &   0.15  & 48.00  &  0.16  &   44.50  &  0.168     \\ 
d &  2450950--2451950 &  2451450 &    $\cdots$    &   $\cdots$  &  $\cdots$ &  $\cdots$  &   $\cdots$  &  $\cdots$     \\ 
e &  2451450--2452450 &  2451950 &    54.84  &   0.60  & 60.01  &  0.58  &   63.24  &  0.621     \\ 
f &  2451950--2452950 &  2452450 &    50.01  &   0.65  & 45.37  &  0.64  &   51.03  &  0.660     \\ 
g &  2452450--2453450 &  2452950 &    50.00  &   0.68  & 48.56  &  0.67  &   51.01  &  0.701     \\ 
h &  2452950--2453950 &  2453450 &    50.00  &   0.67  & 40.76  &  0.67  &   47.00  &  0.715     \\ 
i &  2453450--2454450 &  2453950 &    42.00  &   0.64  & 44.00  &  0.61  &   45.00  &  0.678     \\ 
j &  2453950--2454950 &  2454450 &    $\cdots$    &   $\cdots$  &  $\cdots$ &  $\cdots$  &   $\cdots$  &  $\cdots$     \\ 
k &  2454450--2455450 &  2454950 &    46.43  &   0.63  & 45.87  &  0.63  &   44.92  &  0.637     \\ 
l &  2454950--2455950 &  2455450 &    34.61  &   0.73  & 36.03  &  0.72  &   33.51  &  0.742     \\ 
m &  2455450--2456450 &  2455950 &    33.93  &   0.61  & 34.11  &  0.60  &   32.97  &  0.627     \\ 
n &  2455950--2456950 &  2456450 &    32.47  &   0.75  & 32.32  &  0.73  &   41.00  &  0.760     \\ 
o &  2456450--2457450 &  2456950 &    $\cdots$    &   $\cdots$  &  $\cdots$ &  $\cdots$  &   $\cdots$  &  $\cdots$     \\ 
p &  2456950--2457950 &  2457450 &    $\cdots$    &   $\cdots$  &  $\cdots$ &  $\cdots$  &   $\cdots$  &  $\cdots$ \\ 
q &  2457450--2458450 &  2457950 &    $\cdots$    &   $\cdots$  &  $\cdots$ &  $\cdots$  &   $\cdots$  &  $\cdots$    \\ 
                   \hline                          
E1 & 2449450--2451400 &  2450425 &   42.76   & 0.31  &   45.83  &  0.31  & 41.06  &  0.32 \\
E2 & 2452600--2453100 &  2452850 &   38.79   & 0.68  &   38.30  &  0.66  & 51.12  &  0.72 \\
E3 & 2455100--2455800 &  2455450 &   59.60   & 0.65  &   37.23  &  0.70  & 52.99  &  0.69
 \enddata
 \tablecomments{Due to poor time samplings and/or lack of variability features, the measurements in Epochs d, j, o, p, q have been omitted.}
\end{deluxetable*}

In addition, the IR-to-optical variability amplitude AMP (bottom panel of
Figure~\ref{fig:lag_evolution}) does not show notable correlations with the
time lag, indicating the amount of dust does not change drastically on
relatively short time scales following the AGN optical variability.  However,
the AMP value increases gradually over a 10-year scale consistent with the
conclusion in Section~\ref{sec:torus-growth}.

As another test,  we focus on three time windows, E1, E2 and E3, that cover the
optical flux peaks at 1997, 2003 and 2011, to test if the torus structures are
the same when the AGN activity is in a high state. The torus sizes for these
three windows do not change within the uncertainties. The values of AMP show a
steadily increasing trend, consistent with the growing torus picture.

\subsection{Comparison with Previous Reverberation-Mapping}
\label{sec:comparison}

NGC 4151 has been the subject of extensive near-IR dust reverberation mapping 
programs. The {SAI} group has studied it in this way from the 1970s through the
2010s with J, H, K, and L band data and has reported a large range of time lag
measurements from 18 days to 104 days \citep[e.g.,][]{okny99, okny14, okny19}.
A high-cadence V-K monitoring effort over 2001-2006 was carried out as part of the
MAGNUM program \citep{Minezaki2004, Koshida2009}. \cite{Koshida2009}
reported a time-lag variation and suggested the possibility of dust destruction
and fast-reformation. Later, study by \citet{Schnulle2013, Schnulle2015}
presented the results of z, Y, J, H, K band monitoring between 2010 and 2014
and reported a decreased dust lag during the observed epoch.

However, our re-analysis of the entire data set does not support  
a strong evolution of the torus inner size. The diverse results reported in the
literature are likely a result of different methodologies of cross-correlation
analysis, and/or inadequate time-sampling of the light curves.

The evidence for a wavelength dependence of the torus size is mixed. \cite{okny99} found that
the L-band lag was significantly larger than that at K-band before 1996. After
2008, the lags in the K and L-bands became similar \citep{okny14, okny19}.
\cite{Schnulle2015} found similar lags in the J, H, and K bands but rejected evidence in their data for a possible  secondary longer delay in these bands during 2010-2014. Their
observations did not include L-band where the longer delay becomes prominent, as shown in Figure~\ref{fig:newmodel-event-C}.  The contribution of the longer lag may not be obvious in JHK,
especially when the time-samplings of the light curves are not good enough or
the intrinsic variation does not produce enough light curve features. Their
result therefore does not contradict our finding of two time lags of $\sim40$
and $\sim90$ days.  

\subsection{Reverberation at 10 $\mu$m and Longer Wavelengths}\label{sec:Nband}

\subsubsection{N-Band (10 $\mu$m): Yearly Timescales}\label{sec:yrlyNband}

\cite{sem87} reported six ``additional observations'' in IRAS Band 1
(12~$\mu$m) over a 25 day period in 1983 and found no variations. 
Insight to any variability on a year-to-year timescale at N-band (10.6~$\mu$m) is provided by the five
measurements in each of 1975 and 1976 from \citet{Rieke1981}. There
are no significant changes within each five-measurement sequence (see
Table~\ref{Table 1}).  The weighted averages for the two years are $2032 \pm
49$ and $2002 \pm 38$ mJy, respectively.  This behavior can be compared with
that of the nucleus in the B band, from \citet{lyuty99}, with the galaxy
contribution subtracted as described in that paper; the nuclear output
decreased from an average of $\sim$ 26 mJy to $\sim$ 12 mJy. To compare with
the behavior in the K band, for 1975 we take data from \citet{allen76,
stein76}.\footnote{ The second reference only gives L-band photometry, which we correct
to K-band by multiplying the flux density by 0.506, derived from the {SAI}
photometry when the source was of similar brightness. } After correction for the
galaxy contribution, making use of the multi-aperture photometry in
\citet{mca83} as well as the discussion in this paper, the average of the
measurements is 195 mJy, with good agreement (peak-to-peak differences of
$\pm$15\%). For 1976, we take measurements from \citet{kemp77, odell78} with
aperture corrections and removal of the galaxy contribution as before. The
average nuclear output was 124 mJy with similar scatter. The K-band
measurements are not simultaneous with the N-band ones, but the 1--2
month smoothing of the K-band signal makes them representative for the times of
the 10 $\mu$m observations (see Figure~\ref{fig:lc_full}). 

While the brightness of the nucleus decreased by more than a factor
of 2 at B and of 1.6 at K between 1975 and 1976, the ratio of fluxes at 10
$\mu$m was $0.985 \pm 0.030$, i.e. there was no change to within a few percent.
Alternatively, from Table 1, this change can be expressed as 30 $\pm$ 62 mJy.
If the 70 mJy change seen in K-band has a 1500K black body spectrum, a change
of order 40 mJy would be expected at 10 $\mu$m, consistent with the
observations but illustrating that the hot dust that dominates at K- and L-band
is not accompanied by large amounts of only  somewhat cooler dust sharing
its variations. That is, the 10 $\mu$m emission is generated by a different
source component, not a continuation of the one that dominates at 1--4 $\mu$m.
Consequently, the very hot dust will not contribute a
significant silicate emission feature at 10 $\mu$m because it contributes only
a very small fraction of the total output at this wavelength.

\subsubsection{N-Band: Decadal Timescales}
\label{sec:decadalNband}

\begin{figure*}[hbtp]
\center
\includegraphics[width=0.58\hsize]{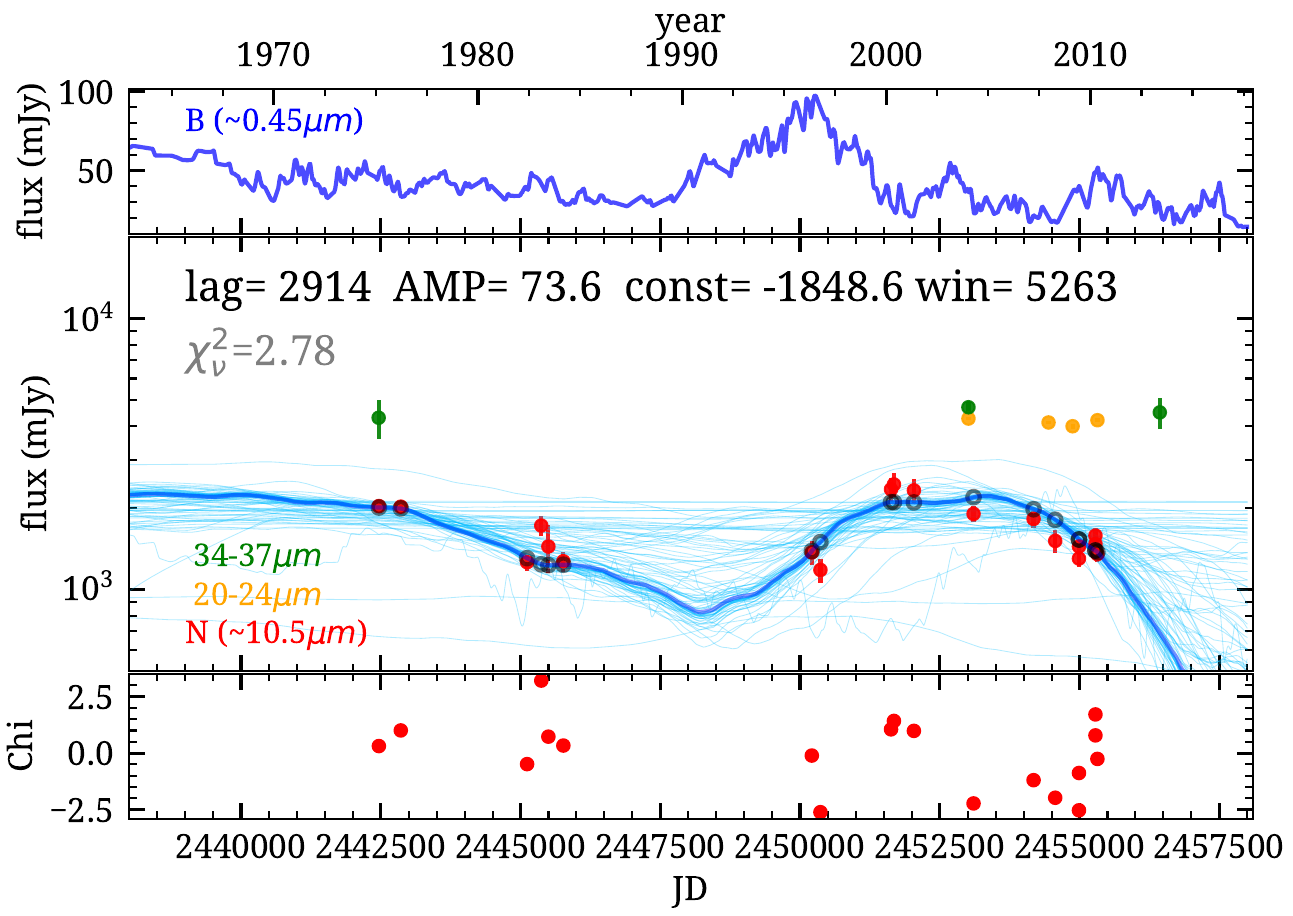}
\includegraphics[width=0.41\hsize]{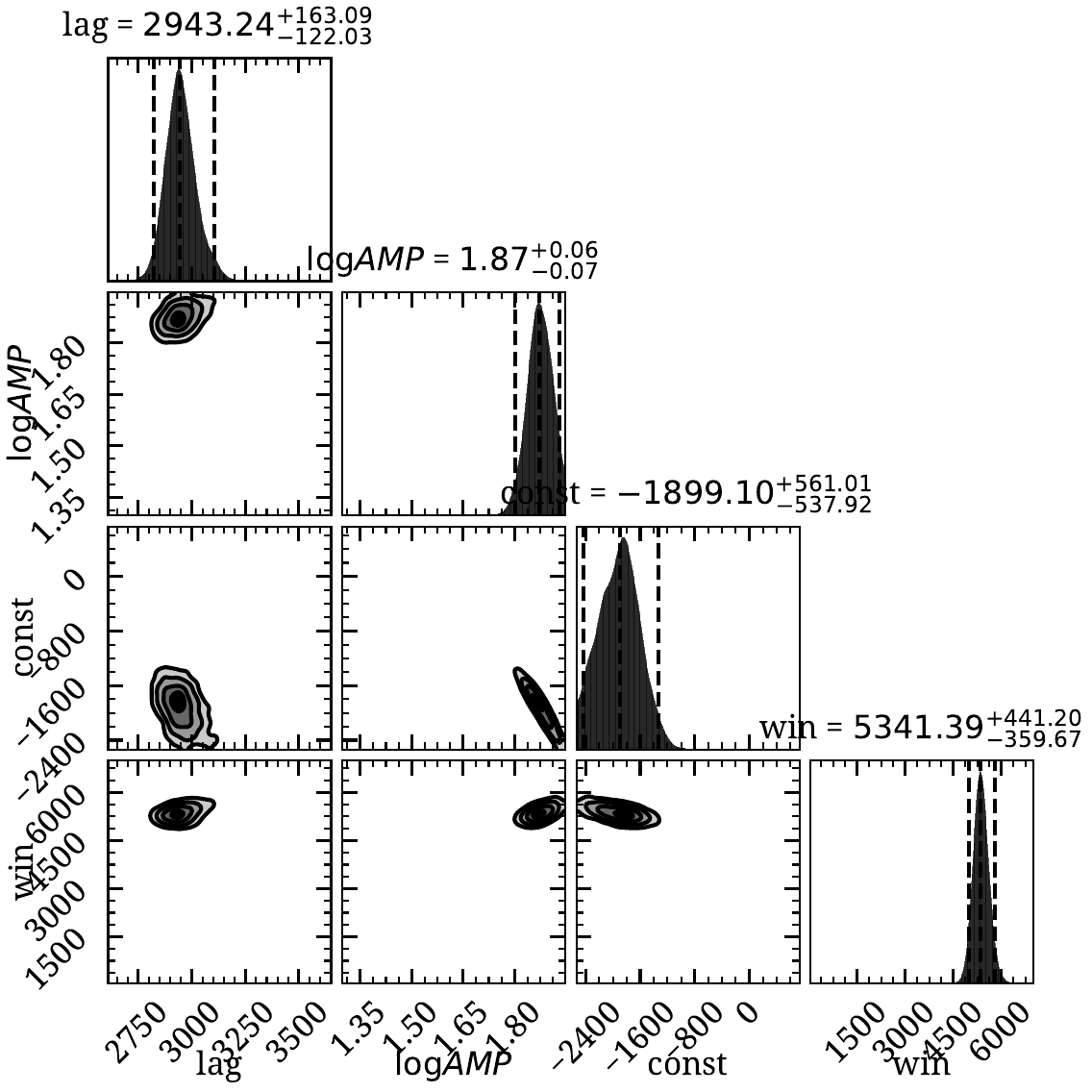}
\caption{Single-lag dust reverberation model fitting of the N-band light curve
	of NGC 4151. On the left, the top panel shows the optical B-band light
	curve (blue solid line); the middle panel presents N-band (red dots),
	20--24~$\mu$m (orange dots) and 34-37~$\mu$m (green dots) light curves,
	and the best-fit reverberation model of the N-band light curve (light
	blue curves); the bottom panel shows the residual of the N-band light
	curve fitting. Panels on the right are the marginalized posterior
	probability distributions of the fitting parameters to the N-band
	data.}
\label{fig:N-band}
\end{figure*}

{ The relatively low luminosity and
short reverberation timescales for NGC 4151 suggest that changes might be
detectable at much longer wavelengths. Simple thermal equilibrium suggests lags $\gtrsim$ 3 years at 10 $\mu$m, so with
historical observations of NGC 4151 in this band over
the past 35 years, there is a chance to detect dust reverberation
signals. 
The top panel of Figure~\ref{fig:N-band} shows a 30-day smoothed
optical B-band light curve extracted from \cite{okny16}. There is a broad
maximum between 1988 and 1998 with a flux increase by a
factor of three. The N-band flux clearly
shows a bump between 1995 and 2010 (the middle panel), as would be expected from these arguments.\footnote{
To confirm this variability,
we have carried out a $\chi^2$ test  \citep[e.g.,][]{deDiego2010} 
of the N-band data points. 
For a number of measurements with the flux $f_i$ and uncertainty $\sigma_i$, the $\chi^2$ is defined as 
\begin{equation}
    \chi^2 = \sum_i \frac{(f_i - \bar{f})^2}{(\sigma_i)^2} ~~,
\end{equation}
where $\bar{f}$ is the average value of all the measurements. 
From the $\chi^2$-distribution, a $p$-value 
can be computed to describe the significance of the measurements randomly selected from a normal Gaussian 
distribution, i.e., the AGN is not variable. Since there is no evidence for variations on a yearly timescale, we combined measurements in the same year as described in Appendix B.  We then obtained $\chi^2$ (for 18 degrees of freedom, dof=18) 
for the N-band measurements to be 283.5 with a $p$-value of $\sim10^{-50}$.
Even if we arbitrarily increase all flux uncertainties by a factor of three, 
we still have a $p$-value of 0.007, too small for the flux variation to be due to 
random measurement errors. Thus, the N-band variations over decadal timescales are undoubtedly real.
}
}

{ We used the  B-band light curve of Figure~\ref{fig:N-band} to fit the N-band light curve with a 
single-lag model as in \citet{Lyu2019}. Given the large extent of the mid-IR 
emission \citep{burt13},
we used a top-hat window to smooth the B-band light curve and left the
window size as a free parameter. }
The best-fit model, shown in Figure~\ref{fig:N-band},  has a reduced $\chi^2$ of
2.78 and a smoothing window size of about 5300 days.  The indicated time lag 
is 2943$^{+163}_{-122}$ days. 

To quantify how sensitive the best-fit N-band time lag is to the smoothing
effects, we arbitrarily fixed the smoothing window size to values from 100 to
3000 days and redid the one-lag reverberation fitting. The results are
summarized in Table~\ref{tab:N-lag}.  The best-fit N-band lags range from 2743
to 3500 days, within the range of the reported uncertainties. { If we take the extremes 
in the fits, a time lag between 2600 and 3600 days is indicated, corresponding to 
physical scales between 2.2 and 3.1 pc. }

\begin{deluxetable}{cccc}[hbt!]
\tabletypesize{\scriptsize}
\tablecaption{Best-fit Parameters of One-lag Fitting at 10.6 $\mu$m\label{tab:N-lag}}
\tablewidth{1.0\hsize}
\tablehead{
 \colhead{Smoothing Window}  & \colhead{Lag}  &  \colhead{log(AMP)} &  \colhead{$F_{\rm const}$} \\
 \colhead{(days)}  & \colhead{(days)}  &  \colhead{} &  \colhead{}
 }
\startdata
 100 (fixed) &  3481$^{+21}_{-21}$ & 1.34$^{+0.05}_{-0.06}$ & 631$^{+161}_{-156}$ \\
 300 (fixed) &  3488$^{+72}_{-88}$ & 1.29$^{+0.06}_{-0.06}$ & 773$^{+146}_{-155}$ \\
 500 (fixed) &  3427$^{+103}_{-152}$ & 1.30$^{+0.06}_{-0.07}$ & 770$^{+151}_{-155}$ \\
1000 (fixed) &  3317$^{+111}_{-121}$ & 1.35$^{+0.07}_{-0.07}$ &  670$^{+174}_{-185}$  \\
2000 (fixed) &  2794$^{+673}_{-151}$ & 1.49$^{+0.07}_{-0.08}$ & 334$^{+243}_{-252}$  \\
3000 (fixed) &  2996$^{+115}_{-154}$ & 1.61$^{+0.06}_{-0.07}$ & $-163^{+291}_{-324}$ \\
5341$^{+441}_{-360}$ &  2943$^{+163}_{-122}$ & 1.87$^{+0.06}_{-0.07}$ & $-1899^{+561}_{-538}$
 \enddata
\end{deluxetable}

\subsubsection{Longer Wavelengths: 20--40~$\mu$m}\label{sec:20var}

There is no convincing evidence for reverberation-type variations at
wavelengths longer than 10 $\mu$m.  At $\sim$20 $\mu$m, ground-based
measurement errors tend to be relatively large and there are not enough
measurements distributed appropriately to make a strong statement.  Although
the Infrared Astronomical Satellite (IRAS) has a filter near 25~$\mu$m, the
$\sim0.5'$ beam is too large, making it impossible to trace AGN variability due
to the contamination from the host galaxy.  However, the
post-2000 space-based observations from {\it Spitzer}, {\it WISE} and {\it
AKARI} have small beam sizes of 6--12\arcsec  at $\sim$22--24~$\mu$m. We converted the measured fluxes into the
equivalent values for the MIPS~24 $\mu$m filter,  making the necessary color
corrections with synthesis photometry on the {\it Spitzer}/IRS spectrum of NGC
4151.  We found 4.27$\pm$0.05 Jy on JD 2453013 from the {\it Spitzer}/IRS
spectrum, 4.13$\pm$ 0.05 Jy on JD 2445498 from {\it Spitzer}/MIPS-24~$\mu$m
images,\footnote{We adopted the PSF photometry from the SEIP source catalog,
which is measured based on a mosaic image from two sets of observations made in
May 30, 2007 and June 23, 2008.  We took the median JD of these two dates.}
4.00$\pm$0.08 Jy on JD 2454876 from an {\it AKARI}/IRC L18W
image,\footnote{This measurement is based on aperture photometry, which could
underestimate the source flux.} and 4.21$\pm$0.07 on JD 2455321 from the {\it
WISE} 22-$\mu$m image. These values agree to within 3$\sigma$, even without
taking account of systematic errors (e.g., in the conversion to the MIPS band).
Figure~\ref{fig:N-band} shows this light curve does not share the
falling-with-time trend in the N-band.

The
measurement at 34 $\mu$m by \citet{Rieke1975} of 4.3$\pm$0.7 Jy on JD 2442462
agrees well with the value of 4.5$\pm$0.6 Jy at 37 $\mu$m on JD 2457436 from
\citet{Fuller2019} (the radial profile in the latter reference shows that
virtually the entire flux would be captured in the beam used for the former
one).  One additional data point can be obtained by computing synthetic {\it
SOFIA}/{\it FORCAST} photometry at 37 $\mu$m from the {\it Spitzer}/IRS
spectrum of NGC~4151, which is 4.7$\pm$0.2 Jy on JD 2453013. Comparison with
Figure~\ref{fig:N-band} shows that these measurements span the entire duration
of the 10 $\mu$m ones and that the delayed and smoothed B-band light curve is
about a factor of four lower for the most recent measurement than for the
first. Again there is no sign of variability.  

We conclude that IR emission at these longer wavelengths comes from a more extended dust component than that for the N-band.

\section{Properties of the AGN Circumnuclear Dust Structures}\label{sec:discussion}

To summarize, the torus emission in the J, H, K and L bands (1 $-$ 4 $\mu$m) is contributed
by three distinct dust populations. Two of these populations (at $T\sim$900--1000K
and $\sim$1500--2000K respectively) are at nominal  radii of $\sim$
0.033 and 0.076 pc, with temperatures consistent those of sublimating graphite and silicate dust grains. A third, more
persistent component has a characteristic temperature of $\sim$ 700K. Over a
timescale of $\sim$20--25 years, the variable component of 
hot dust emission has increased gradually by $\sim$ 4\% per year
compared with the B-band signal, indicating a growing inner edge to the torus. We also detect a $\sim 10~\mu m$
dust reverberation signal, indicating a fourth 
component at a nominal radius of about 2.2 - 3.1 pc. The lack of 
variability at 20--24~$\mu$m and 34--37~$\mu$m indicates even
more extended dust structures.

In this section, we discuss these results in a broader context. We (1) compare them with other indications of the circumnuclear structure (\S~\ref{sec:resolved_study}); (2) use them to constrain the vertical distribution of material (\S~\ref{sec:torus-vert}); (3) consider them in the context of the AGN spectral energy distribution (\S~\ref{sec:dust_temp}); (4) use the results to confirm constraints on the dust grain sizes  (\S~\ref{sec:two-lag-physics}); and (5) highlight concerns about current models (\S~\ref{sec:models}).

\subsection{ Comparison of Reverberation Results with Other Observations of Circumnuclear Structures}\label{sec:resolved_study}

{ From the reverberation behavior, the relative sizes of the three variable dust components are 
75 (N-band warm dust):2.25 (sublimating silicate dust):1 (sublimating graphite dust). 
Given the proximity of NGC 4151, we can compare these values
with size constraints on its circumnuclear dust structures from
imaging and interferometry.}

In the near-IR, NGC 4151 has been studied at K-band with the Keck
interferometer \citep{Kishimoto2009, Kishimoto2011, Kishimoto2013, Pott2010,
Honig2014}.  \citet{Kishimoto2013} and \citet{Honig2014} summarize the results
at 2.2 $\mu$m. The discussion in the latter paper demonstrates that the
constraints on the location of the inner dust rim of the torus from
reverberation mapping and those from interferometry are consistent within the
uncertainties.  \citet{Kishimoto2013} present a single measurement that implies
the rim has increased in size; we cannot directly test this inference, but it
is of modest statistical significance and removed from any features in the
B-band light curve that might be expected to accompany such an event.  It does
not significantly undermine our conclusion that there are no substantial
changes of this nature synchronous with increases in nuclear luminosity.

At $\sim$10~$\mu$m, we can compare the source size derived from reverberation mapping in the N-band band
with the dust size measurement with the VLTI/MIDI interferometry observations of NGC 4151 by \cite{Burtscher2009}. 
Based on a comparison of
the data to a simple Gaussian model, they inferred the emission at these
wavelengths has a FWHM diameter of {(2.3$\pm$0.5)} pc\footnote{The original paper
adopted a distance of 14 Mpc for NGC 4151 and reported the size to be
(2.0$\pm$0.4) pc.}.  To compare this measurement with the size from
reverberation mapping, a number of  corrections are needed: (1) from the
geometry of the NLR light cones, \citet{Crenshaw2010, Fischer2013} estimate
that the circumnuclear torus is inclined by $\sim$ 45$^\circ$ into the sky; (2)
the position angle of the jet, 77$^\circ$ \citep{Mundell2003}, implies that the
circumnuclear disk is only $\sim$ 13$^\circ$ from north-south
\citep{Kamali2019}; and (3) the fitted Gaussian FWHM significantly
underestimates the diameter of a disk.  The first two points indicate that any
disk would be fore-shortened in the direction along the interferometer
baseline, which was close to east-west. With regard to the third point, the
full diameter of a disk with a 70\% central hole is more than 1.5 times the
FWHM of a fitted Gaussian. Taken together, the FWHM from the interferometry
would underestimate the diameter of a simple disk by about a factor of two. The
complexities of the true nuclear structure - disk, wind, or cone -  make this
only a rough estimate. However, it implies that the radius of the 10 $\mu$m
source is about equal to the quoted FWHM, which makes it similar to the radius
deduced from the reverberation behavior at this wavelength.

Additional size limits on the torus size of NGC 4151 have been measured in the
radio bands. With VLBA and the JVLA at 21 cm, \cite{Mundell2003} found the
circumnuclear absorbing gas of NGC 4151 to be distributed in a thin layer of
clumpy gas between the molecular and ionized gas, with the transition at
{$\sim$3.9$(\sin^{-1}\theta)$~pc}.\footnote{The original paper assumed NGC 4151 is
at a distance of 13.3 Mpc and reported a size of  $3.3(\sin^{-1}\theta)$~pc.}
This value can be treated as the upper-limit of the diameter of the compact
dust torus.

\subsection{ Circumnuclear Material Vertical Structure}\label{sec:torus-vert}

\begin{figure}
\center
\includegraphics[width=1.0\hsize]{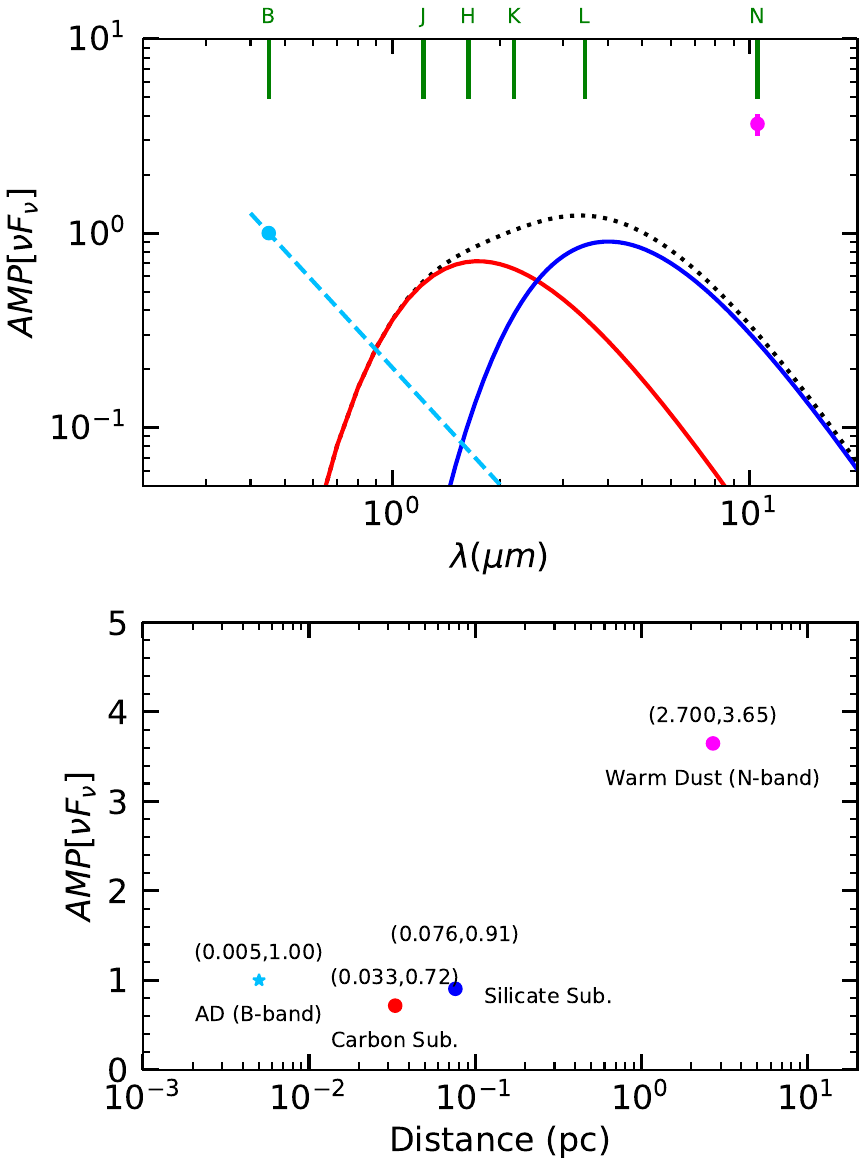}
\caption{
Top panel: The relative scaling factor AMP for luminosity ($\nu F_\nu$) as a
	function of wavelength.  The B-band is shown as a light blue dot with
	the wavelength dependence of AMP of the accretion disk as a dashed blue
	line. The red and blue solid lines are the two hot dust components
	(graphite and silicate) with their combination as a black dotted line.
	The N-band is shown as a purple dot.  Bottom panel: The relative
	scaling factor AMP for luminosity ($\nu F_\nu$) as a function of
	distance.  We show the radial distance from the nucleus for each
	component. 
	}
\label{fig:amp_sed}
\end{figure}

Besides the source sizes and dust temperatures, our reverberation mapping also
reveals information on the vertical structure.  Figure~\ref{fig:amp_sed}
presents the  variation amplitudes of the IR dust reverberation signals
relative to the B-band variability, as a function of  component
distance\footnote{We have converted the optical-to-IR relative flux density
variation amplitude AMP (see Equation ~\ref{eqn:drm-model-2lag}) into
optical-to-IR relative luminosity variation amplitude and denoted it as
AMP$(\nu F_\nu)$ or $f_{\rm B, DRM}$. This parameter describes the amount of
dust emission that responds to the optical variations.}. More and
more accretion disk emission is reprocessed into the IR bands as the dust
distance increases. In addition, as shown in Section~\ref{sec:resolved_study},
the radius of the dominant emission component at 10 $\mu$m deduced from
reverberation mapping is similar to that measured with interferometry. If the 10
$\mu$m emission region were heated via radiative transfer, its reaction would
be significantly slower than speed-of-light behavior
\citep[e.g.,][]{Guo2002}. Therefore, these results together indicate a vertical
structure that intercepts directly the nuclear emission that powers the 10
$\mu$m component. { This vertical height could take the form, for example, of a
flared disk with clouds distributed generally above and below into the polar regions 
\citep[e.g.,][]{siebenmorgen2015} or of a dense wind feeding the NLR from the inner edge 
of a thin disk \citep[e.g.,][]{Honig2013}.} At our current level of
understanding, the difference is largely semantic; for brevity in the
following, we will describe it as a flared disk.

\begin{figure}
    \centering
    \includegraphics{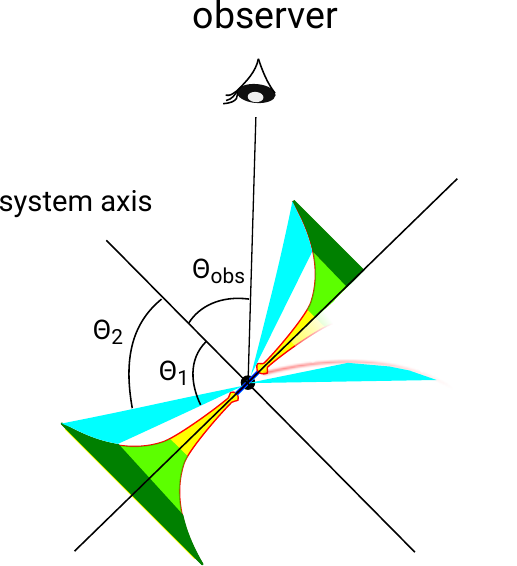}
    \caption{Simple geometric models to illustrate the torus
	vertical structures. We denote the black hole and its accretion as a
	big black dot crossed-over by a blue line. { In the upper right case, the torus is connected to
	the outer part of accretion disk as a flared disk with a puffed-up inner edge  (yellow shaded
	region).} The system is inclined to the observer at angle $\theta_{\rm
	obs}$. To aid our derivation, we highlight a dust component in the
	torus  (dark green shaded region) with a half-opening-angle from
	$\theta_1$ to $\theta_2$. { The inner puffed-up edge shields much of the 
	intervening torus from direct nuclear radiation. Only between these angles can the accretion
	disk emission (light blue shaded region) reach this dust component.
	See text for details. To the lower right, a wind model is illustrated. The wind is lifted from the inner, puffed-up edge and intercepts the nuclear light preferentially in the zone that dominates the emission in the mid-infrared. The puffed-up edge is the source of most of the near-infrared emission and also shields the outer zones of the narrow torus disk from direct illumination from the nucleus.}}
    \label{fig:torus_vert}
\end{figure}

With some simple assumptions, we can estimate the vertical profile
quantitatively.  Figure~\ref{fig:torus_vert} illustrates a simple model of a flared
torus (yellow shading) surrounding a BH accretion disk (a black dot crossed
over by a dark blue line). The complementary side of the sketch shows an equivalent (for our purposes) sketch of a wind model. The system has an inclination angle $\theta_{\rm
obs}$ relative to the observer.  We assume that the accretion disk emission is
anisotropic and that it can be described by the model proposed in
\citet{Netzer1987}, 
\begin{equation}
L(\theta) = L_{\rm AD,0} \cos{\theta}(2\cos{\theta}+1)/3
\end{equation}
For a specific dust component within the torus (dark green shaded region), the
accretion disk can provide heating $L_{\rm AD, DRS}$ (``DRS" represents
dust reverberation signals) only through a limited range of angles [$\theta_1$,
$\theta_2$] (the light blue shaded area), i.e.,
\begin{equation}
L_{\rm AD, DRS}=\int_{\theta_1}^{\theta_2} dF(\theta)
\end{equation}
Assuming $L_{\rm AD, DRS}$ is fully reprocessed into the IR, the corresponding total IR luminosity of this dust component is
\begin{equation}
L_{\rm dust, DRS} = L_{\rm AD, DRS} 
\end{equation}

The observer sees the B-band luminosity, whose strength is
proportional to the accretion disk luminosity through the B-band bolometric
correction, $L_B = 0.2L_{\rm AD}$ \citep{Richards2006}, thus
\begin{equation}
L_{\rm B, obs}= 0.2 \times  L(\theta_{\rm obs})
\end{equation}
Meanwhile, the dust reverberation signals are linearly connected to the
observed B-band luminosity, 
\begin{equation}
L_{\rm IR, obs} = f_{\rm B, DRS} L_{\rm B, obs}
\end{equation}
and the $L_{\rm IR, obs}$ is linearly related to $L_{\rm dust, DRS}$ by
introducing a bolometric correction to a black body spectrum associated with
this dust component. 
From fitting observations of the narrow-line region, the inclination angle of
NGC 4151 is estimated to be $\sim45^{\circ}$ \citep{Crenshaw2010}. Combining
all the above equations, finally we have
\begin{equation}
    \cos{\theta}(2\cos{\theta}+1) \Big|_{\theta_1}^{\theta_2}= 0.2\times(1+\cos{45^\circ})\times f_{\rm B, DRS} \left(\frac{L_{IR, obs}}{L_{\rm dust, IR}}\right)^{-1} 
    \label{eqn:dust-angle}
\end{equation}

The results are summarized in Table~\ref{tab:torus-vert}. With increasing distance (and resulting decrease in dust temperature), the dust components receive
the accretion disk emission from smaller inclination angles,  consistent with
the flared torus picture.  The minimum value of $\theta$ for the warm
dust component defines the half-opening angle of the torus to be
$\sim$23$^{\circ}$. This value is very approximate;  for example, it depends on
the assumed anisotropy of the accretion disk emission. Nonetheless, it is
consistent with the average value of the 15--33$^\circ$ half-opening angle of
the narrow line region bi-cone in NGC 4151 as inferred by \citet{Das2005,
Fischer2013}. In addition, assuming the torus obscuration is the main driver
for the different AGN types, this number corresponds to the  $\sim$30\%
fraction of type-1 AGN, which is roughly consistent with observations for an
AGN with $L_{\rm bol}=10^{43.8}$ erg/s as in NGC 4151
\citep[e.g.][]{Schmitt2001}.

\begin{deluxetable}{ccccc}
    \tabletypesize{\footnotesize}
    \tablewidth{1.0\hsize}
    \tablecolumns{5}
    \tablecaption{Estimations on the Torus Vertical Structures 
    \label{tab:torus-vert}
    }
    \tablehead{
	\colhead{Component} &
        \colhead{T(K)} &
	\colhead{f$_{\rm B, DRS}$} &
        \colhead{$L_{\rm IR, obs}/L_{\rm dust, IR}$} &
        \colhead{$\theta(\deg)$} \\
        \colhead{(1)} &
        \colhead{(2)} &
        \colhead{(3)} &
        \colhead{(4)} &
        \colhead{(5)} 
}
\startdata
Carbon Sub.     &  2000           &   0.72     & 1.00  &   76.4 -- 90  \\
Silicate Sub.   &   900           &   0.91     & 1.00  &   65.5 -- 76.4 \\
Warm dust       &   285$\pm$12K   &   3.65     & 1.48 &   23.5 -- 65.5 \\
Cold dust       &   77$\pm$5      &  $\cdots$  & $\cdots$  &   $\cdots$
\enddata
\tablecomments{Col. (2) inferred dust temperatures from reverberation analysis
	and/or SED decomposition of normal AGN template; Col. (3) the fraction
	of energy ($\nu F_\nu$) between the IR and the optical B-band
	variability signals as seen by the observer; Col. (4) the bolometric
	correction of the observed IR band to the integrated black-body
	spectrum with dust temperature in Col. (3); Col. (5) the range of dust
	half-opening angles as derived in Equation~\ref{eqn:dust-angle}.}
\end{deluxetable}

\subsection{ Linking Dust Reverberation Signals to the IR SED}\label{sec:dust_temp}

\begin{figure*}[htbp]
    \centering
    \includegraphics[width=0.8\hsize]{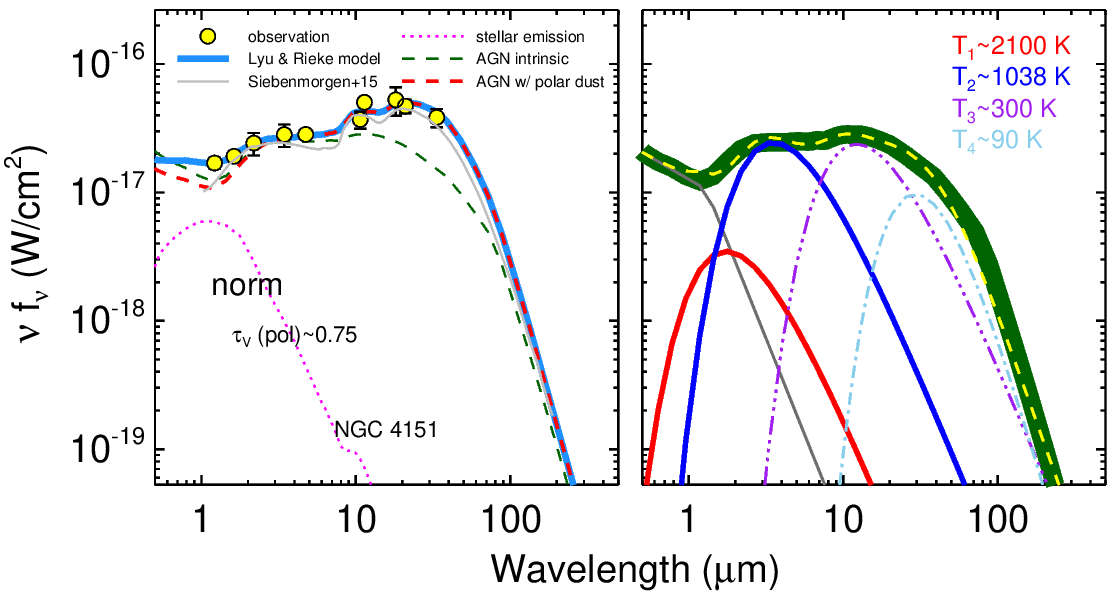}
    \caption{Left panel: Best-fit results of the AGN IR SED of NGC 4151 with
	the semi-empirical model presented in \citet{Lyu2018}. The photometry
	data (yellow dots) was taken from \cite{ah2003}.  The SED model (blue
	thick solid lines) is composed of the AGN component (red dashed line)
	and the stellar component (magenta dotted lines). We also plot the
	corresponding intrinsic AGN template before polar dust obscuration
	(green dashed line). As an comparison, the grey line is the dust radiation transfer model 
	SED for NGC 4151 in \cite{siebenmorgen2015}, shifted down slightly so it 
	is not obliterated by our empirical fit. Right panel: Dust component decomposition of the
	normal AGN intrinsic template. The final model (yellow dashed lines)
	includes a broken power-law component to represent the accretion disk
	emission (gray solid line), three black body spectra (red solid line;
	blue solid line; purple dashed-dotted-dotted-dotted line) and one gray
	body spectrum (light blue dashed-dotted line) to characterize the torus
	dust emission. We have fixed the temperature of the hottest dust
	component at 2100 K and summarize the best-fit temperatures in the
	top-right corner of the panel.  }
    \label{fig:sed}
\end{figure*}

{ We now test whether the picture of the torus derived above is consistent 
with the spectral energy distribution (SED) of the AGN, shown in Figure~\ref{fig:sed}. 
The SED is generally measured with sufficiently large fields of view to capture the 
emission from polar dust as well as from the torus. The 2--5 $\mu$m spectral bump 
is characteristic of a {\it normal} infrared SED in the terminology of \cite{Lyu2017a} 
and indicates that the polar dust contribution at these wavelengths is small \citep{Honig2017, Lyu2018}. 
However, it can be larger, even dominant, 
at the longer wavelengths.}

\citet{Radomski2003} resolved 506 mJy of polar dust with spatial resolution
$\sim$ 0\farcs55 (diameter); if we apply the photometric correction for their
wide spectral band, this becomes 630 mJy.  To estimate a total, we have assumed
that the polar infrared emission is proportional to that in the [O {\sevenrm
III}]$\lambda$5007\AA ~line and used the slit fluxes of \citet{Das2005} to
determine how much additional emission is likely inside of 0\farcs3 ({radius}),
leading to a total flux of 0.9 Jy.  The model of \citet{Burtscher2009} finds
0.70 $\pm$ 0.16 Jy in a $\sim$ 30 mas ($\sim$ 2.3 pc) FWHM  Gaussian and 0.2 Jy in a central
point source; if we combine these values with the 0.9 Jy measured by
\citet{Radomski2003} (corrected) in a more extended component, we account for
the $\sim$ 1.8 Jy total flux at the time of the \citet{Burtscher2009}
measurements (JD 2454579; see Table~\ref{tab:nfluxes}). This polar 
dust component extends to $\sim$200 pc \citep{Radomski2003}; only the 
inner $\sim$ half of the flux ($\sim$ 0.9 Jy) can  contribute to the N-band 
variations.

The left panel of Figure~\ref{fig:sed} shows the SED of NGC 4151
and our best-fit empirical model, which  reconciles the  mid-IR polar emission with the range
of type-1 AGN IR SEDs observed at $z\sim$0--6 and with the commonly seen
UV-optical obscuration \citep{Lyu2018}. The
AGN-heated dust emission can be described by two components: (1) the
traditional relatively compact torus (without dust along the polar
direction) whose face-on SED is described by a small set of AGN intrinsic SEDs
\citep{Lyu2017a}; and (2) extended infrared optically-thin polar
dust with its emission SED peaked around 25-30~$\mu$m and strength typically 
represented by an effective optical depth $\tau_{V}$. The validity of this
model is confirmed by the close match of 10~$\mu$m polar emission strengths
from our SED model and those from mid-IR interferometry of galaxies where it is
available.  
The SED decomposition suggests polar dust emission of optical
depth $\tau_{\rm V}\sim0.75$ as in \cite{Lyu2018}. The
nuclear/torus contribution is 56\% of the total 10$\mu$m flux density, or for
the measurements used in the fit, accounts for 1 Jy of the output, with $\sim$
0.8 Jy from the polar dust. These values are consistent with those derived above from the observations.

To identify the major dust components in the torus, in the right panel of
Figure~\ref{fig:sed}, we decompose the IR emission of the normal AGN intrinsic
SED into several black/grey body spectra after subtracting a broken power-law
component that represents the accretion disk. 
The greybody spectra are a useful approximation because the disk temperature can  
drop precipitously inside the inner edge of the disk in full radiative transfer 
models (see, e.g., Figure 3 in \citet{venanzi2020}), which is combined with the T$^4$ dependence 
of the emitted luminosity. This behavior is supported in our work by the lack of response at 10 $\mu$m to
variations detected in the K-band.  Thus, where a disk inner edge is set by grain sublimation, the
greybodies are not only a reasonable approximation to the total emission, but also are 
characteristic of dust close to the sublimation temperature. 

The fit fixes the hottest dust emission at a temperature of $T\sim2100$K, with the 
best-fit values for the other dust components
of $\sim$890, 285 and 85 K, confirming that the 1--4~$\mu$m emission is dominated by two distinct
dust components close to the sublimation temperatures of graphite and silicates\footnote{ The $T\sim2100$~K temperature of the hottest dust component is higher than the typical
value of $\sim1800$~K assumed for the dust sublimation of graphite (e.g.,
\citealt{Barvainis1987}; \citealt{Mor2009}). However, the higher temperature may 
be a result of stochastic heating of transient very small graphite particles as discussed in the next section.  Alternatively, this temperature may result because the dust sublimation
zone is likely adjacent to the gas in AGN broad-line regions \citep{baskin2018}.}. For the component that dominates the emission at 8--20~$\mu$m,
we obtain a dust temperature of 285K. \citet{Burtscher2009}
reported a similar color temperature from the 0.3\arcsec-aperture ($\lesssim23$~pc) MIDI
mid-IR spectrum of NGC 4151. We conclude that the N-band dust reverberation signals are dominated by this
component. If we use the wind temperature relation from \citet{venanzi2020}, put $r_{sub}$ at 0.076 pc 
from the reverberation behavior, and assign a temperature there of 1000 K, we find that the temperatures 
dominating at 10 $\mu$m are at 2.0 (310K) to 3.6 (250K) pc. This behavior agrees perfectly with expectations 
from the N-band lag determination. Of course, this emission is only one component of the polar emission; 
the much more extended flux \citep{Radomski2003} must arise from stochasticaly heated grains or 
grains heated locally, e.g., from hot gas.

The cold dust component in the far-IR is characterized by a gray body at
$T\sim85$K. Given the similar far-IR SED shapes of various types of  AGN, this
emission is likely to be optically-thin and very extended \citep{Lyu2017}.
It probably originates in very extended polar dust. The lack of variability at
20--24~$\mu$m and 34--37~$\mu$m (see Section~\ref{sec:20var}) supports the
conclusion that this spectral range is dominated by these very extended dust
components.

In conclusion, the AGN-heated torus emission of NGC 4151 can be characterized
by five dust components.  The first two correspond to graphite and
silicate grains in the dust sublimation zone of the AGN torus, the third provides  
persistent emission at an apparent temperature of $\sim$ 700 K, the fourth is at $\sim$ 285 K 
and between 2.2 and 3.1 pc from the nucleus, and the fifth is at $\sim$ 85 K and is 
extended up to $\sim$ 200 pc; it contributes the far infrared emission. 
{ These results are only rough estimates. In more realistic cases, one would
need to build a detailed dust radiative transfer model to match the dust
reverberation constraints as well as the SED and consider the optical thickness
of the clouds, the inclination angle effect, the anisotropy of the accretion
disk emission, etc. The left panel of Figure~\ref{fig:sed} shows that the SED from an existing 
radiative transfer model agrees closely with our empirical one, but the new information 
from the lag analysis can support further refinement. Such modeling efforts are beyond 
the scope of this paper.}

\subsection{Properties of Dust Grains}\label{sec:two-lag-physics}

{ 
It is likely that the dust near an AGN can only survive in large grains \citep{Laor1993, Maiolino2001,baskin2018}.
Our observations, which locate the sublimating grains in distance from the accretion disk, support this conclusion. }
We take the bolometric luminosity of NGC 4151 
to be $7\times10^{43}$ erg~s$^{-1}$ from the $\lambda F_\lambda(5100$\AA$)$ flux
\citep{Kaspi2005}. Adopting $F_{\rm UV} = 0.165 F_{\rm tot}$ \citep{Risaliti2004} 
and putting the dust grains at their typical sublimating temperatures
($T_{sub, S}\sim1000$K and $T_{sub, C}\sim1500$K), equations~\ref{eqn:sub_c},
\ref{eqn:sub_s} give minimum survival grain sizes and can be written as 
\begin{equation}
	R_{\rm sub, C} = 0.044 \left(\frac{a_{\rm C}}{0.05~\mu m}\right)^{-0.5}~{\rm pc} ~ ~,
\end{equation}
\begin{equation}
	R_{\rm sub, S} = 0.092 \left(\frac{a_{\rm S}}{0.05~\mu m}\right)^{-0.5}~{\rm pc}~ ~,
\end{equation}
The two time lags as 
measured in Interval C correspond to physical scales of about 0.033 pc and
0.076 pc, which would suggest grain sizes of $\gtrapprox$ 0.07--0.08~$\mu m$ for both
graphite and silicate grains. 
These inferred grain sizes are an order of magnitude larger than typical small
grains in the classical diffuse ISM \citep[e.g.,][]{Wein2001}. 

Erosion of these large grains will yield a transitory population of very small
refractory grains, composed both of carbon and of very small grains of refractory 
oxides such as
FeO and MgO \citep{mann2007}. Such grains can be heated stochastically above
their thermal equilibrium values \citep{man99,dra01}, consistent with the high 
temperature for the hottest dust in the SED fitting in the previous section.

{ 
\subsection{ Variability/Lag Behavior Compared  with Torus Models}
\label{sec:models}

Given the constraints on the AGN dust structures in NGC4151
obtained above, we now examine the assumptions and basic features of 
some popular torus models. The following discussion will 
highlight areas where refinement is recommended in the future modeling work.

\subsubsection{Challenges to the Clumpy Torus Models}\label{sec:clumpy_model}

Many previous studies of circumnuclear tori have used {\it geometric/ad hoc
models}, in the nomenclature of \citet{Ramos2017}.  Such models fit the
infrared SED by assuming a quasi-static dust distribution (e.g., smooth,
clumpy, or a combination) and a grain composition. The most popular class of
model posits the torus to be in the form of many individual gas/dust clouds
with little material in between, hence the term ``clumpy'' models.  Such tori
were modeled extensively by \citet{nen08a, nen08b}, whose results (and
subsequent elaborations) are widely used to analyze various AGN IR
observations. { As summarized in the abstract of \citet{nen08a} --- ``in a
clumpy medium, a large range of dust temperatures coexist at the same distance
from the radiation central source.'' --- a major feature } of these models is
that individual optically-thick clumps emit at a broad range of temperatures
due to the effects of radiative transfer, so that the emission at different
wavelengths can come from the same clumps, making a relatively compact torus
\citep{nen08a}.

The behavior of NGC 4151  appears to be inconsistent with the predictions of these models in two ways:

\vspace{1mm}
\noindent
{ 10--30 $\mu$m Reverberation Behavior}: 
\citet{alm17, alme17,
alme20} investigated  extensively the infrared reverberation response of the
\citet{nen08a, nen08b} clumpy models from a theoretical perspective, providing
a useful comparison for our study. In particular, their work
focused on NGC 6418, a type 1 AGN quite similar to NGC 4151, with about 1/3 the
luminosity and a reverberation lag time at 3.6 $\mu$m of about 37 days
\citep{vaz15}, i.e. about 40\% that for NGC 4151 at this wavelength (roughly
consistent with the expected luminosity scaling).  Given their overall
similarity, the conclusions for NGC 6418 should also apply with suitable
scaling to NGC 4151.

Given the general characteristic of clumpy models that each clump contains a
broad range of temperatures and hence emits over a broad range of infrared
wavelengths, the relative delays from one infrared wavelength to another are
small in these models. The highest fidelity models, for anisotropic torus
illumination and full radiative transfer, show simulated lags of 31.46, 33,
35.97, and 55.91 days respectively for 2.2, 3.6, 4.5, and 10 $\mu$m
\citep{alm17}. Although the response curves are smoothed increasingly with
increasing wavelength, over the 2.2--10 $\mu$m range, they track each other
well. Other than short periods of mismatch due to different response times, the
fractional change in the baseline flux at 10 $\mu$m is 80--100 \% of the
fractional change of the baseline at 2.2 $\mu$m \citep{alm17}. 

The behavior we have found for NGC 4151 is in poor agreement with these
predictions. For example, in Figure~\ref{fig04} we compare the K-band
photometry (delayed by 30 days roughly as predicted by the models) with the 10
$\mu$m measurements. They strongly disagree.  In Section~\ref{sec:dust_temp} we
show that about half of the 10 $\mu$m flux originates in the compact source
responsible for the reverberation behavior. It is not plausible that the
behavior shown in Figure~\ref{fig04} could hold, even if half of the 10 $\mu$m
signal is varying synchronously with the K-band signal and with 80--100\% of
its relative (to the baseline flux) amplitude, while the other half of the
signal is constant, e.g. from very extended polar dust.

The set of 10 $\mu$m measurements in 1975 and 1976 can be used for another test
of clumpy models. Taking the flux from the nucleus (i.e., eliminating the
contribution of any polar dust, see Section ~\ref{sec:dust_temp}) to be 1000
mJy, the measurements put an upper limit (3 $\sigma$) of 220 mJy on any
decrease from 1975 to 1976. The change between these two years in K-band is a
decrease of $\sim$ 36\%. Assuming that the fractional change at 10 $\mu$m is
$\ge$ 0.8 as large \citep{alm17}, we would expect a change of $\ge$ 290 mJy,
well above the 3 $\sigma$ upper limit. 

\begin{figure}
\center
\includegraphics[width=1.0\hsize]{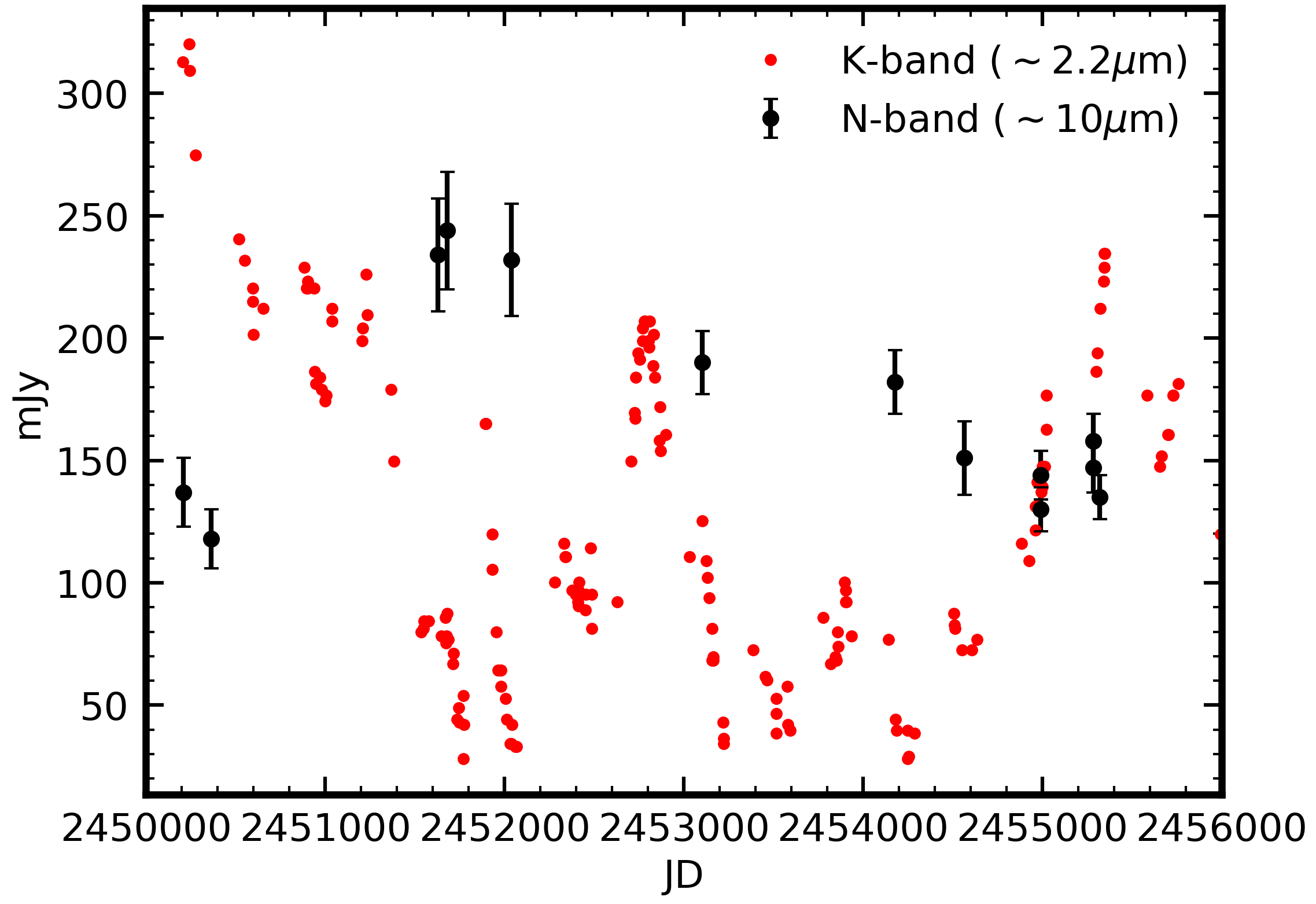}
\caption{
	Light curve at K (red) compared with 10 $\mu$m measurements (black; the
	former delayed by 30 days and the latter renormalized to bring them
	onto the same scale). Although models for the reverberation behavior
	predict that the behavior at these two wavelengths should be very
	similar \citep{alm17}, the measurements show very little resemblance.
	}
\label{fig04}
\end{figure}

In \citet{Lyu2019}, we reported the lack of variability at 24 $\mu$m over
multi-year timescales for low-redshift quasars. We found that this behavior is
also likely to be inconsistent with pure clumpy torus models, although this
conclusion was backed by less detail than for NGC 4151. Combining the studies,
we conclude that these models have significant issues for modeling the global
infrared emission of AGNs, although many  aspects of them are still likely to be
relevant. A likely possibility is that they leave out components that are responsible 
for the longer wavelength infrared emission, which might be extended and diffuse
such as winds and polar dust (see next section). 

\vspace{1mm}
{ Treatment of the Dust Sublimation Zone}:
The \citet{nen08a, nen08b} clumpy models
have a simple and sharp cut of the dust distribution without
discerning the different sublimation radii of graphite and silicate grains.
Such pure {\it clumpy} models
have difficulty fitting the 1--4 $\mu$m output of AGNs \citep{Mor2009,
Hernan2016}. \citet{Mor2009} introduced a hot black body
component (associated with  sublimating
graphite dust) to achieve adequate matching. 
Recently,  \cite{alme20} found the predicted torus sizes at 2.2~$\mu$m 
from the \cite{nen08a, nen08b} models are typically
a factor of $\sim$2 larger than the constrains from previous K-band reverberation
mapping and argued the existence of a graphite-dominated torus innermost region.
Our finding of two distinct dust time lags in the near-IR emission
of NGC~4151 strongly support the existence a graphite-only region, which should be considered 
in future improvements of clumpy models.

\subsubsection{Other Torus Models}\label{sec:smooth_models}

The polar opposite to clumpy models is smooth ones, exemplified by those of
\citet{Fritz2006}. A number of issues with these models led to the development
of the clumpy ones \citep{nen08a,nen08b}. In addition to contradicting
observations of clumpy structure in some galaxies, very generally these models
have difficulty fitting the mild silicate (emission or absorption) features in
AGN SEDs. The behavior of NGC 4151 at 10 $\mu$m raises another concern. In the
\citet{Fritz2006} models, the SED is generated by a range of temperatures
throughout the torus, with outer radial zones heated by radiative transfer, which 
should be significantly slower than direct radiation.
This picture is difficult to reconcile with the variability at 10 $\mu$m, and
particularly with the apparent agreement in the size of the 10 $\mu$m source
from the reverberation behavior with that measured through interferometry, as
discussed further in Section~\ref{sec:resolved_study}.

The mixed clumpy/smooth torus models
of \citet{Stalevski2012} appear to retain the advantages of the clumpy torus
models but with the additional smooth component needed to fit the hot dust
emission.  The overall success of these models relative to our results depends
on the structure that dominates the reverberation behavior at 10 $\mu$m. In the
models, much of the 10 $\mu$m flux is generated in the clumps (see their Figure
5) and will have reverberation behavior similar to that at shorter wavelengths.
We find that the variable 10 $\mu$m flux arises from a much more extended
component, posing a challenge for the \citet{Stalevski2012} models as well. 

In addition, the predicted torus inner size of such two-phase models based on SED analysis of NGC 4151
is systematically
larger than the observational constraints from near-IR dust-reverberation mapping and interferometry
observations. 
\citet{swain2021} have introduced an additional
graphite ring-like structure between the torus and the accretion disk, and 
carried out a full 3D radiative transfer model
of the NGC 4151 IR SED. They conclude that the graphite ring has
an inner radius of $\sim$ 0.04 pc (= 48 light days) and that there is an outer disk of normal ISM 
dust with an inner edge at $\sim$ 0.1 pc (= 120 light days), i.e. within uncertainties 
in excellent agreement with our reverberation analysis 
values of 0.033 and 0.076 pc, respectively. Nevertheless, the detailed 
geometry of torus inner structures
can be more complicated. For example,
it is proposed that the inner ring might be puffed up due
to radiation pressure \citep[e.g.,][]{Krolik2007, Honig2019}. 

Besides the classical perception of a compact torus, the winds 
feeding the polar dust \citep{Honig2013} are also candidates to account for the 10 $\mu$m reverberation behavior. 
The physics of these winds is described by \citet{venanzi2020}. 
A suite of radiative models has been presented by \citet{Honig2017} that 
predict the mid-infrared spectra of these winds and specifically fit the 
SED of the wind-dominated case of NGC 3783. Although the emphasis in these models is polar dust at relatively large distances from the nuclear engines, they should contain dust at the appropriate distance and directly illuminated by the accretion disk, as needed to account for the 10 $\mu$m variations.

These models are paramaterized in terms of $\alpha_{IR}$, 
the spectral index between 3 and 6 $\mu$m, and $\alpha_{MIR}$, 
the index between 8 and 14 $\mu$m. The latter is $\sim -$1.5 from the IRS spectrum \citep{weed05}. 
The former is more difficult to constrain; simultaneous  observations (to avoid effects from variability) 
are generally not available at either wavelength, and the one exception, the ISO SWS spectrum, is of poor quality at 6 $\mu$m. A value of about $-$1 to $-$1.2 is indicated. These values place NGC 4151 in the region where strong silicate emission is expected \citep{Honig2017}, whereas none is seen in NGC 4151 \citep{weed05}. Further analysis should clarify the role in the 10 $\mu$m variability of optically thin dust clouds in the wind.
}

\subsection{ Other Aspects of Torus Behavior}

\subsubsection{Variations not Described by Geometric/Ad Hoc Models}

The three {\it geometric/ad hoc} model types just discussed all describe 
a quasi-static torus, i.e., they do not include mechanisms to account for changes
on decadal timescales. However, as presented in Section~\ref{sec:torus-growth}, the hot dust
emission strength has gradually increased over two decades, whereas in 
the previous $\sim$ 4 years it had been decreasing \citep{okny99}.  

Possible physical insights into this behavior can be gained from {\it physical}
models. In fact, the AGN torus is a violent environment
featuring both inflow and outflow, a dynamic aspect that emerges in various
hydrodynamical simulations \citep[e.g., see references in][]{Netzer2015}.  After
considering both stellar and AGN feedback, the simulations of \citet{Wada2012} 
establish the circumnuclear torus as being dominated by turbulence and transitory
density enhancements (i.e., ``clumps'') embedded in a smoother gas
distribution. The density contrast between the components is a factor of a few.
They predict turbulence and winds continuously lifting significant amounts of
material off the torus, which would naturally explain long term trends such as
the growing (and variable) hot dust emission, as well as the maintenance of the
polar dust.

Evidence for a turbulent torus with outflow/inflow
signatures has been seen in a few nearby AGNs at submm and mm wavelengths. With
the high-spatial resolutions achieved by long-baseline interferometry, water
maser emission in Circinus galaxy and NGC 3079 shows a compact structure with
possible signatures of outflows \citep{Greenhill2003, Kondratko2005}.  ALMA
observations have revealed a turbulent circumnuclear structure with complicated
dynamical motions in NGC 1068 \citep[e.g.,][]{Garcia-Burilo2016,
Garcia-Burilo2019, Gallimore2016, Imanishi2018} and some other AGNs as well
\citep[e.g.,][]{ah2018}.

\subsubsection{The lack of adjustment to changes in nuclear luminosity}\label{sec:size-evo}

It is well-established
that the time lag for hot dust at the inner edge of the torus is strongly
correlated with AGN bolometric luminosity, following $\Delta
t\propto\sqrt{L_{\rm AGN}} ~$\citep[e.g.,][]{okny01, kosh14, Lyu2019, Minezaki2019}.
However, we found in Section~\ref{sec:receding-torus} that no significant 
changes of this type are seen. Evidently the relaxation to the usual
relation with luminosity proceeds slowly (at least more slowly than over a
decade).

In fact, due to its high optical
depth the underlying disk in the dusty torus will be relatively robust against
erosion. Since the torus inner edge is optically thick, changes will propagate at the sound speed
of turbulent gas within the torus, which is expected to be 20--100 km s$^{-1}$ \citep[e.g.,][]{Hopkins2012}. 
A size change of 10 light days would typically require 16--80 years to re-establish the equilibrium structure.
In addition, much of the hot dust emission may come from dusty
clouds lifted by turbulence above the torus. Their destruction is only on a timescale of a decade or longer under the
direct exposure to AGN radiation \citep{Namekata2014}. In addition, 
to first order the process of turbulence lifting dust out of the 
torus will continue roughly continuously (e.g., described as
a ``fountain-like'' structure in \citet{Wada2012}), further reducing
the chance to detect notable structure changes.

Only on these longer timescales should we expect the underlying dusty 
disk to retreat under increased nuclear luminosity and its infrared 
emission to reach the relation $\Delta t\propto\sqrt{L_{\rm AGN}}$.

\subsubsection{The Similarity of AGN Torus 1--5 $\mu$m SEDs Results from  Dust Properties} 

The torus emission at 1--2.5 $\mu$m is dominated by carbon dust near its
sublimation temperature,  while the emission at 2.5--5
$\mu$m adds a substantial contribution from silicate dust. Because the torus is 
optically thick, the emission will be dominated by dust just behind the sublimation 
radii and close to the sublimation temperatures. The relative abundances of these materials
should also be similar for the circumnuclear environments of most, if not all,
AGNs. As a result of the similar temperatures, positions, and compositions of
the emitting dust, there should be a generally similar SED for AGNs from 1--5
$\mu$m. This result is confirmed by observation \citep{Lyu2017}; the variety in
SEDs can be traced to (1) the amount of emitting material in the inner torus,
with hot-dust-deficient (HDD) SEDs resulting from low-dust-mass cases; and (2)
the relative amounts of material at large radii, with warm-dust-deficient (WDD)
SEDs resulting when the outer parts of the torus and/or the winds feeding the polar dust are reduced.

\section{Conclusion and Summary}
\label{sec:insight}

We have conducted a comprehensive study of the dust reverberation signals at
$\lambda\sim$1--40~$\mu$m from the famous type-1 AGN in NGC 4151, covering a
time-frame of 30-40 years. Although a number of reverberation studies in the
near-IR have been published previously, they are each based on a limited subset
of the available data and only probe the very inner part of torus.  By analyzing
all the light curve data at 1--4 $\mu$m and literature measurements at
10--40~$\mu$m, we have been able to identify two lags corresponding to 
the sublimation radii of carbon and silicate dust, 
respectively, plus a third lag for cooler dust at $\sim$ 285 K. This approach 
has led to a a deeper understanding of the  circumnuclear dust structures. 

We have identified several issues that have might affect previous optical-IR correlation analyses, i.e.,

\begin{itemize}

    \item{The deduced reverberation delays are sensitive to the observation cadence, 
    and can be affected even in some of the individual highest-cadence studies: \S \ref{sec:cadence}.}
    
    \item {Although the optical and blue photometric bands are frequently used as 
    proxies for the UV spectral component that heats the circumnuclear dust, these 
    bands sometimes show behavior that is not reflected in the behavior of the torus: \S \ref{sec:first-look}.}

\end{itemize}

To eliminate effects of the first type, we have focused our analysis on a a seven-year period (June 2000 - August 2007) when multiple compaigns can be combined to provide well sampled light curves in both B-band and in the near infrared. To mitigate issues of the second type, we first measured relative dust time
lags of the J, H, K and L bands (1.25 -- 3.6 $\mu$m) relative to each other as
the dust emission variability at these wavelengths shares the same physical
mechanism. We then measure the lag of K relative to B, since these two bands
have the highest measurement cadences. In addition, we have also developed
physical guidelines to fit all the IR light curves simultaneously to reveal the
physical properties of the underlying dust components.

{ 
As the results from this work cover a large range of scales and behavior, we will organize the summary of our scientific conclusions around Figure~\ref{fig:cartoon}.

\begin{figure*}[htbp]
    \centering
    \includegraphics[width=0.8\hsize]{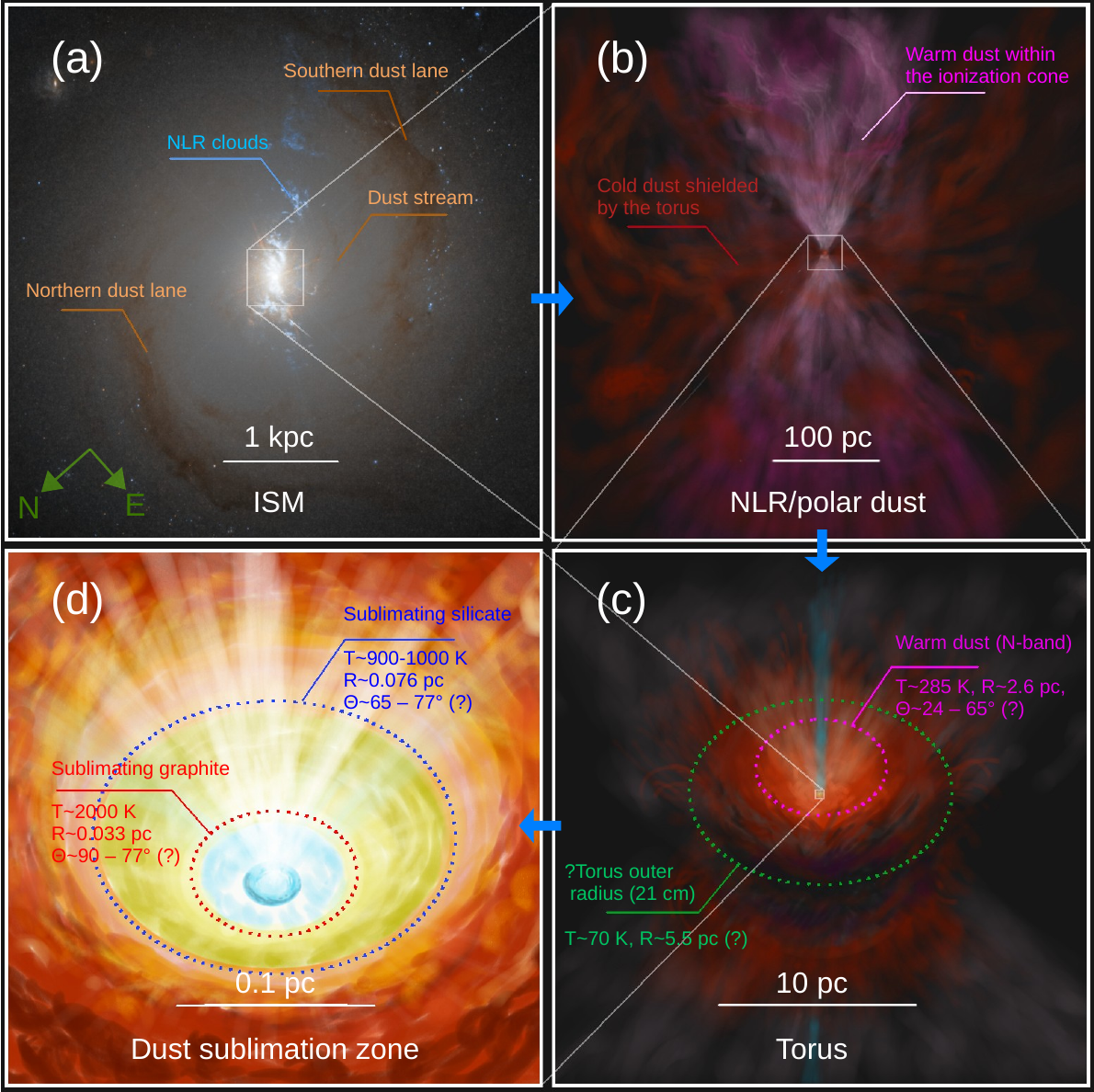}
    \caption{Illustration of the circumnuclear dusty environment of NGC 4151 at
	different scales. Panel (a): The inner 5-kpc region of NGC 4151 as
	revealed by HST WFC3 images. We can directly see the narrow-line-region
	(NLR) clouds near the central bright AGN, projecting along two
	directions, and two dust lanes as well as numerous small dust streams
	at different locations. Panel (b): Dust components at 100 pc scales,
	featuring the narrow line region and polar dust in the AGN ionization
	cone.  Outside this cone, relatively cool dust is shielded by the torus
	from the direct heating of the AGN.  Panel (c): Dust structures at
	$\sim$1--10 pc scales, featuring a flared torus with possible
	turbulent/outflowing signatures. We denote the torus size constraints
	in the N-band from our dust reverberation mapping and in the 21cm from
	radio-band observations in \citet{Mundell2003} (The latter value
	assumes a system inclination angle of 45$^\circ$). Panel (d): The
	innermost region of the torus over $\sim$0.01--0.1 scales, featuring a
	likely turbulent disk. We have labeled different radii ($R$), dust
	temperatures ($T$) and ranges of dust half-opening-angles ($\theta$)
	for sublimations of graphite and silicate dust grains in Panel (d) and
	AGN warm dust in Panel (c).}
    \label{fig:cartoon}
\end{figure*}

\subsection{Panel \rm{(a)}, {\it the Surrounding Galaxy}}\label{sec:hostgalaxy}

\begin{itemize}

    \item {There are apparent outbursts in the B-band that do not create a reaction of reradiated infrared emission; they might arise from supernovae or other activity in the surrounding galaxy: \S\ref{sec:first-look}.}
   
    \item{Even with imaging data, there are significant discrepancies in estimates of the galaxy infrared flux that needs to be subtracted to get a pure nuclear flux. We demonstrated a J $-$ H vs. H color-magnitude technique that allows accurate estimation of the galaxy flux: \S\ref{sec:host_sub}.}
    
    
    
\end{itemize}

\subsection{Panel \rm{(b)}, {\it the NLR and Polar Dust}}\label{sec:nlrpolar}

\begin{itemize}
    
    \item {The lack of variations on decadal timescales at 20 $-$ 40 $\mu$m (\S \ref{sec:20var}) indicates this emission originates far out in the polar dust, which is observed to extend to a $\sim100$ pc scale and also accounts for up to about half the 10 $\mu$m emission \citep{Radomski2003, Lyu2018}: \S \ref{sec:torus-vert}. }
   
\end{itemize}

\subsection{Panel \rm{(c)}, {\it the Circumnuclear Torus}}\label{sec:crcmtorus}

\begin{itemize}
    \item {The variability of about half of the 10 $\mu$m flux (\S\ref{sec:decadalNband}) 
    is consistent with it arising between about 2.2 and 3.1 kpc from the nucleus, possibly 
    in the direct wind leaving the inner edge of the torus, although there could be a contribution 
    from a large flared torus: \S \ref{sec:torus-vert}.}
     \item From the variability amplitiudes, we estimated the half-opening angles of the different dust components, with sublimating graphite dust at 		$\theta\sim$76--90$^\circ$, sublimating silicate dust at 		$\theta\sim$66--76$^\circ$, and $\sim10~\mu$m warm dust component at $\theta\sim$24--66$^\circ$, i.e., a torus half-opening angle of $\sim23^\circ$. This estimate agrees well with the half-opening angle of the narrow-line region bi-cone in NGC 4151, which is 15--33$^\circ$, with an average value equal to the minimum half-opening angle of this warm dust component \cite{Das2005, Fischer2013}: \S \ref{sec:torus-vert}. 
    \item {The range of time lag behavior between 1.2 and 10 $\mu$m and the lack
	of IR variability at 20--40~$\mu$m show that the hot, warm and cold
    dust emission originates over a large range of radius, not all
	from individual clumps as in pure clumpy torus models. A similar issue has been found for a large sample of quasars \citep{Lyu2019}. Although such models have been popular for fitting torus spectral energy distributions, improvements are needed to make them consistent with the results of reverberation mapping: \S \ref{sec:clumpy_model}. }
    
\end{itemize}

\subsection{Panel \rm{(d)}, {\it the Sublimation Zone}}\label{sec:sublimation}

\begin{itemize}
\item The 1--4 $\mu$m torus emission is dominated by carbon and silicate dust, at or close to sublimation temperatures of $\gtrsim$ 1500K and $\sim$ 900--1000K, and at distances from the central engine of about $\sim0.033$ pc and $\sim0.076$ pc, respectively: \S \ref{sec:fullreverb}, \S \ref{sec:dust_temp}. Our finding of two distinct lag times in the near infrared is as predicted by
a number of studies \citep[e.g.,][]{baskin2018,swain2021} and consistent with general 
expectations given the silicate/carbon composition of interstellar dust.

\item The radii of the inner edges of these torus components confirm previous conclusions that the grains are an order of magnitude larger than the small grains in the interstellar medium. Nonetheless, still smaller grains as sublimation products are expected to be stochastically heated and have transient lifetimes, consistent with  the 		relatively high color temperature of the excess between J- and H-bands: \S \ref{sec:two-lag-physics}.

 \item The emission by hot dust in the variable sublimation-temperature component is growing, at $\sim$ 4\% per year: \S \ref{sec:torus-growth}.
    
\item Underlying the variable emission from the sublimation-temperature dust, there is a relatively non-variable component with a
temperature of $\sim$ 700K: \S \ref{sec:non-var-comp}. 
    
\item The inner edge of the torus, identified by the 1500--2500~K dust, does not retreat or otherwise respond significantly to changes in 	nuclear luminosity on decadal time scales: \S \ref{sec:receding-torus}.

\item The robustness of the torus inner edge is consistent with : (1) the sound-speed, which suggests relaxation of the edge should occur only over decadal timescales; and (2) the timescales for destruction of dusty clouds lifted above the torus:  \S \ref{sec:size-evo}

\end{itemize}
}

The properties of dust structures in NGC 4151 appear to be generally similar to
those for other AGNs and quasars, which have  remarkably similar SEDs in the near infrared  for $z\sim$0--6 and $L_{\rm
AGN}\sim10^8$--$10^{13}~L_\odot$ in \cite{Lyu2018}.  From our study of NGC 4151, this similarity probably arises because: (1) the dust composition is dominated by carbon and silicates, which account for similar portions of the interstellar dust from one AGN to another; (2) the sublimation temperatures place these two dust types at the same temperature regimes regardless of the AGN luminosity; and (3) radiative transfer in the optically-thick circumnuclear tori results in a steep temperature drop beyond the sublimation radius, so the output SEDs resemble each other regardless of details of the torus structure.

\section{acknowledgements}
We thank Tom Soifer for tracking down the origin of a 11 $\mu$m measurement,
Almudena Alonso-Herrero and Triana Almeyda for comments on the clumpy torus
models, Hengxiao Guo for the discussion on the robustness of time lag measurements, Raphael Hviding for feedback on the presentation of this paper. We particularly thank Sebastian H\"onig for an extensive discussion. We also 
thank the referees for their reports. This research has made use of the VizieR catalog access tool, CDS, Strasbourg, France. This work was supported by NASA grants NNX13AD82G and 1255094. 

\software{Dynesty \citep{dynesty}, JAVELIN \citep{Zu2013},
Matplotlib \citep{matplotlib}, Astropy \citep{astropy2013, astropy2018}, PyCCF \citep{Peterson1998, Sun2018}, I$^2$CCF (Guo et al. in prep)}

\bibliographystyle{apj.bst}

\appendix

\section{Host Galaxy K Flux}\label{app:host}

\citet{kosh14} quote work by \citet{Minezaki2004} who analyzed images of the
galaxy to derive a net host galaxy K-band contribution of $44.22 \pm 3.83$ mJy
in the $8\farcs3$ aperture used in the work reported in both papers.
\citet{kot92} report a detailed analysis of JHK images that should be
particularly useful for measuring the galaxy since they were obtained when the
nucleus was very faint \citep{okny16}.  We have started with their measurements
in the J band since the contribution of the nucleus is small there. We first
subtract their estimate of the nuclear flux (16 mJy) from the averages of all
three sets of measurements in the full range of apertures. We then fit the run
of J-band flux with aperture diameter with a quadratic (the residuals are
negligible) to derive a flux of 77.9mJy in an $8\farcs3$ aperture. Using the
standard 2MASS galaxy colors, this value translates into 75.4~mJy at K (using
the 2MASS calibration by \citet{cohen03}).  \citet{tar13} report two additional
estimates of the galaxy contribution, in this case within the 12$''$ aperture
used for the {SAI} photometry. One approach was to assume that the J
photometry included only light from the galaxy at the minimum flux. This
minimum was at J magnitude 10.73, which translates (using the standard galaxy
colors and the aperture effect derived from \citet{kot92}) to a K flux density
in a 8\farcs3 aperture of $\sim$ 66 mJy. The second estimate is derived by
comparing the variations in the various bands and finds the galaxy to be 0.26
magnitudes fainter, leading to a K-band flux of $\sim$ 44 mJy within 8\farcs3.
The deviations in these estimates need to be reconciled.

We therefore made an independent determination of the host galaxy flux. If we
make the assumption that the infrared colors of the variable component remain
constant and only its amplitude changes (see \citet{tar13} for a supporting
discussion for this case), than the run of $J-H$ with brightness, e.g. H
magnitude, puts tight constraints on both the galaxy flux into the aperture and
on the $J-H$ color of the variable component. The infrared colors of a normal
galaxy are critical for this analysis. Fortunately, these colors are virtually
independent of galaxy type for ellipticals through mid-type spirals
\citep[e.g.][]{frog78,per79, will84, glass84}. For the JHK colors, we averaged
the values in \citet{per79} and those in \citet{glass84}, and transformed each
to the 2MASS photometric system as in \citet{car01}. In the first case, the
result was $<J-H>$ = 0.667 and $<H - K>$ = 0.238 with errors of $\sim$ 0.01 and
in the second case, $<J-H>$ = 0.684 and $H-K>$ = 0.242, with errors almost as
small. We adopted $<J-H>$ = 0.674 and $H-K>$ = 0.240. To compare with the SAI photometry, we used the {SAI}
standard star list \citep{shen11}, transformed onto the 2MASS system (next paragraph).  Although K-L
has less data for deriving transformations, fortunately their accuracy is not
critical for our analysis, so we use $<K-L>$ = 0.25 from \citet{will84}.

The transformations from the {SAI} JHK photometry to 2MASS were based on the
standard stars listed in \citet{shen11}. These stars are all much too bright to
be observed in the unsaturated mode with 2MASS, so they were compared with
heritage photometry from multiple sources, transformed to the 2MASS system as
in \citet{car01} or with our own transformations of heritage photometry
for members of the Bright Star Catalog. The results are:
\begin{equation}
(J-H)_{\rm 2MASS} = (J-H)_{SAI} + (0.0018 \pm 0.0175) \times (J-H)_{\rm SAI}  + (0.0352 \pm 0.0088)
\end{equation}
\begin{equation}
(H-K)_{\rm 2MASS} = (H-K)_{SAI} + (0.0084 \pm 0.0120) \times (J-K)_{\rm SAI} + (-0.0084 \pm 0.0074)
\end{equation}

\begin{equation}
J_{\rm 2MASS} = J_{\rm SAI} + (-0.0327 \pm 0.0043) \times (J-K)_{\rm SAI} + (0.0722 \pm 0.0027)
\end{equation}

\begin{equation}
H_{\rm 2MASS} = H_{\rm SAI} + (-0.0417 \pm 0.0161) \times (J-K)_{\rm SAI} + (0.0423 \pm 0.0099)
\end{equation}

\begin{equation}
K_{\rm 2MASS} = K_{\rm SAI} +  (-0.0399 \pm 0.0073) \times (J-K)_{\rm SAI} + (0.0533 \pm 0.0045)
\end{equation}

The resulting run of $J-H$ with $H$ in the 2MASS photometric system is shown as
Figure~\ref{fig:jhhcolor} in the main text, where we discuss how it allows an
accurate determination host galaxy fluxes at K of $73 \pm 5$ mJy in the 12$''$
aperture used for the {SAI} photometry and $59 \pm 4$ mJy in the 8\farcs3
aperture used in the \citet{kosh14} data. 

\section{Reconciliation of 10 \lowercase{$\mu m$} Measurements}\label{app:10um}

In this section of the appendix, we describe some of the details of the
reconciliation of the $\sim$ 10 $\mu$m measurements to a consistent
calibration. 

There was a series of ``additional observations" with IRAS (the normal survey
of IRAS did not cover NGC 4151 because of exhaustion of the liquid helium)
obtained in June, 1983. The calibration of the additional observations at the
flux level of NGC 4151 is uncertain \citep{hel88}.  We averaged two values at
12 $\mu$m both derived from these observations: 1.72 Jy \citep{sem87} and 2.01
Jy \citep{san03}. We subtracted 0.12 Jy as an estimate of the host galaxy
contribution, determined by fitting a template SED to the far infrared
measurements that are expected to be powered by star formation.

For the test of the photometry, we made use of the {\it Spitzer} IRS spectrum of the galaxy
\citep{weed05} both to carry out bandpass corrections to convert measurements
in the very broad N filter to the equivalent monochromatic values, and to
correct measurements at other wavelengths to 10.6 $\mu$m. In all cases where
only statistical errors were reported, we added a 7\% photometric error by RSS,
except in cases where unconventional photometry (e.g., through a scanning slit)
suggested larger errors, and in these cases we typically used 10\%. The images
of \citet{Radomski2003} indicate faint extensions to the nuclear source that
account for $\sim$ 27\% of its 10.8 $\mu$m output. We have applied the
indicated correction to the measurement by \citet{soi03}, consistent with their
suggestion. To capture this component, we also re-reduced the imaging results
summarized by \citet{Asmus2014} to use aperture photometry with aperture
diameters of $4\farcs62$ at 12.6 $\mu$m and $3\farcs45$ at the other bands.

\section{Additional Figures from The Bayesian Analyses}

\begin{figure*}[htp]
\center
\includegraphics[width=0.30\hsize]{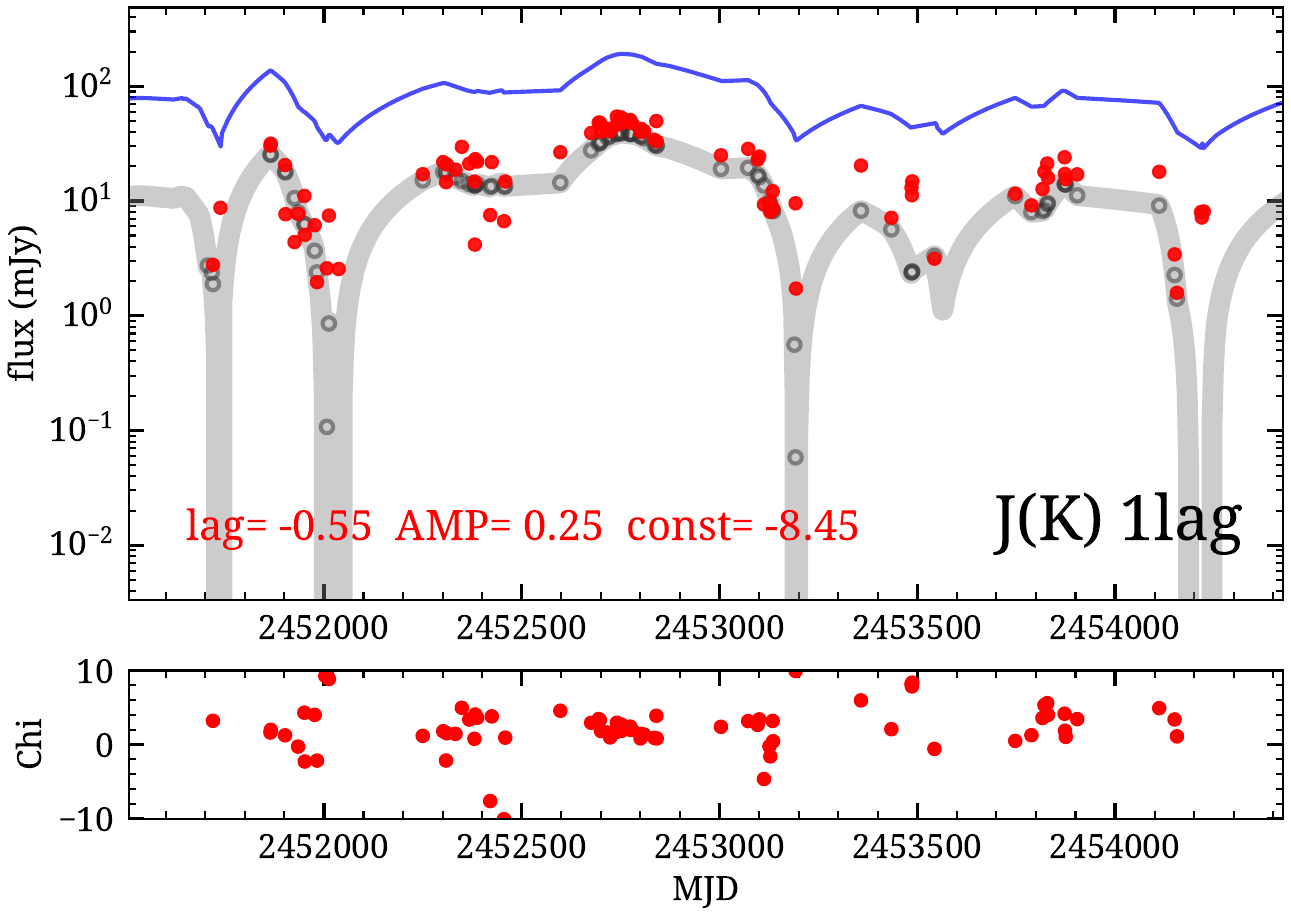}
\includegraphics[width=0.30\hsize]{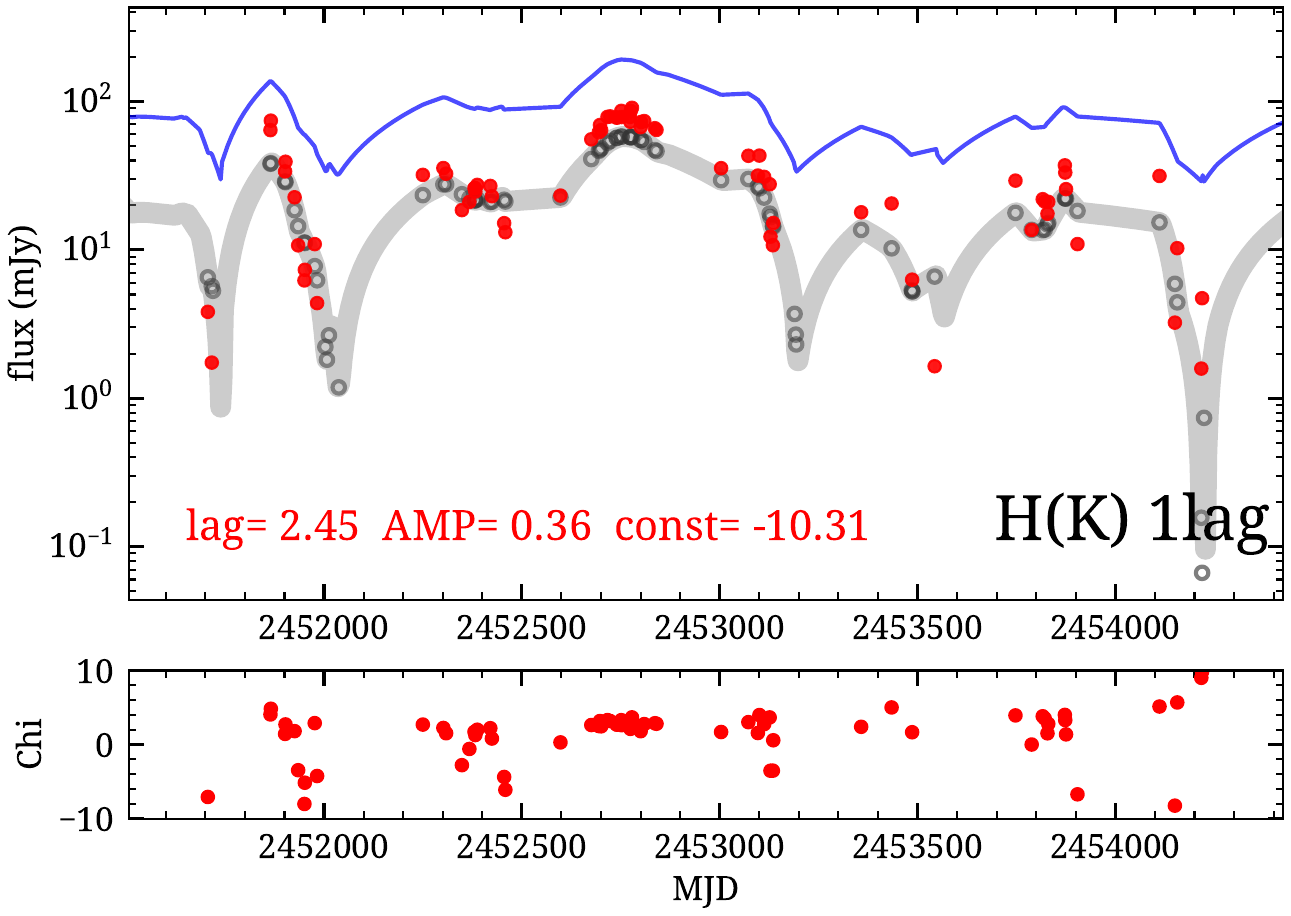}
\includegraphics[width=0.30\hsize]{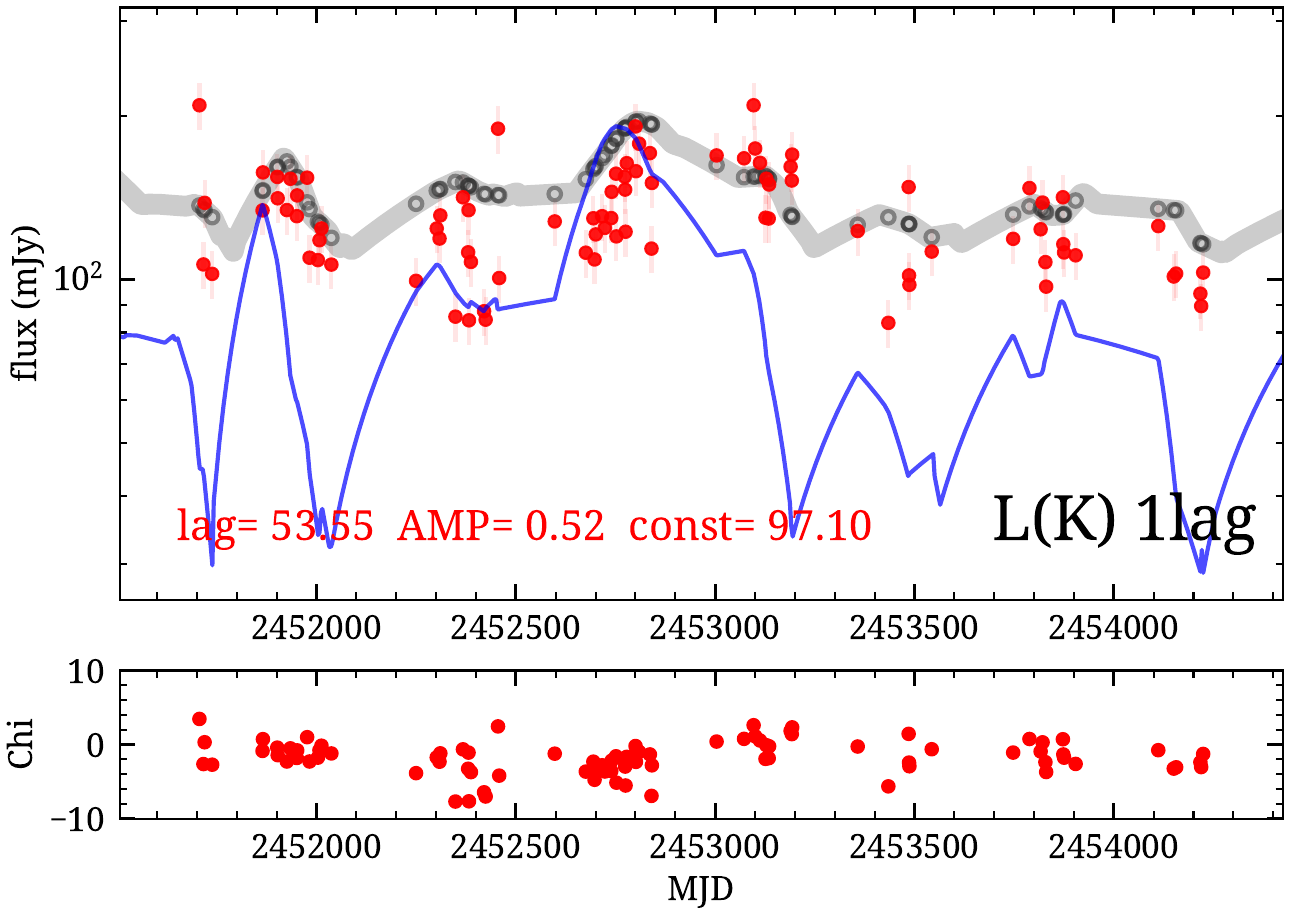}
\includegraphics[width=0.30\hsize]{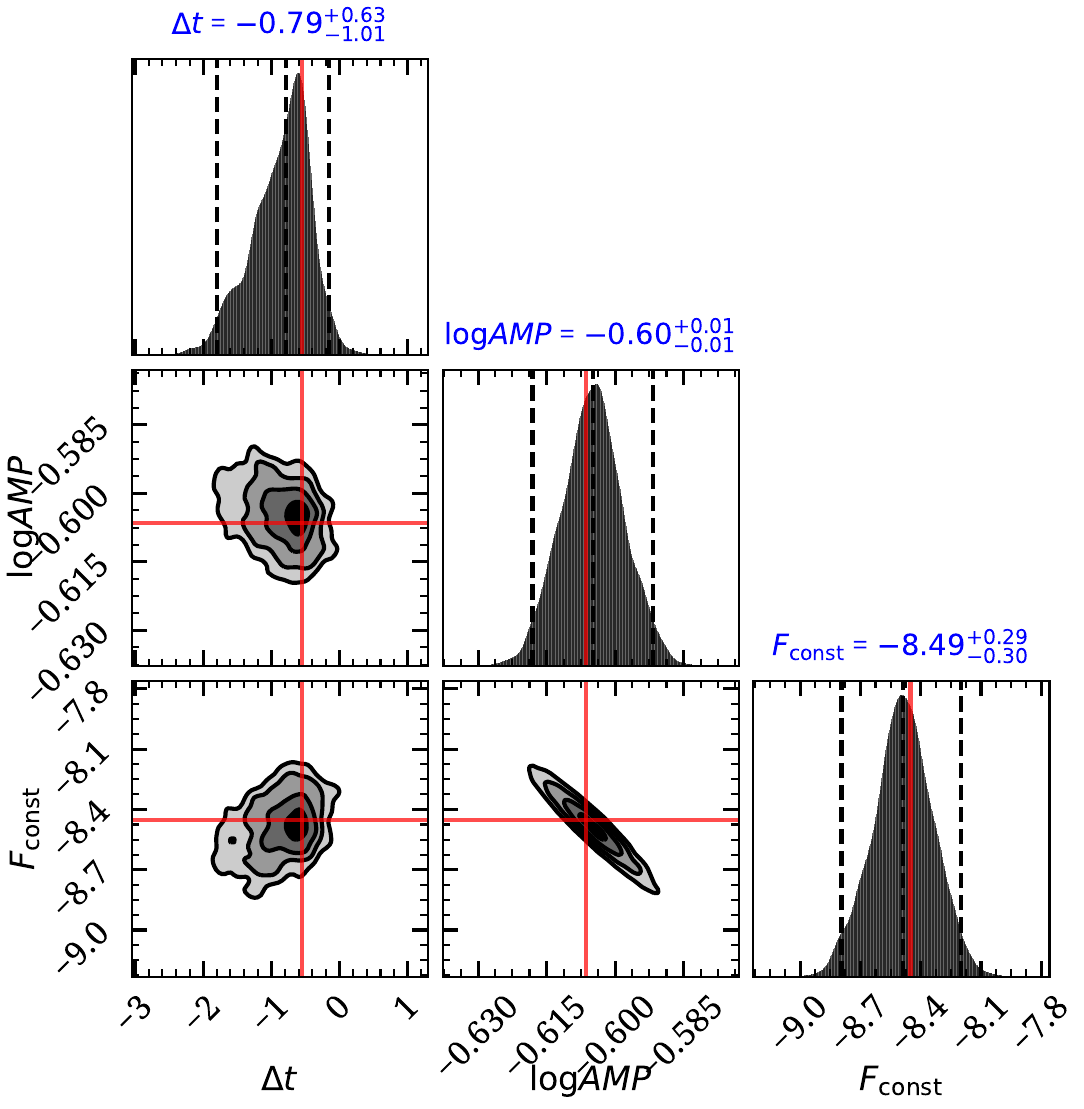}
\includegraphics[width=0.30\hsize]{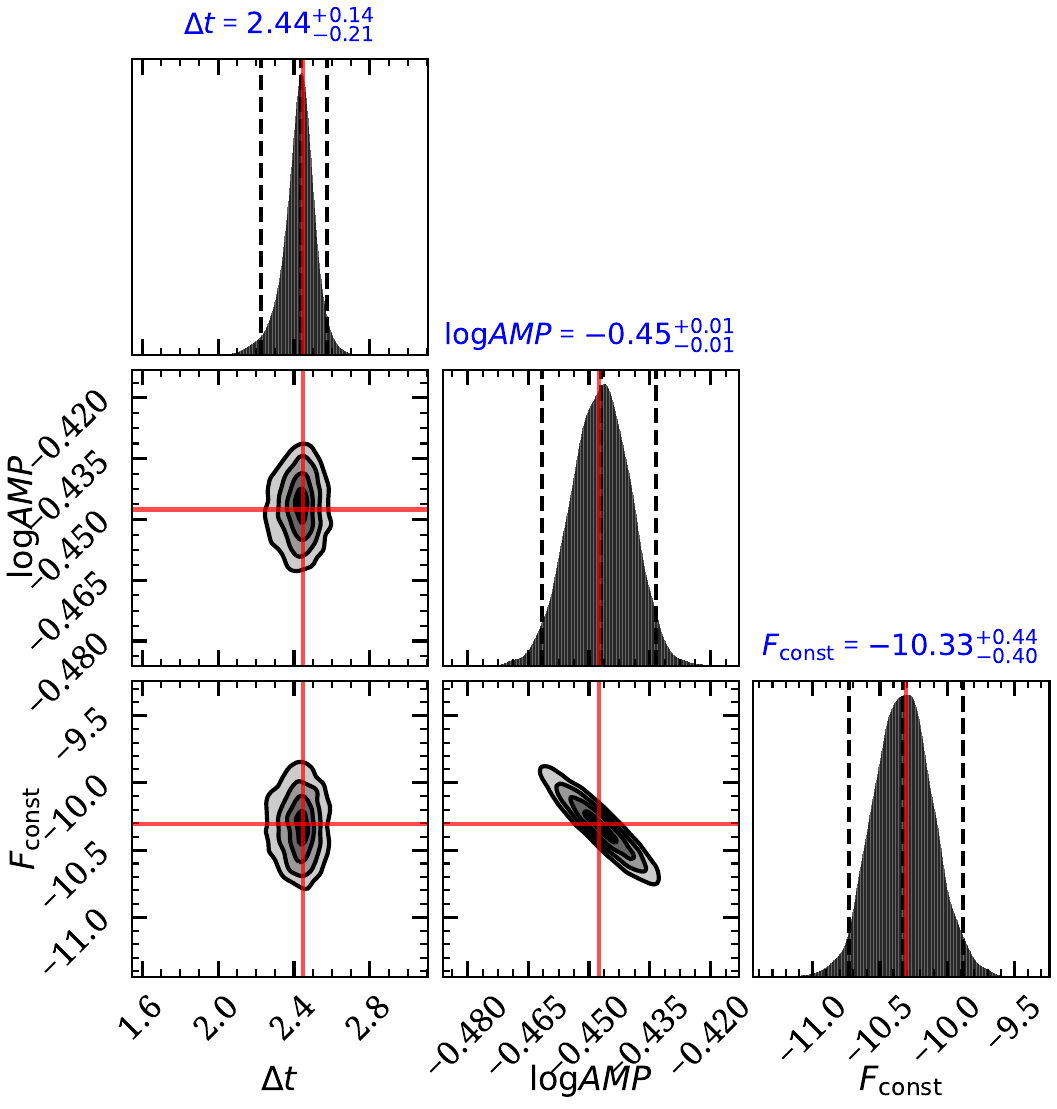}
\includegraphics[width=0.30\hsize]{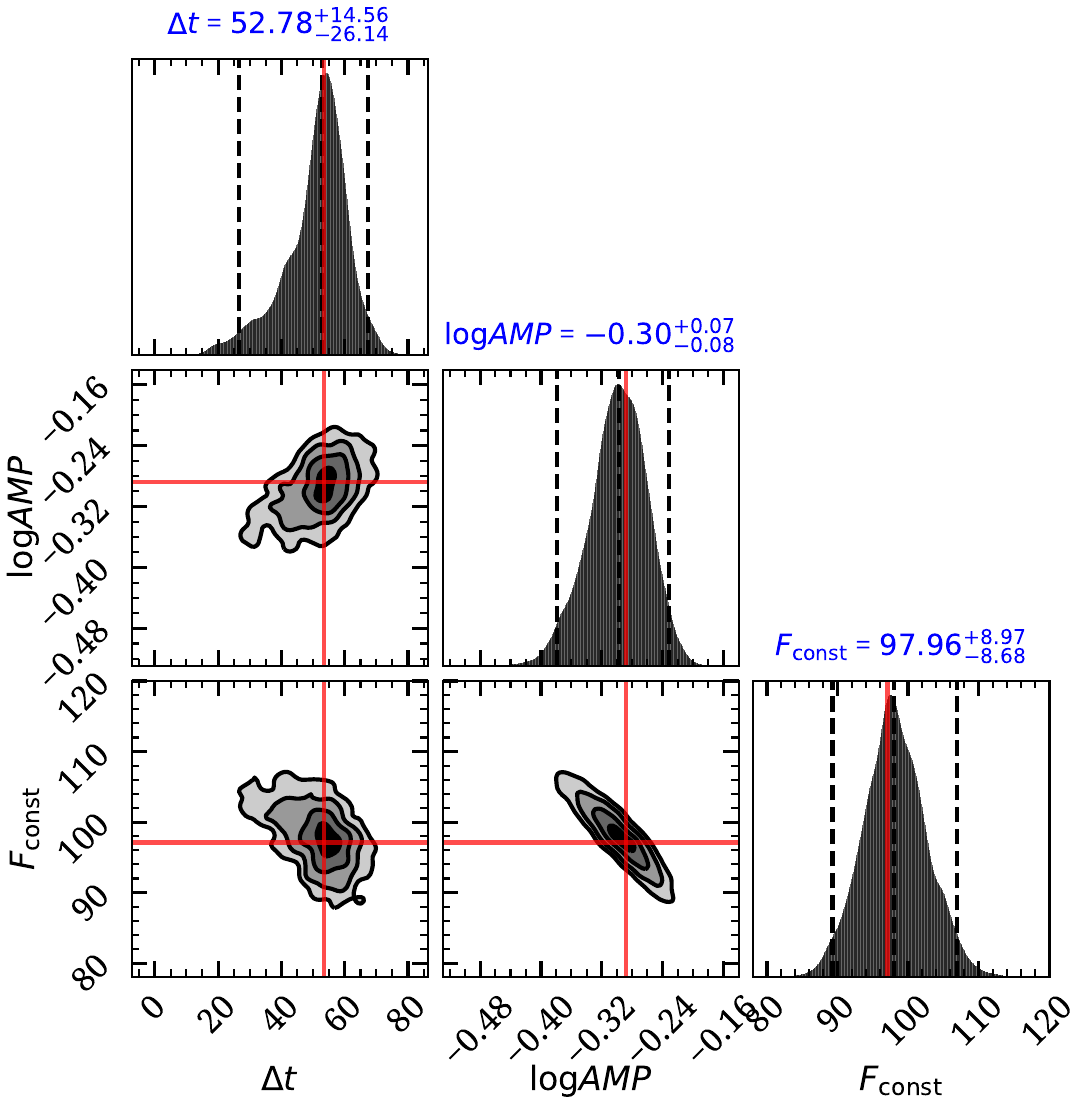}
\caption{Cross-correlation analysis between the J, H, and L light curves to the
	K-band light curve of NGC 4151 for Interval C. The results for J, H, L
	bands are shown from left to right. In the top panels, we show the J,
	K, L light curve data after subtracting the accretion disk variability
	(and the dust variability traced by K-band subtracted from the L band, based on a $\sim$1800 K black
	body spectrum) as red dots, the best-fit one-lag model as grey
	thick lines, and the K-band light curve interpolated by a DRW model
	after subtracting the accretion disk variability as blue thin lines. We
	denote the value of the best-fit parameters in the MAP
	sample in red. In the bottom panels, we present the marginalized
	posterior probability distributions of the fitting parameters, which are the lag time ($\Delta$ t), the amplitude of the relevant component (AMP), and a constant term (e.g., a non-variable component, $F_{const}$). On the
	top of each histogram, we denote the median value of the fitted
	parameters with 2-$\sigma$ ``uncertainties'' (i.e., 2.5\%, 50\% and
	97.5\% quantiles) that define the 95\% confidence intervals in blue. In the
	probability distribution plots, we use red lines to denote the best-fit
	values in the MAP sample.}
\label{fig:cc-event-C}
\end{figure*}

\begin{figure*}[htp]
\center
\includegraphics[width=0.30\hsize]{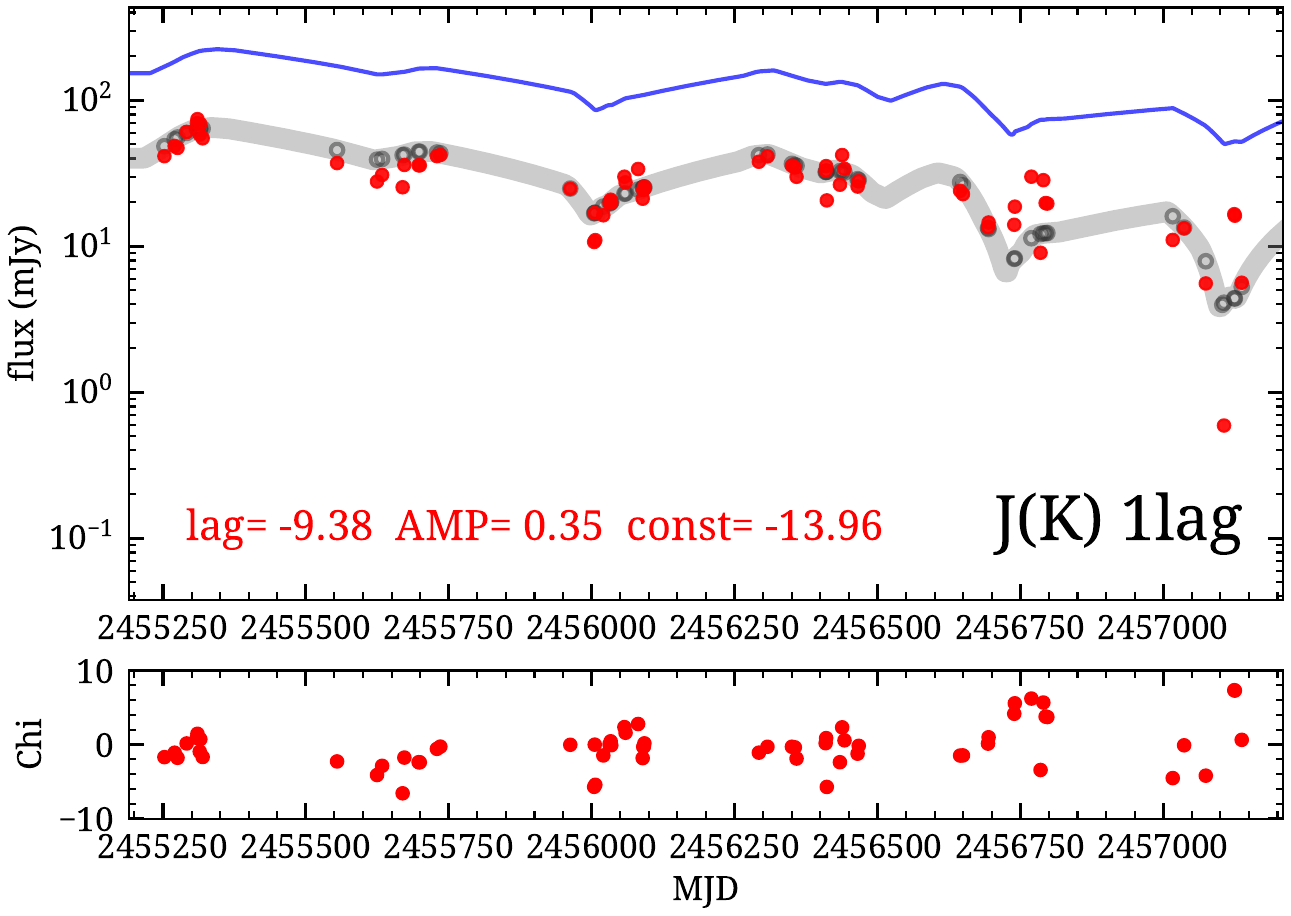}
\includegraphics[width=0.30\hsize]{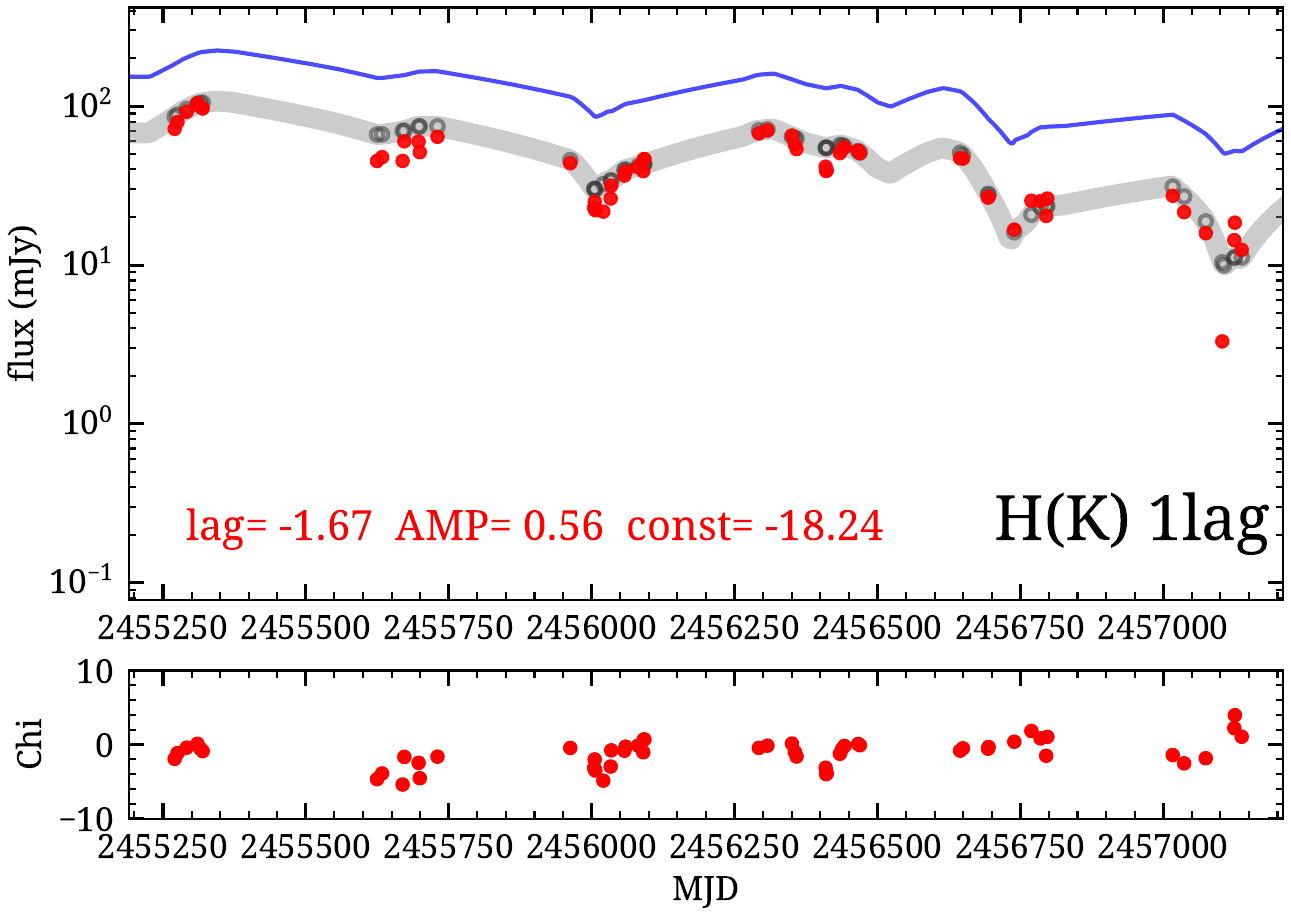}
\includegraphics[width=0.30\hsize]{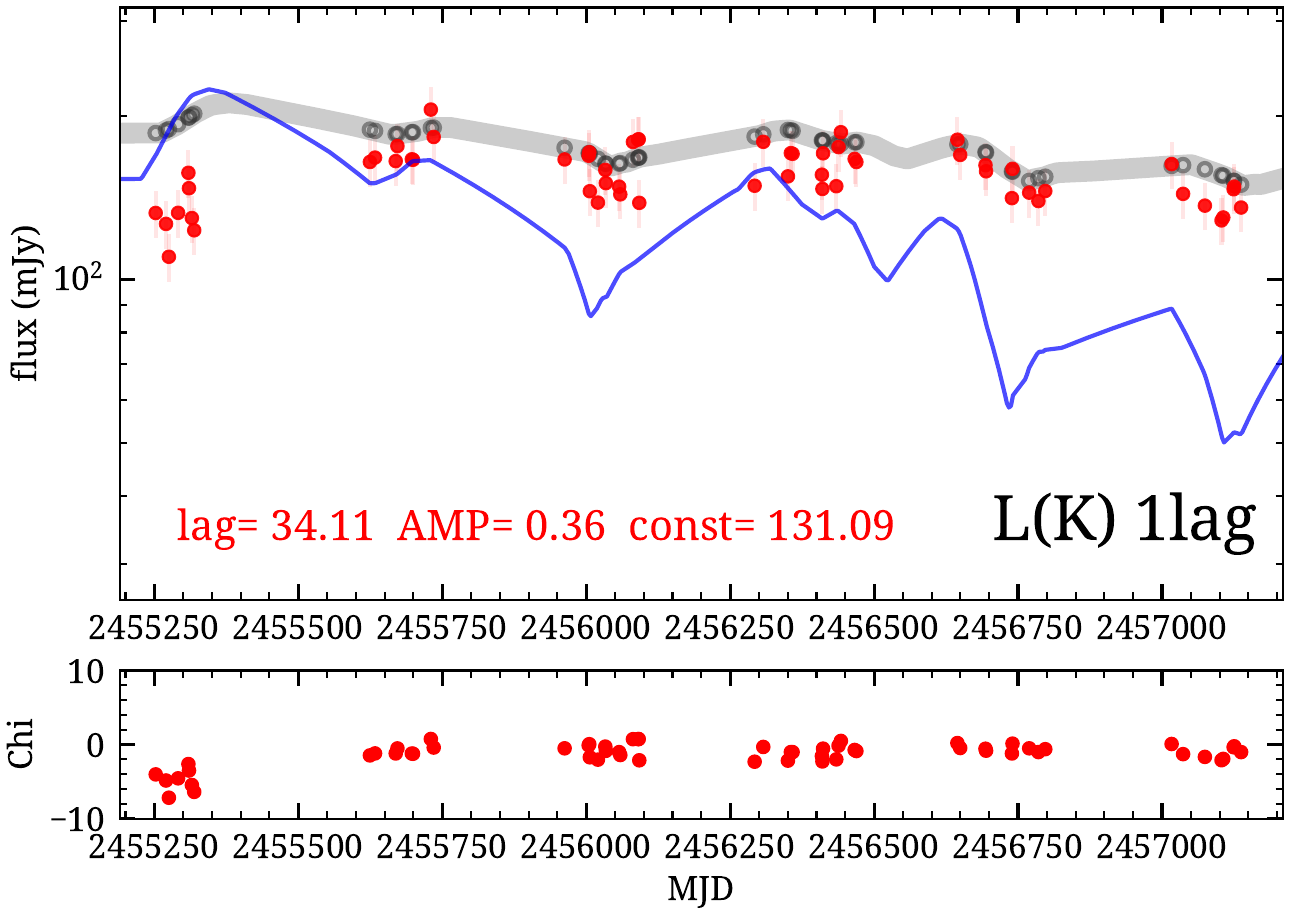}
\includegraphics[width=0.30\hsize]{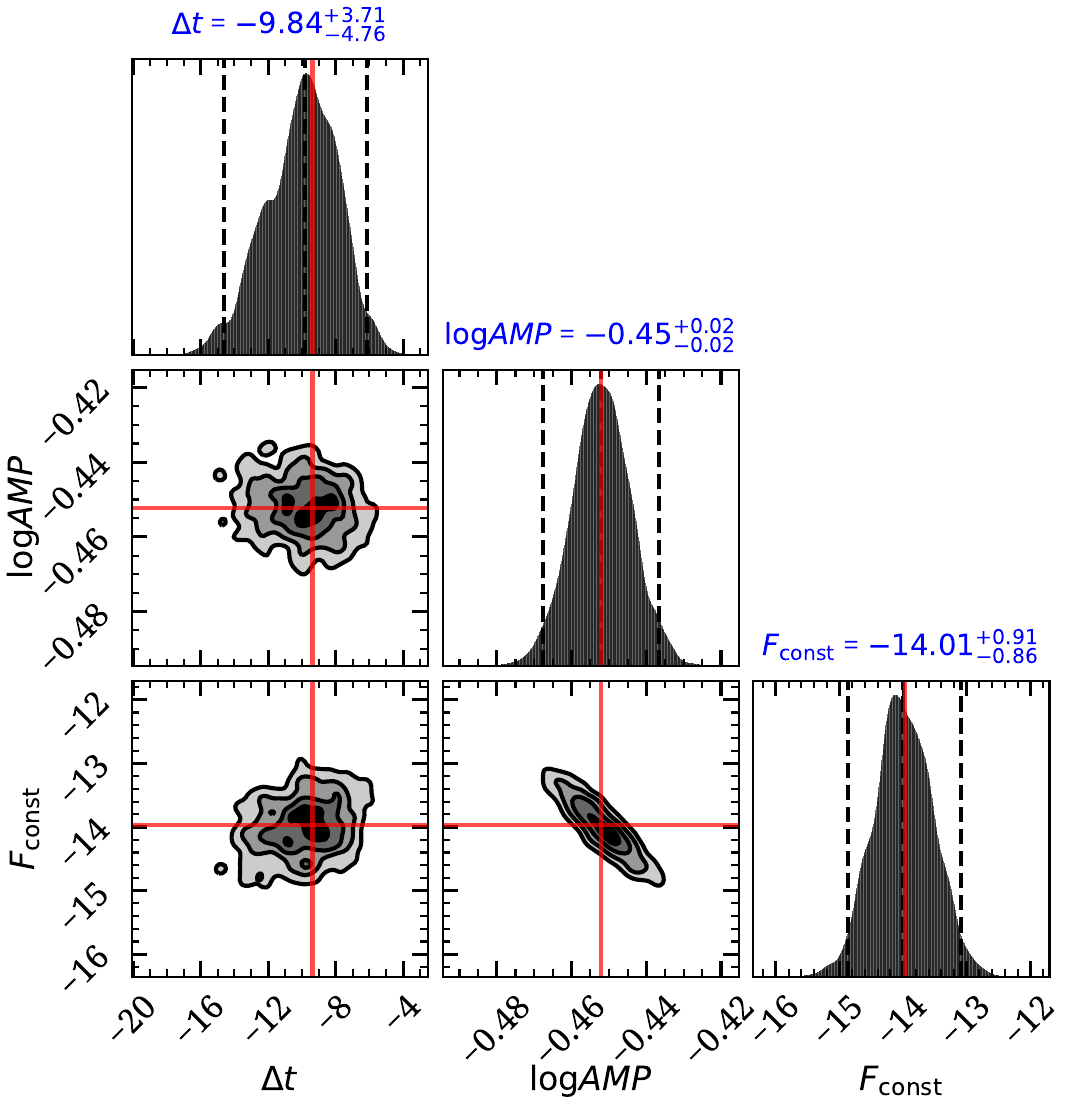}
\includegraphics[width=0.30\hsize]{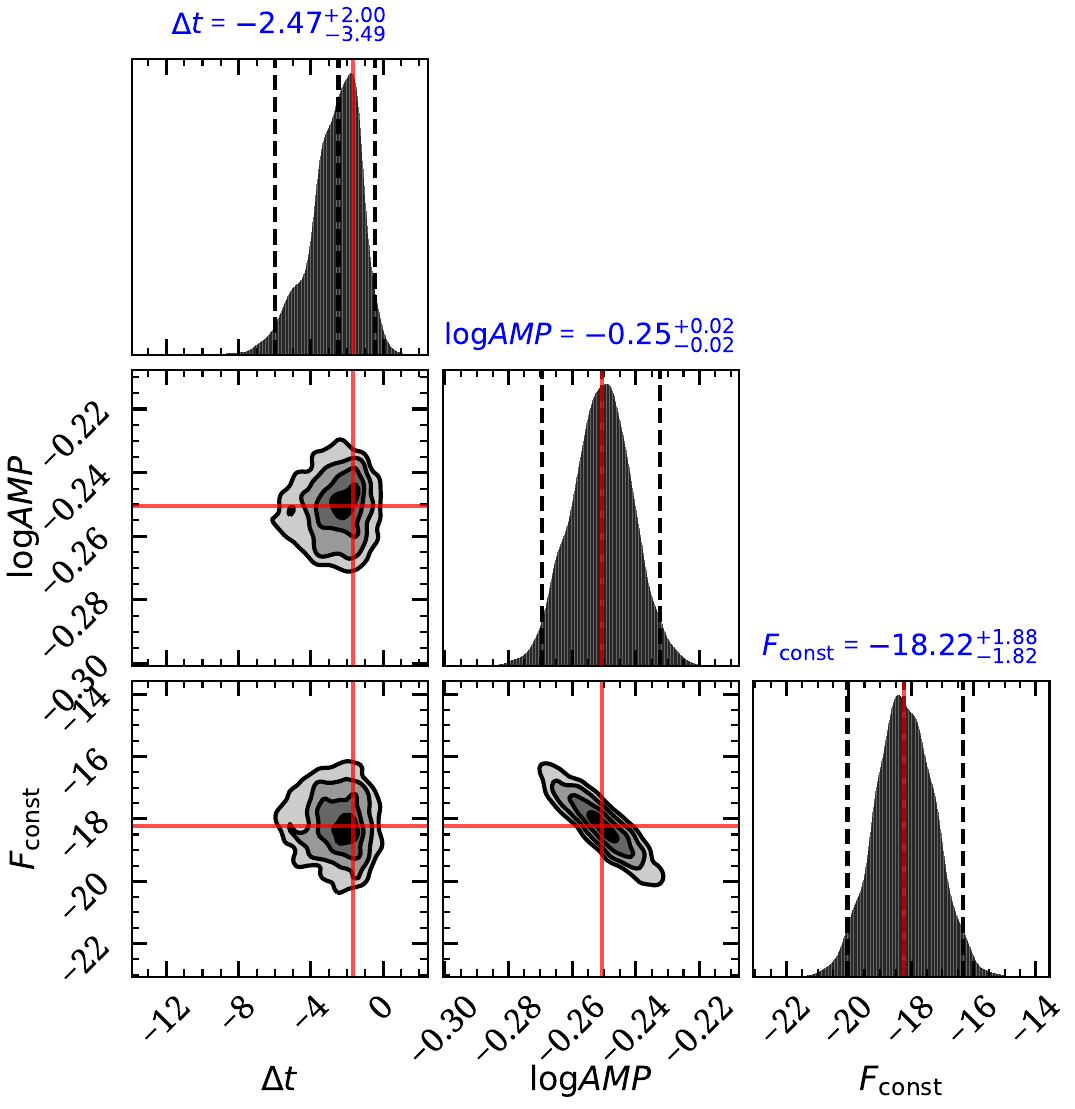}
\includegraphics[width=0.30\hsize]{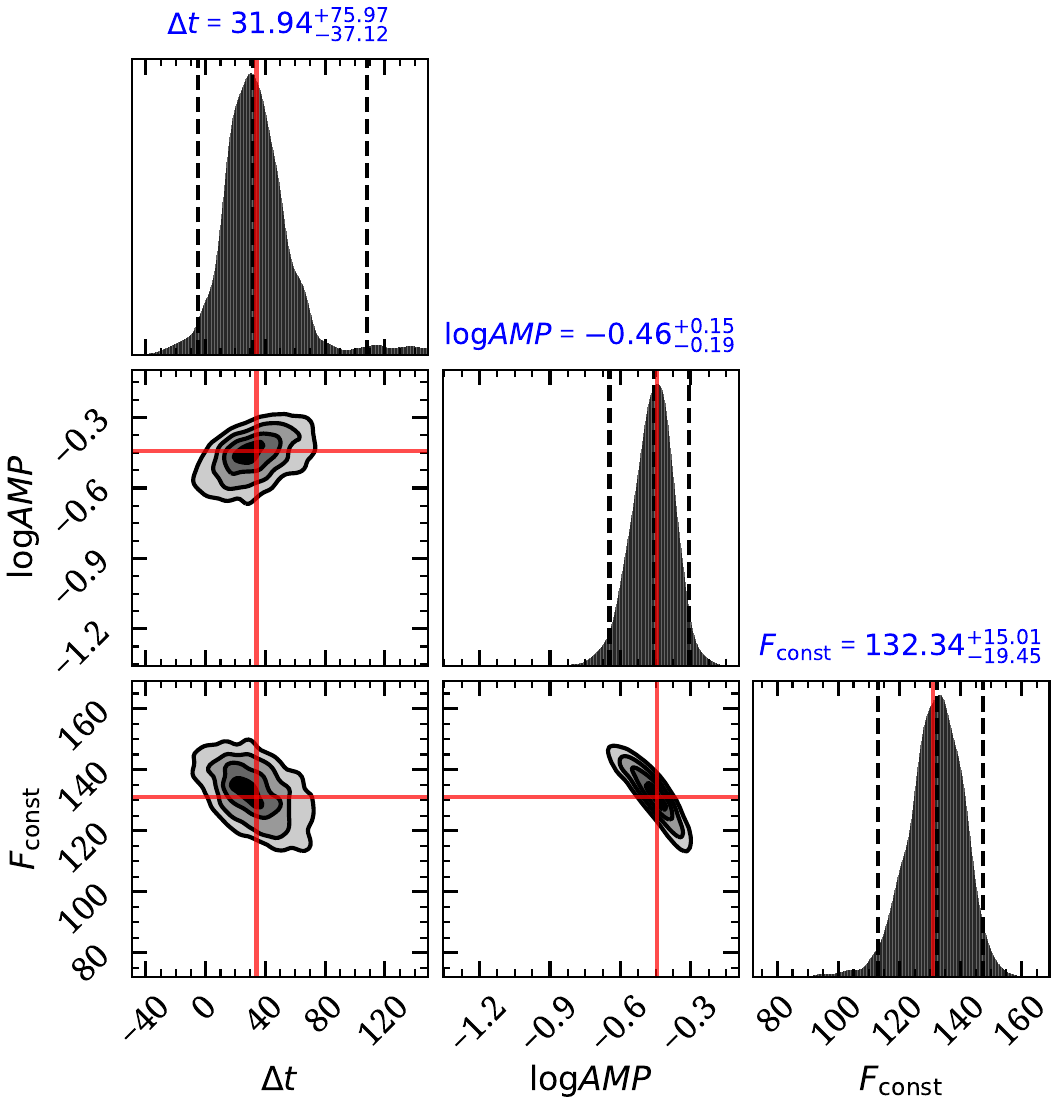}
\caption{Same as Figure~\ref{fig:cc-event-C}, but for Interval D. }
\label{fig:cc-event-D}
\end{figure*}

In this part of the appendix, we present the outputs of the {\it Dynesty} Bayesian fitting of the various time lags. These parameters are indicated in appropriate tables in the main text. Figure~\ref{fig:cc-event-C} shows the analysis of the J, H, and L light curves in Interval C relative to the K one.  Figure~\ref{fig:cc-event-D} shows similar results for Interval D. Although these latter fits are of lower weight than Interval C due to poorer sampling, they are consistent with the findings for that period. In both cases, we  assumed priors for the time lags ($\Delta t/{\rm day}=[-50, 350]$). The following two figures illustrate the fits for the infrared lag relative to the B-band light curve, where we assumed the additional prior of the relative optical-to-IR variability amplitude ($\log{\rm AMP} = [-2, 3]$) for the two lags. 

\begin{figure*}[htp]
\center
\includegraphics[height=0.28\hsize]{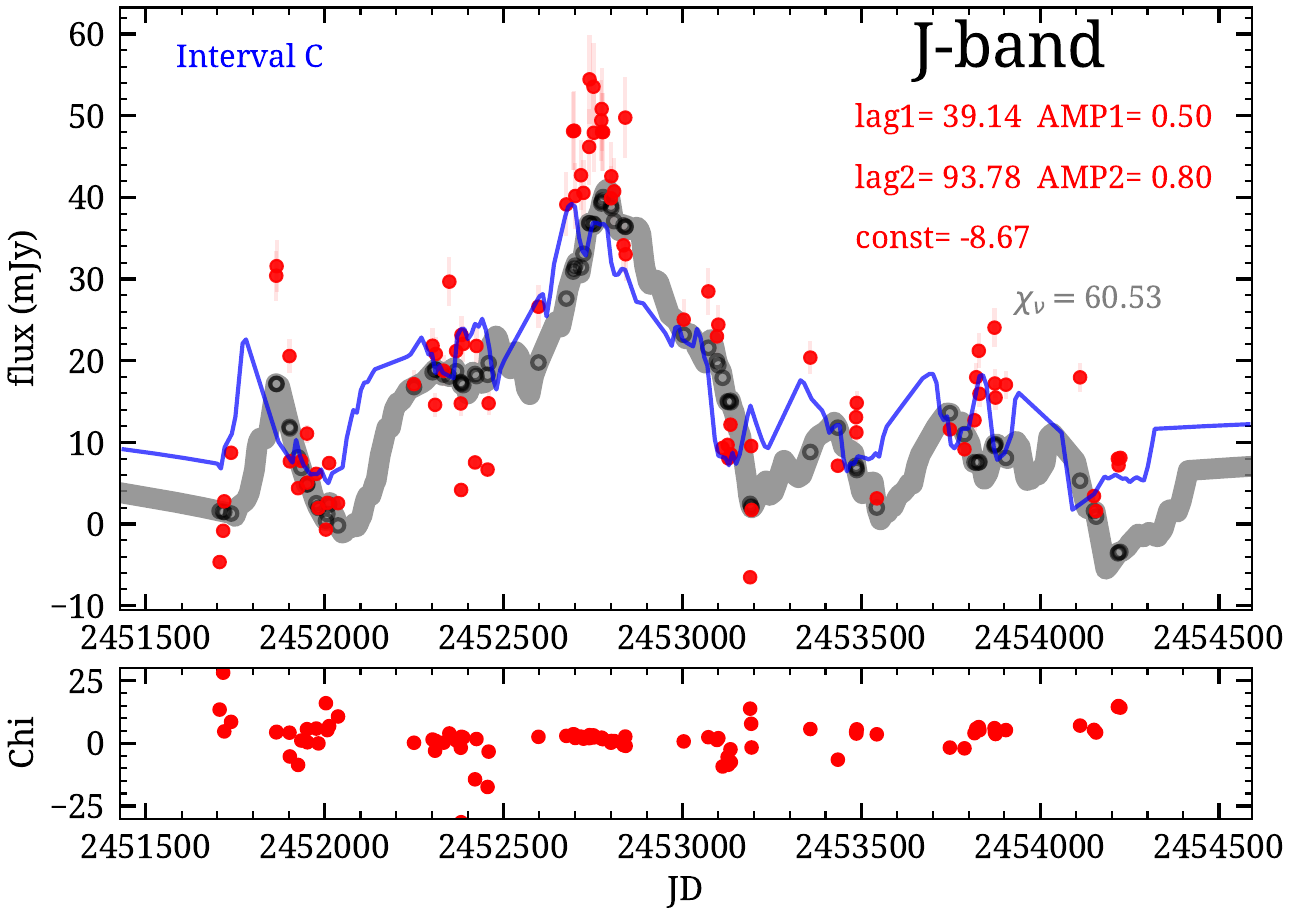}
\includegraphics[height=0.28\hsize]{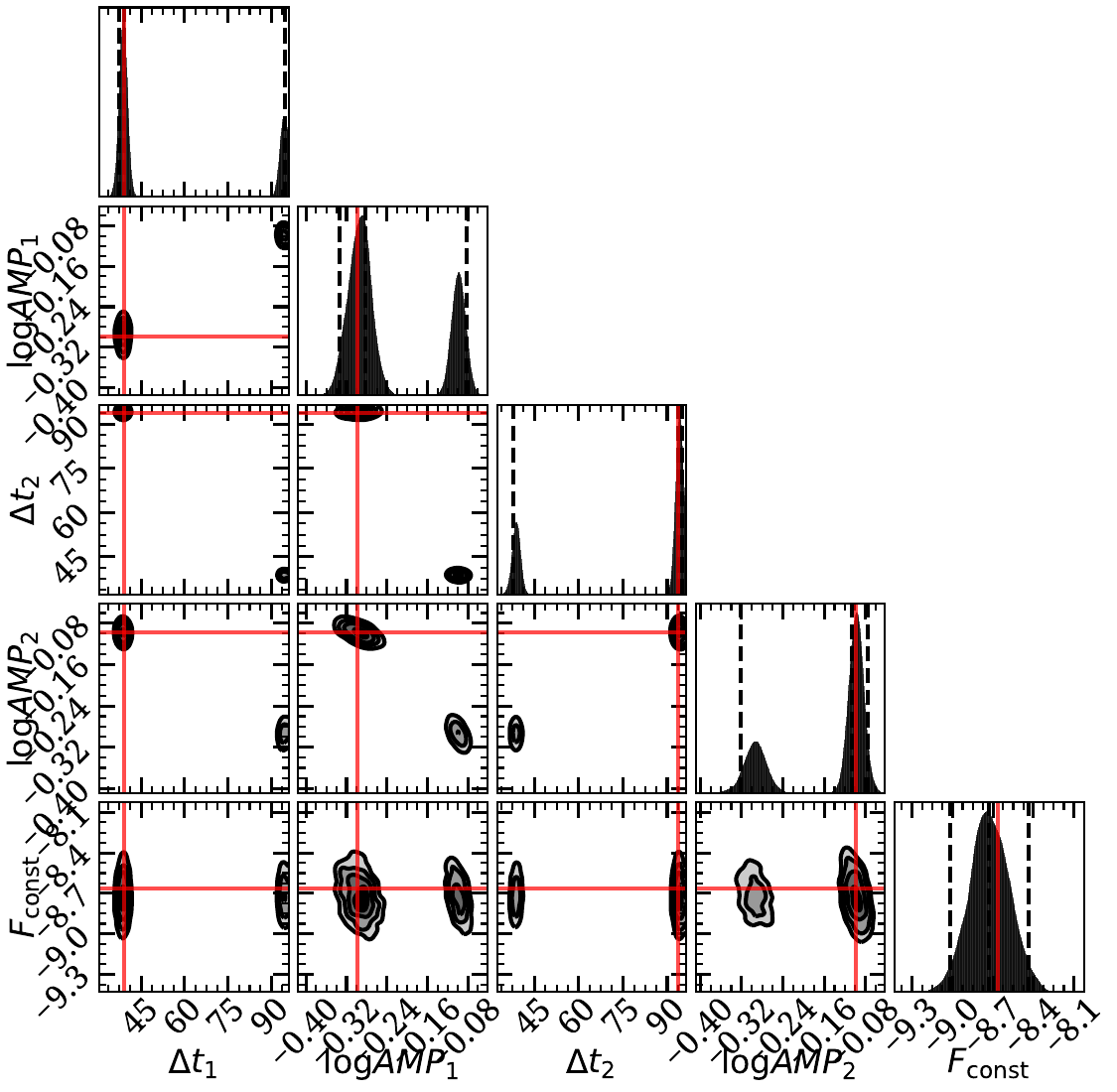}
\includegraphics[height=0.28\hsize]{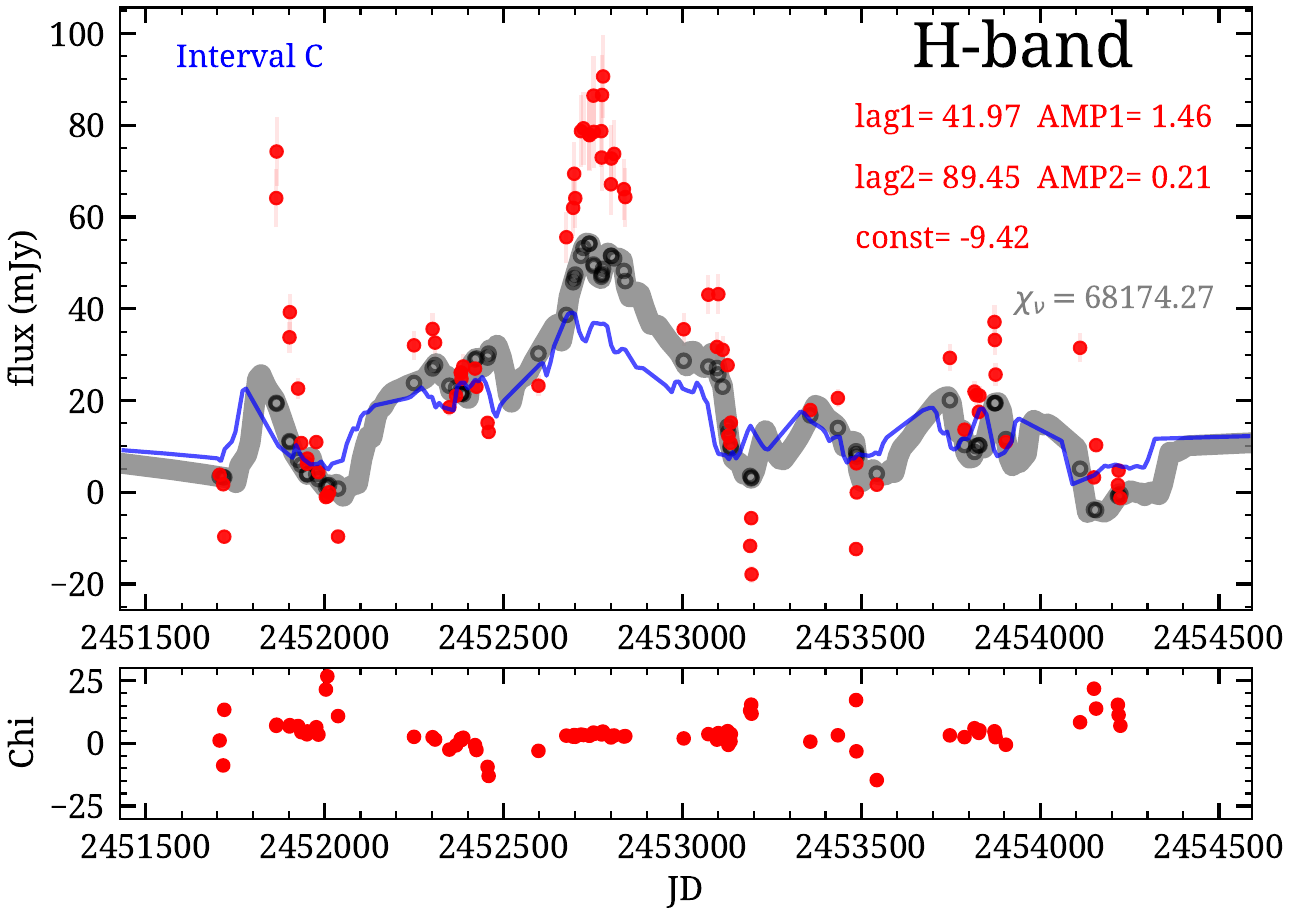}
\includegraphics[height=0.28\hsize]{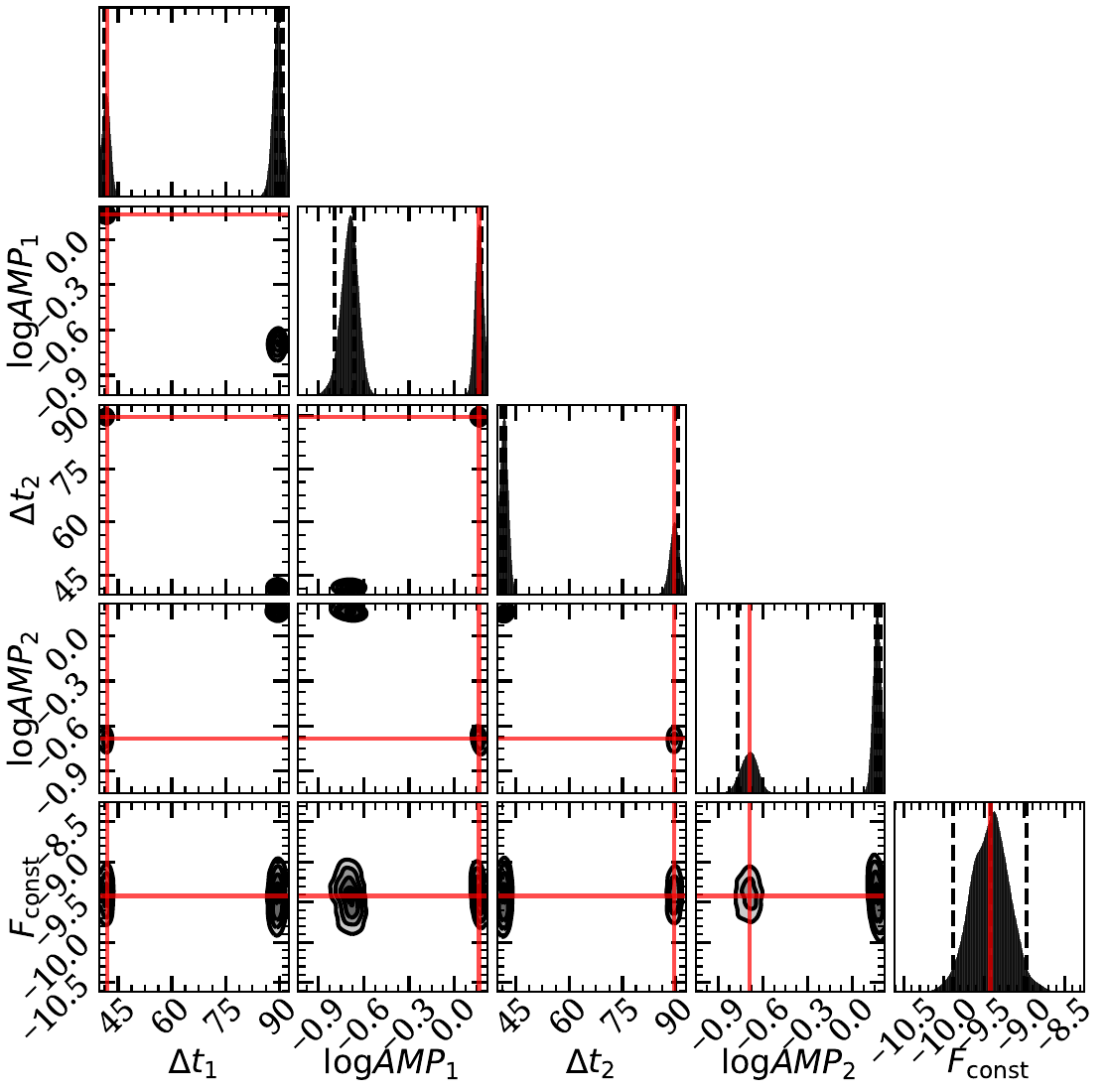}
\includegraphics[height=0.28\hsize]{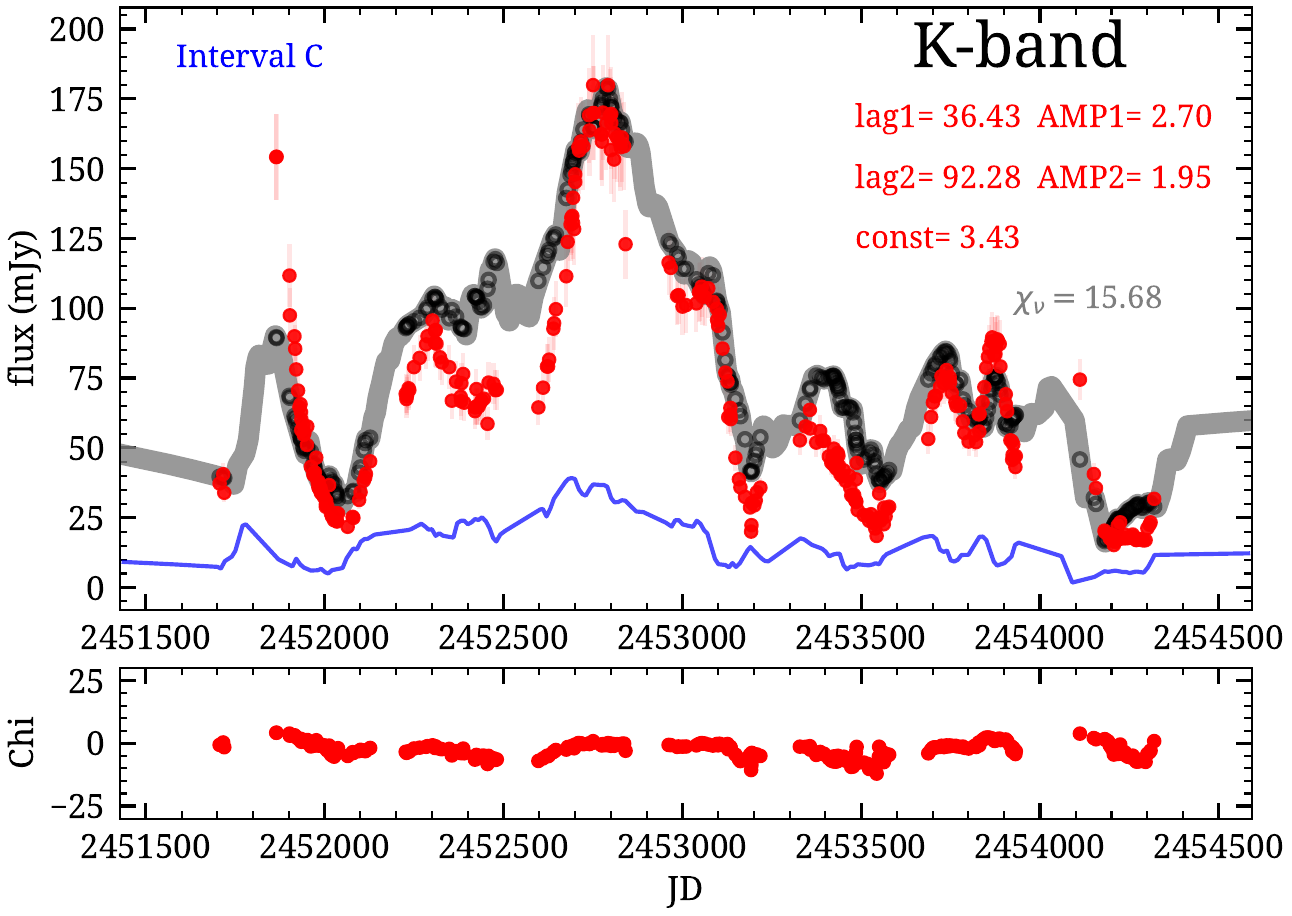}
\includegraphics[height=0.28\hsize]{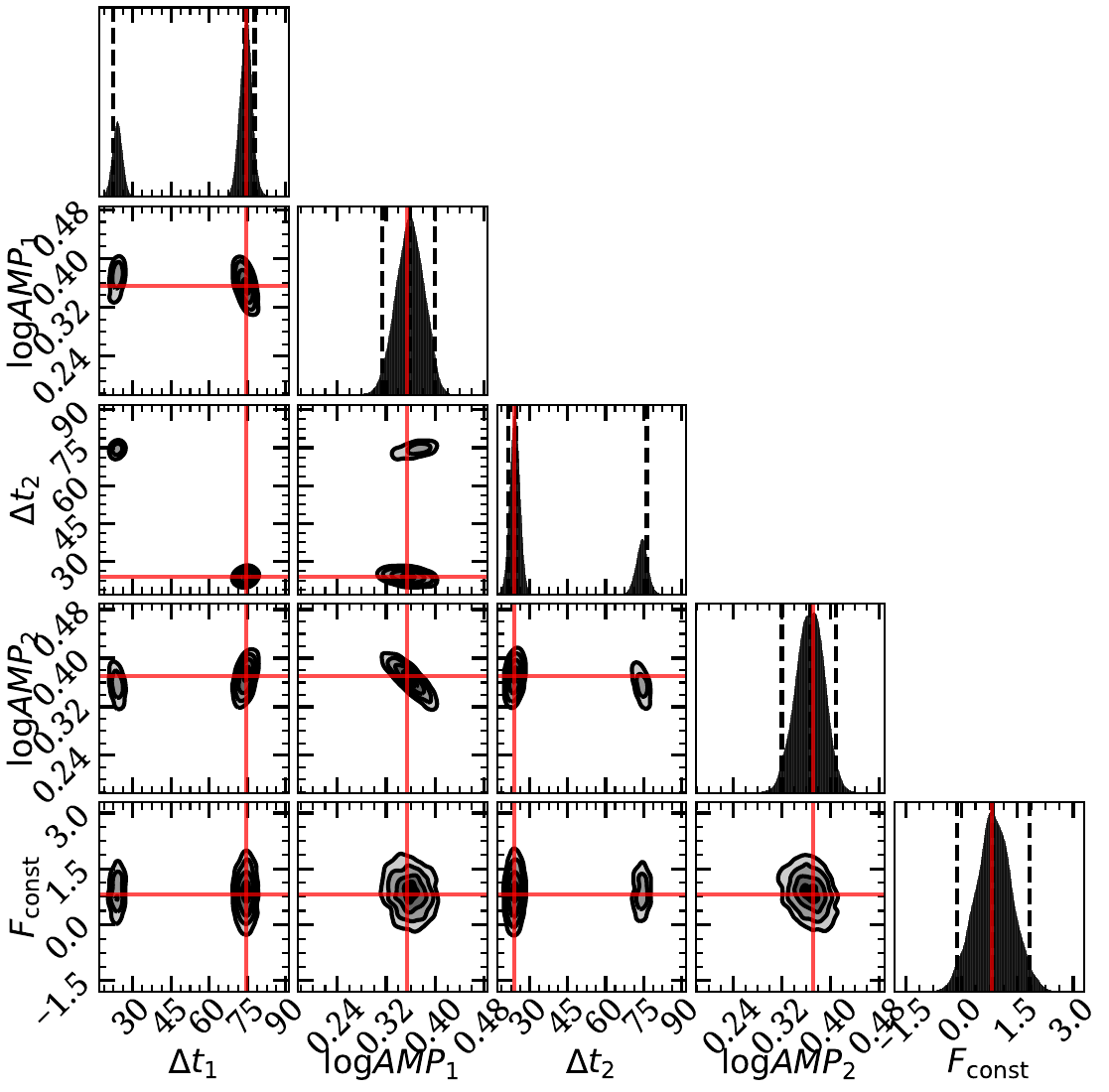}
\includegraphics[height=0.28\hsize]{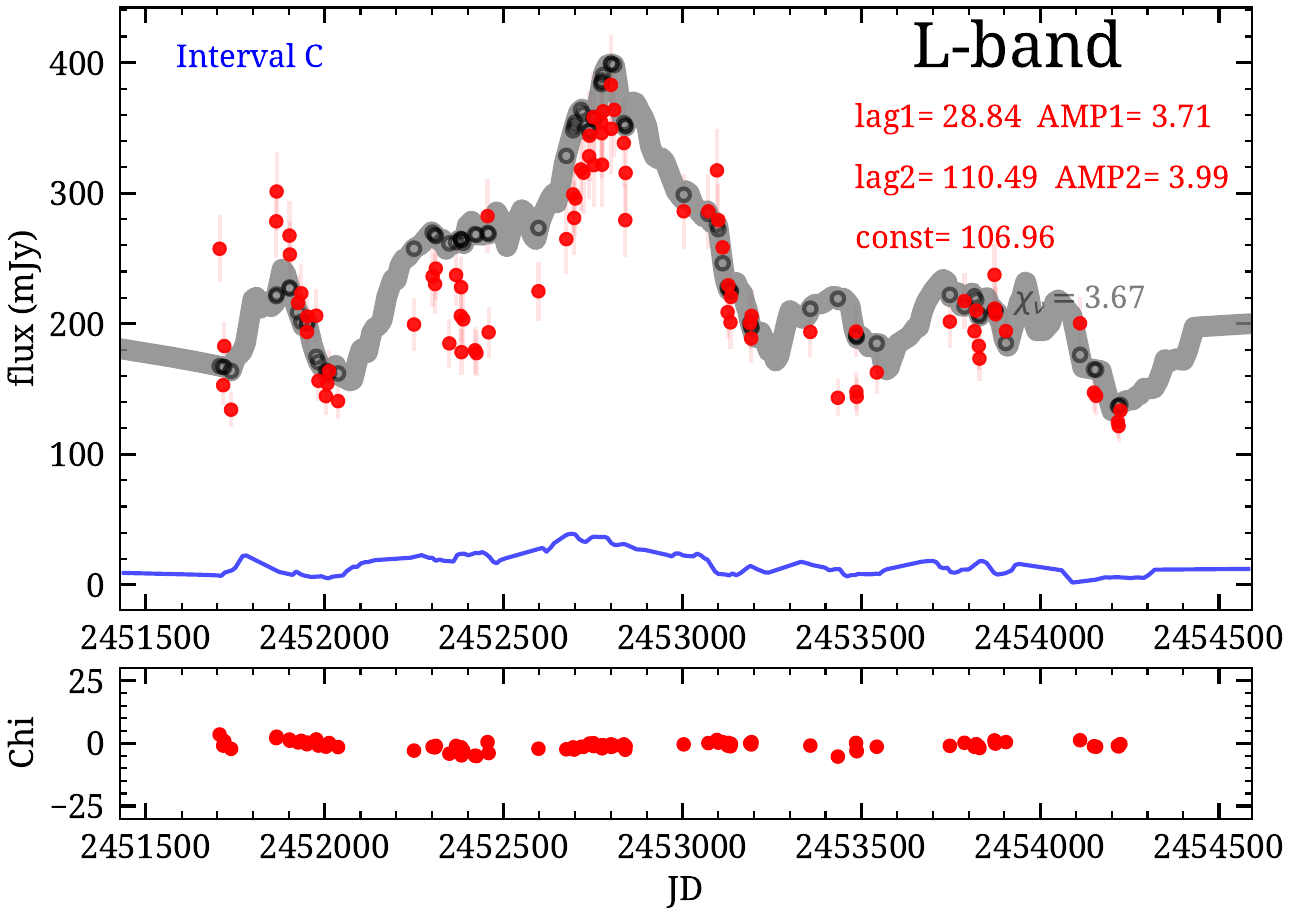}
\includegraphics[height=0.28\hsize]{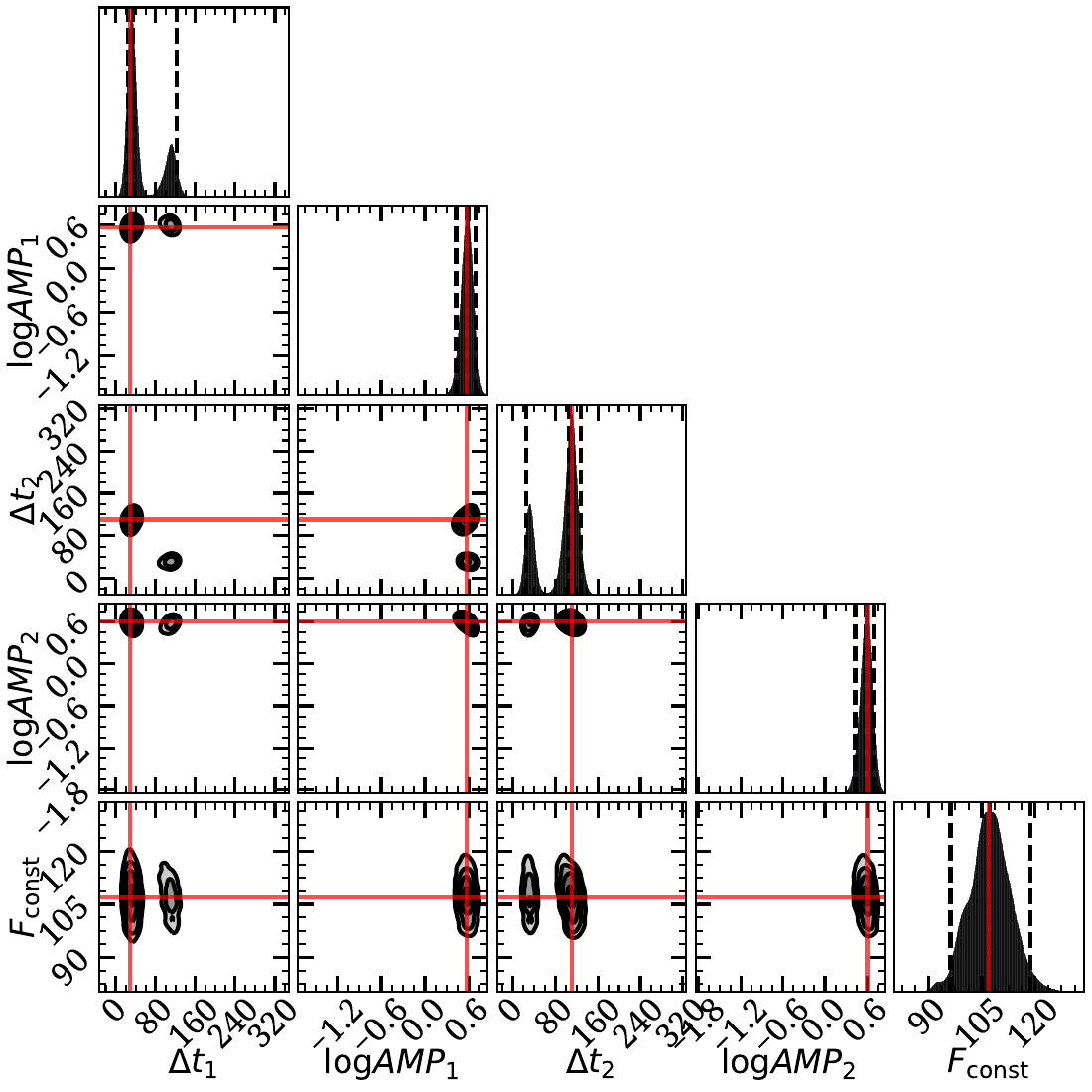}
\caption{Two-lag dust reverberation model fitting results of individual IR
	light curves of NGC 4151 during Interval C (2000/06--2007/05).  In the
	left panels, we show the IR light curve data after subtracting the
	accretion disk variability as red dots, the best-fit two-lag dust
	reverberation model as grey thick lines, and the B-band light curve
	interpolated by a DRW model as blue solid lines. We denote the value of
	the best-fit parameters in the MAP sample in red. In
	the right panels, we present the marginalized posterior probability
	distributions of the fitting parameters. In the probability distribution plots
	and histogram plots,
	we use red lines to denote the best-fit values in the MAP sample.
}
\label{fig:event-c}
\end{figure*}

 Figure~\ref{fig:event-c} shows the {\it Dynesty} results for the lag in Interval C of the individual infrared bands relative to the B-band.

\end{document}